\documentclass[showpacs,preprintnumbers, amsmath,amssymb]{revtex4}
\usepackage{morefloats}
\usepackage{psfrag}
\usepackage{graphicx}
\usepackage{subfigure}
\def\be{\begin{equation}}       \def\ee{\end{equation}}
\def\bea{\begin{eqnarray}}      \def\eea{\end{eqnarray}}

\begin{document}

\title{Persistence of entanglement in thermal states of spin systems}

\author{Gehad Sadiek \footnote{Corresponding author: gehad@ksu.edu.sa}}
\affiliation{Department of Physics, King Saud University, Riyadh 11451, Saudi Arabia and\\
Department of Physics, Ain Shams University, Cairo 11566, Egypt}
\author{Sabre Kais}
\affiliation{Department of Chemistry and Birck Nanotechnology center,
Purdue University, West Lafayette, Indiana 47907, USA}
 
\begin{abstract}
We study and compare the persistence of bipartite entanglement (BE) and multipartite
entanglement (ME) in one-dimensional and two-dimensional spin XY models in an external
transverse magnetic field under the effect of thermal excitations. We compare the threshold
temperature at which the entanglement vanishes in both types of entanglement. We use the
entanglement of formation as a measure of the BE and the geometric measure to evaluate the
ME of the system. We have found that in both dimensions in the anisotropic and partially
anisotropic spin systems at zero temperatures, all types of entanglement decay as the magnetic
field increases but are sustained with very small magnitudes at high field values. Also we
found that for the same systems, the threshold temperatures of the nearest neighbour (nn) BEs
are higher than both of the next-to-nearest neighbour BEs and MEs and the three of them
increase monotonically with the magnetic field strength. Thus, as the temperature increases,
the ME and the far parts BE of the system become more fragile to thermal excitations
compared to the nn BE. For the isotropic system, all types of entanglement and threshold
temperatures vanish at the same exact small value of the magnetic field. We emphasise the
major role played by both the properties of the ground state of the system and the energy gap
in controlling the characteristics of the entanglement and threshold temperatures. In addition,
we have shown how an inserted magnetic impurity can be used to preserve all types of
entanglement and enhance their threshold temperatures. Furthermore, we found that the
quantum effects in the spin systems can be maintained at high temperatures, as the different
types of entanglements in the spin lattices are sustained at high temperatures by applying
sufficiently high magnetic fields.
\end{abstract}


\maketitle

\section{Introduction}
The state of a classical composite system is described in the phase space as a product of its individual constituents separate states whereas the state of a composite quantum system is expressed in the Hilbert space as a superposition of tensor products of its individual subsystems states. Therefore the state of a quantum composite system is not necessarily expressible as a product of the individual quantum subsystems states. This peculiar property of quantum systems is called Entanglement, which has no classical analog \cite{Peres1993}. Recently the interest in studying quantum entanglement was sparked by the development in the fields of quantum information and quantum computing which was initiated in the eighties by the pioneering work of Benioff, Bennett, Deutsch, Feynman and Landauer \cite{Benioff1981,Benioff1982, Bennett1985, Deutsch1985,Deutsch1989, Feynman1982, Landauer1961}. 
Although there is still no complete theory that can quantify entanglement of a general multipartite system in pure or mixed state, there are few cases where we have successful entanglement measures. Most importantly, bipartite system in a pure state and mixed state of two spin $1/2$ possess such measures, also the pure and mixed multipartite systems using geometric measures such as geometric entanglement and relative entanglement\cite{Nielsen2000-book, HorodeckiM2001, HorodeckiP2001, Wootters2001-2, Vedral1998, Wei2003}.
Quantum information processing and quantum computations can only be performed in a many body system with very complicated arrangements concerning the properties of that system \cite{Nielsen2000}. The building unit, smallest for storing information in such a system (qubit), has to be a well defined two state quantum entity that can be easily addressed, manipulated and readout. The basic idea is to define certain quantum degree of freedom to serve
as a qubit, such as the charge, orbital or spin angular momentum. The next step is to define a controllable mechanism to form coupling between two individual qubits in such a way to produce a fundamental quantum computing gate. Furthermore, we have to be able to coherently manipulate such a mechanism to provide an efficient computational process.
On the other hand, quantum phase transitions in many body systems are accompanied by a significant change in the quantum correlations within the system, which led to a great interest in investigating the behavior of quantum entanglement close to the critical points of transitions \cite{Sachdev2001,Osborne2002,Wei2005,Sadiek2010}.

All these facts and developments sparked great interest in studying entanglement properties in many body systems in general and particularly in quantum spin systems in presence of external magnetic fields at zero and finite temperatures \cite{Amico2008, Latorre2009}. There has been special focus on studying entanglement in one-dimensional spin chains, utilizing the possession of exact analytic solution for many of these systems \cite{Kurmann1982, Osborne2002, Roscilde2004, Amico2006, Patane2007, Sadiek2010, Alkurtass2011}. The raised question of the multipartite entanglement (ME) versus bipartite entanglement (BE) and whether they have to coexist and which one is the actual resource for the critical behavior in many body systems has stimulated many investigations. 
To address this problem several works have focused on comparing ME with BE in quantum spin systems. Some of these works made use of the one tangle \cite{Coffman2000, Amico2004} as well as the concurrence \cite{Wootters1998} for that purpose without explicitly evaluating the global entanglement in the system. The one tangle $\tau_1$ represents the entanglement between a single spin with the rest of the system at zero temperature, which is equal to $4 \: det \: \rho^{(1)}$, where $\rho^{(1)}$ is the single site reduced density matrix. On the other hand, the sum of the squared of pairwise concurrences, $\sum_{i\neq j} C_{i,j}^2$, defines another quantity $\tau_2$ representing the weight of the pairwise entanglements in the system. The ratio $R=\tau_2/\tau_1$ was introduced as a measure of the fraction of the total entanglement attributed to the pairwise correlations.

The quantification of the global multipartite entanglement in a many body system is a very hard task as it usually requires the solution of a big set of variational equations and the difficulty of the problem increases non-linearly with the dimension of the Hilbert space. Few different measures of global entanglement have been proposed, the most common among them are the relative entropy of entanglement \cite{Vedral1997,Vedral1998}, the robustness of entanglement \cite{Vidal1999}, polynomial measure \cite{Meyer2002} and the geometric measure \cite{Wei2003}. Particularly, the geometric measure determines the distance between the state under consideration and the closet product state in the Hilbert space. This measure has been used intensively to study the multipartite entanglement in many body systems and especially the one-dimensional spin chain systems utilizing the exact solutions that these systems have \cite{Wei2005, Markham2008, Tamaryan2009, Nakata2009, Hubener2009, Blasone2009, Wei2011}.

Natural systems of interest have strong interaction with its environment, which causes decoherence effects \cite{Zurek1991, Bacon2000}. Particularly, practical many body systems are required to function at finite temperatures, which means that the system will be exposed to thermal excitations and therefore its mixed thermal states should be fully studied and understood. Evaluating the density matrix of mixed thermal states of many body systems is very hard task due to the large size of the Hilbert space of the system in that case. Recently, so much attention has been directed to investigating ME versus BE in thermal states of many body systems and their relative robustness to temperature, exploring the feasibility of achieving hight temperature entangled states. The ME and BE properties in one dimensional $XYX$ spin model in an external field, using Monte Carlo simulation, were investigated \cite{Dmitriev2002}. It was shown that the system possesses a factorized ground state \cite{Kurmann1982} signalled by vanishing $\tau_1$ and $\tau_2$ and a quantum phase transition corresponding to an anomaly in the ratio $R$ in the form of a narrow minimum versus the magnetic field. This suggests that the pairwise correlations suffer a big loss across the quantum critical point in contrary to ME which dominates and as a result ME can be safely considered as the actual resource for the observed quantum phase transition. The minimum of the factor $R$ was suggested as an estimator of the quantum critical points. Also a class of one dimensional $XYZ$ spin systems, with different degrees of anisotropy, was shown to have factorizable ground states \cite{Amico2006} where the pairwise entanglement range diverges while approaching this separable states indicating a long range reshuffling of entanglement. At finite temperature, using $\tau_2$ and concurrence, it was demonstrated that the system may emerge from a separable state into a mixed thermal entangled state with no pairwise entanglement present, i.e. containing only multipartite entanglement. ME of a subsystem of three arbitrary spins in a 1D $XY$ spin chain in an external magnetic field was evaluated \cite{Patane2007}, using the negativity between one spin and the other two \cite{Vidal2002}, and compared with BE of each pair of these three spins, It was shown that ME enjoys a longer range compared with BE through the chain. At finite temperature, it was demonstrated that ME is more robust than BE for a block of three adjacent spins, where ME is still present though there is no pairwise entanglement left in the system. Quite few works have studied the quantification and behavior of global multipartite entanglement in thermal states of many body systems and mainly focused on systems possessing analytic solutions such as one dimensional spin chains (e.g. \cite{Markham2008, Nakata2009, Hide2012}). To overcome the difficulties of evaluating the global entanglement in the thermal mixed states of many body systems, there has been an approach to provide a transition temperature below which the multipartite entanglement is guaranteed in such systems based only on information about the ground state of the system and its partition function \cite{Markham2008}. Using this approach, the robustness of ME in thermal states of one-dimensional spin-1/2 $XY$ system was investigated and the threshold temperature for vanishing entanglement was estimated  \cite{Nakata2009}. It was demonstrated that the threshold temperature increases montonically with the magnetic field in the region of large values of the field. Due to the big computational difficulties, there is a big lack in investigations in two-dimensional (and higher) quantum systems with few notable exceptions \cite{Syljuasen2003, Roscilde2005, Zhou2008, Orus2009, Li2009}. These works have focused on studying entanglement in two-dimensional finite and infinite square lattices using Monte Carlo simulations or the projected entangled-pair states \cite{Murg2007} and used concurrence, one-tangle and fidelity to quantify multipartite entanglement and determine points of separable ground states and phase transitions at zero temperature. 

In this paper, we consider two different systems of finite number of spins, each in presence of an external transverse magnetic field in contact with a heat bath at temperature $T$. We provide an extensive investigation of a two-dimensional $XY$ spin-1/2-star model but also study a one dimensional $XY$ spin-1/2 chain for the sake of comparison with the two-dimensional system and the one-dimensional previous results as well. The number of spins in each system is 7 and the nearest neighbor spins are coupled through an exchange interaction $J$. We investigate and compare the bipartite and the global multipartite entanglement of both systems under the effect of an external transverse magnetic field, thermal excitations and different degrees of anisotropy. We use the entanglement of formation and geometric entanglement as measures of the bipartite and global multipartite entanglements respectively. We show that, for both cases, in the anisotropic and partially isotropic systems at zero temperature the multipartite and bipartite entanglements can be maintained at high magnetic field values where the nearest neighbor and multipartite entanglements assume very small values but still much higher than that of the next to nearest ones.
Also we demonstrate that the threshold temperature, at which the entanglement vanishes, is higher for the nearest neighbor entanglement compared to both of the next to nearest neighbor bipartite and multipartite entanglements. Therefore nearest neighbor bipartite entanglement is more robust to thermal excitations compared to the next to nearest bipartite and global multipartite entanglements in these systems. Also we demonstrated how an impurity can be used to tune and enhance the threshold temperature of all types of entanglement. We also examined the persistence of quantum effects at high temperatures by observing the entanglement behavior and show that we may maintain non zero entanglement at considerably high temperatures by applying strong enough magnetic fields.

This paper is organized as follows. In the next section we present our model. In sec. III we focus on the two-dimensional spin system and evaluate the bipartite entanglement and the thermal energy of the system. The multipartite entanglement and the threshold temperatures for all type of entanglements for the two-dimensional system are evaluated in sec. IV. In sec. V we study impurity effects. In Sec. VI we calculate and compare the bipartite and multipartite entanglements and the corresponding threshold temperatures in the one-dimensional system. We conclude in sec VII.
%%%%%%%%%%%%%%%%%%%%%%%%%%%%%%%%%%%%%%%%%%%%%%%%%%%%%%%%%%%%%%%%%%%%%%%%%
\section{The Model}
\begin{figure}[htbp]
\begin{minipage}[c]{\textwidth}
 \centering
   \subfigure{\label{fig:2D_system}\includegraphics[width=6 cm]{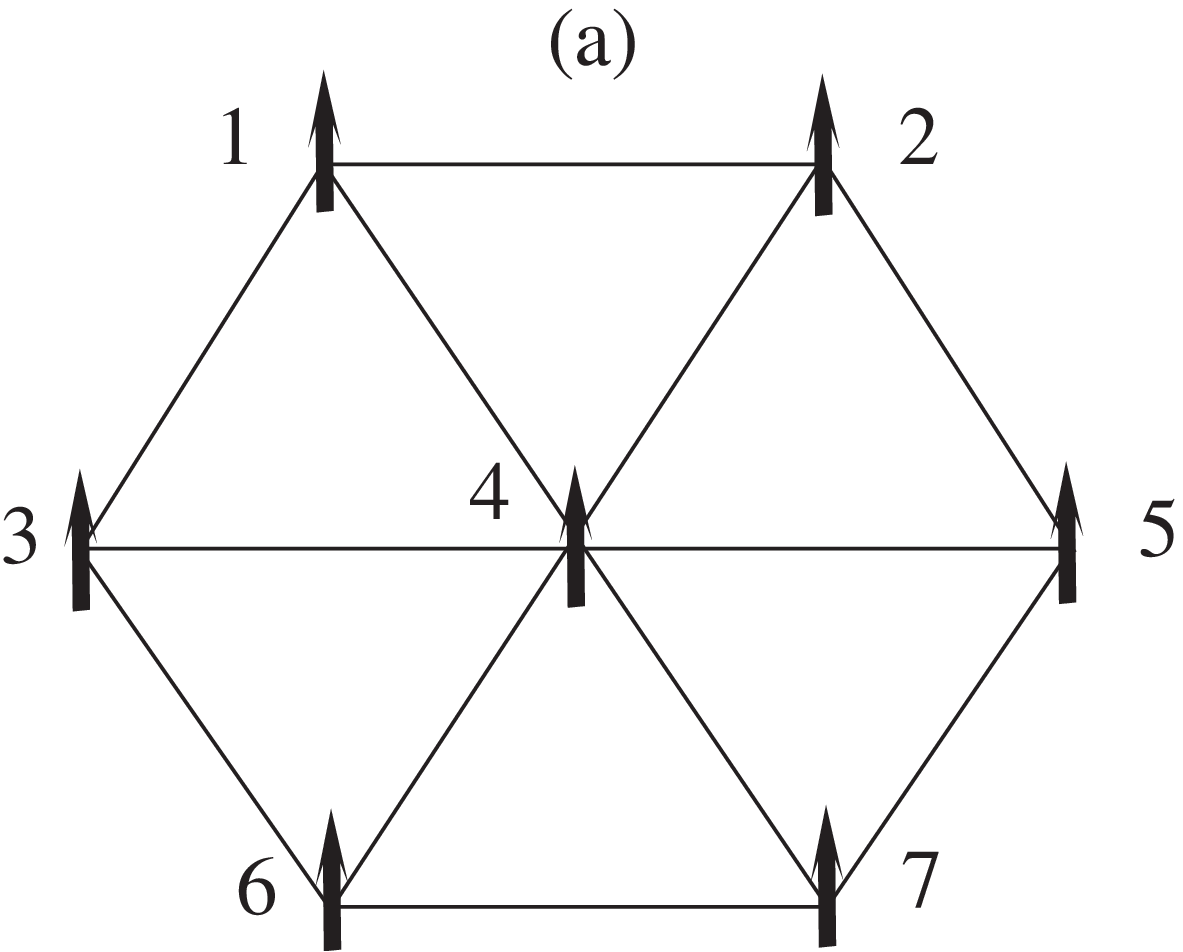}}\quad
   \subfigure{\label{fig:1D_system}\includegraphics[width=7 cm]{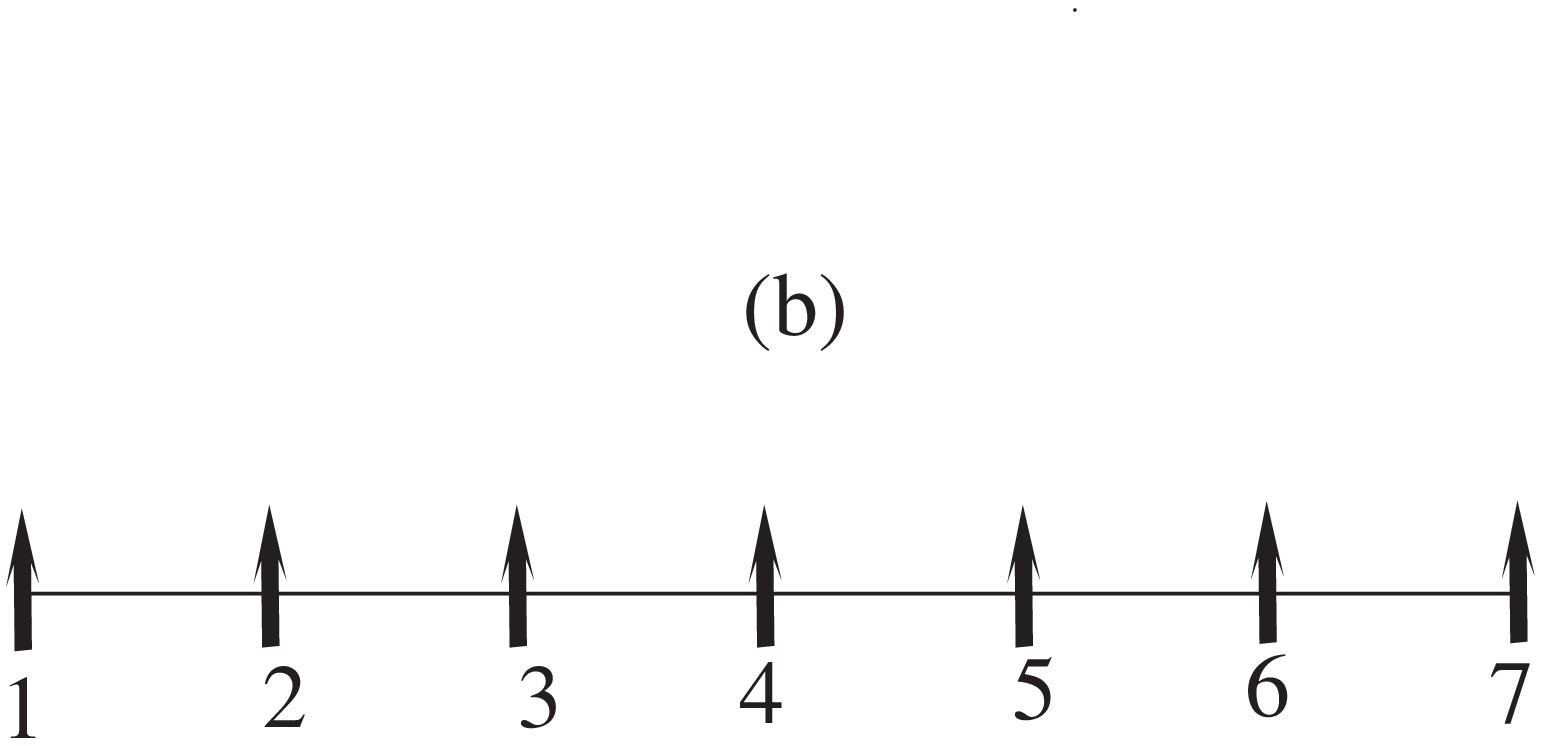}}
  \caption{{\protect\footnotesize (a) Two-dimensional triangular spin lattice; (b) One-dimensional spin chain.}}
 \label{Model}
 \end{minipage}
 \end{figure}

%%%%%%%%%%%%%%%%%%%%%%%%%%%%%%%%%%%%%%%%%%%%%%%%%%%%%%%%%%%%%%%%%%%%%%%%%
We consider two different systems, two and one-dimensional spin-$1/2$ $XY$ model with nearest neighbour exchange coupling $J$ subject to an external magnetic field $h$, with seven spins in each system. The first system is a two-dimensional spin-star model consisting of one central spin and 6 surrounding spins, whereas the second system is a one dimensional spin chain as shown in fig.~\ref{Model} (a) and (b).
The Hamiltonian of the system is given by 
\begin{equation}
\label{Hamiltonian}
H=-\frac{(1+\gamma)}{2}\sum_{<i,j>}J_{i,j}\sigma_{i}^x\sigma_{j}^x -\frac{(1-\gamma)}{2}\sum_{<i,j>}J_{i,j}\sigma_{i}^y\sigma_{j}^y  - h \sum_{i} \sigma_{i}^z,
\end{equation}
%%%%%%%%%%%%%%%%%%%%%%%%%%%%%%%%%%%%%%%%%%%%%%%%%%%%%%%%%%%%%%%%%%%%%%%%%%%%%%%%%%%%%%%%%%%%%%%%%%%%%%%%%
 %fig_1
%%%%%%%%%%%%%%%%%%%%%%%%%%%%%%%%%%%%%%%%%%%%%%%%%%%%%%%%%%%%%%%%%%%%%%%%%%%%%%%%%%%%%%%%%%%%%%%%%%%%%%%%%%%%%%%%%%
where $\sigma_{i}$'s are the Pauli matrices, $\gamma$ is the anisotropy parameter, $<i,j>$ is a pair of nearest-neighbors sites on the lattice and $J_{i,j}=J$ for all sites. For this model it is convenient to study a dimensionless Hamiltonian where we set $J=1$ and define a dimensionless parameter $\lambda=h/J$. The Hilbert space of this spin systems is huge with $2^7$ dimensions, nevertheless it can be exactly diagonalized using the standard computational techniques, yielding the system energy eigenvalues $\{ E_i \}$ and eigenfunctions $\{\psi_i\}$. At absolute zero temperature, the system lies in its ground state $|\psi_0 \rangle$ which is usually entangled with an amount that varies based on the values of the different system parameters. The system is described by the density matrix defined in terms of the pure ground state wavefunction $|\psi_0\rangle$ as
%%%%%%%%%%%%%%%%%%%%%%%%%%%%%%%%%%%%%%%%%%%%%%%%%%%%%%%%%%%%%%%%%%%%%%%%%
\begin{equation}
\label{pure_dens_matrix}
\rho = |\psi_0 \rangle \langle \psi_0 | \; .
\end{equation}
%%%%%%%%%%%%%%%%%%%%%%%%%%%%%%%%%%%%%%%%%%%%%%%%%%%%%%%%%%%%%%%%%%%%%%%%%
Now when the spin system is set into contact with a heat bath at an absolute temperature $T$, the system moves from its initial pure state described by eq. (\ref{pure_dens_matrix}) to a mixed thermal state, which is a mixture of the ground state and a number $N_{e}$ of excited states, represented by 
\begin{equation}
\label{mixed_dens_matrix}
\rho_T = \frac{1}{Z} \{ e^{-\beta E_0} |\psi_0 \rangle \langle \psi_0 | + \sum_{i=1}^{N_e} e^{-\beta E_i} |\psi_i \rangle \langle \psi_i | \}  \; ,
\end{equation}
where $\beta= 1/kT$, $k$ is Boltzmann constant and $Z$ is the system partition function. The number of excited states involved depends on temperature, where more states are added as the temperature is raised. This mixing of excited states with the ground state act as a destructive noise that reduces the amount of entanglement contained in the system. When the temperature reaches certain value, which varies based on the system characteristics and parameters values, the amount of noise created by the excites states due to thermal fluctuations is sufficient to turn the system into a disentangled state. This temperature is known as the threshold temperature, denoted by $T_{th}$ , where below it the system is guaranteed to be entangled \cite{Markham2008}.
%%%%%%%%%%%%%%%%%%%%%%%%%%%%%%%%%%%%%%%%%%%%%%%%%%%%%%%%%%%%%%%%%%%%%%%%%%%%%%%%%%%%%%%%%%%
%%%%%%%%%%%%%%%%%%%%%%%%%%%%%%%%%%%%%%%%%%%%%%%%%%%%%%%%%%%%%%%%%%%%%%%%%%%%%%%%%%%%%%%%%%%
\section{Thermal Bipartite Entanglement in Two-dimensional Spin System}
%%%%%%%%%%%%%%%%%%%%%%%%%%%%%%%%%%%%%%%%%%%%%%%%%%%%%%%%%%%%%%%%%%%%%%%%%%%%%%%%%%%%%%%%%%%
%%%%%%%%%%%%%%%%%%%%%%%%%%%%%%%%%%%%%%%%%%%%%%%%%%%%%%%%%%%%%%%%%%%%%%%%%%%%%%%%%%%%%%%%%%%
To study the bipartite entanglement in the system, we confine our interest to the entanglement between only two spins, at any sites $i$ and $j$ \cite{Osterloh2002}. All the information about the considered two sites $i$ and $j$ is contained in the reduced density matrix $\rho_{i, j}$ which can be obtained from the entire system density matrix by integrating out all the spins states except $i$ and $j$. We adopt the entanglement of formation, as a well known  measure of entanglement where Wootters \cite{Wootters1998} has shown that, for a pair of binary qubits, the concurrence $C$, which goes from $0$ to $1$, can be used to quantify entanglement. The concurrence between two sites $i$ and $j$ is defined as 
%%%%%%%%%%%%%%%%%%%%%%%%%%%%%%%%%%%%%%%%%%%%%%%%%%%%%%%%%%%%%%%%%%%%%%%%%%%%%%%%%%%%%%
\begin{equation}
\label{concurrence}
C(\rho_{i, j})=max\{0,\epsilon_1-\epsilon_2-\epsilon_3-\epsilon_4\},
\end{equation}
%%%%%%%%%%%%%%%%%%%%%%%%%%%%%%%%%%%%%%%%%%%%%%%%%%%%%%%%%%%%%%%%%%%%%%
%%%%%%%%%%%%%%%%%%%%%%%%%%%%%%%%%%%%%%%%%%%%%%%%%%%%%%%%%%%%%%%%%%%%%%%%%
\begin{figure}[htbp]
\begin{minipage}[c]{\textwidth}
 \centering
   \subfigure{\label{fig:C_G_1_a}\includegraphics[width=7.5 cm]{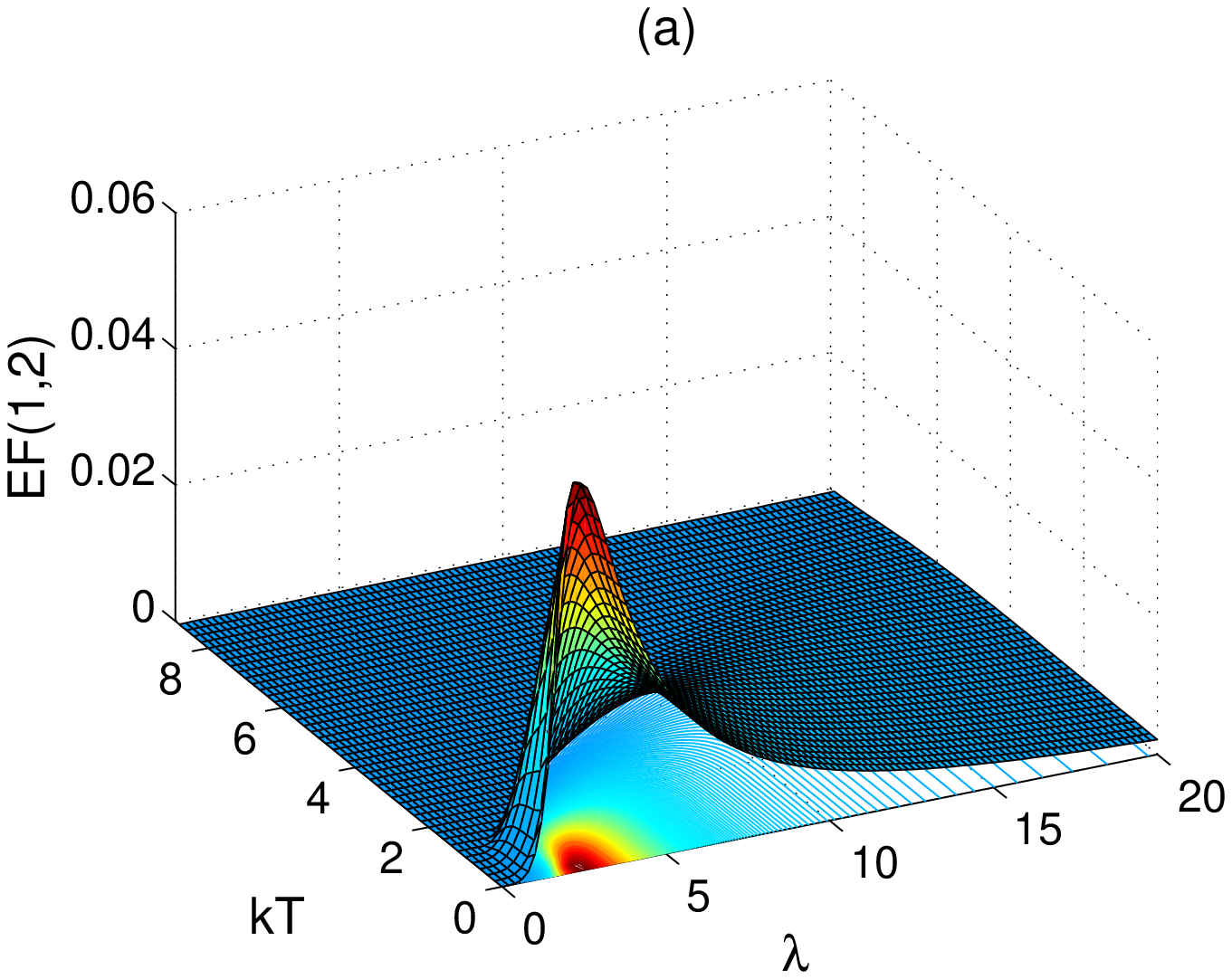}}\quad
   \subfigure{\label{fig:C_G_1_b}\includegraphics[width=7.5 cm]{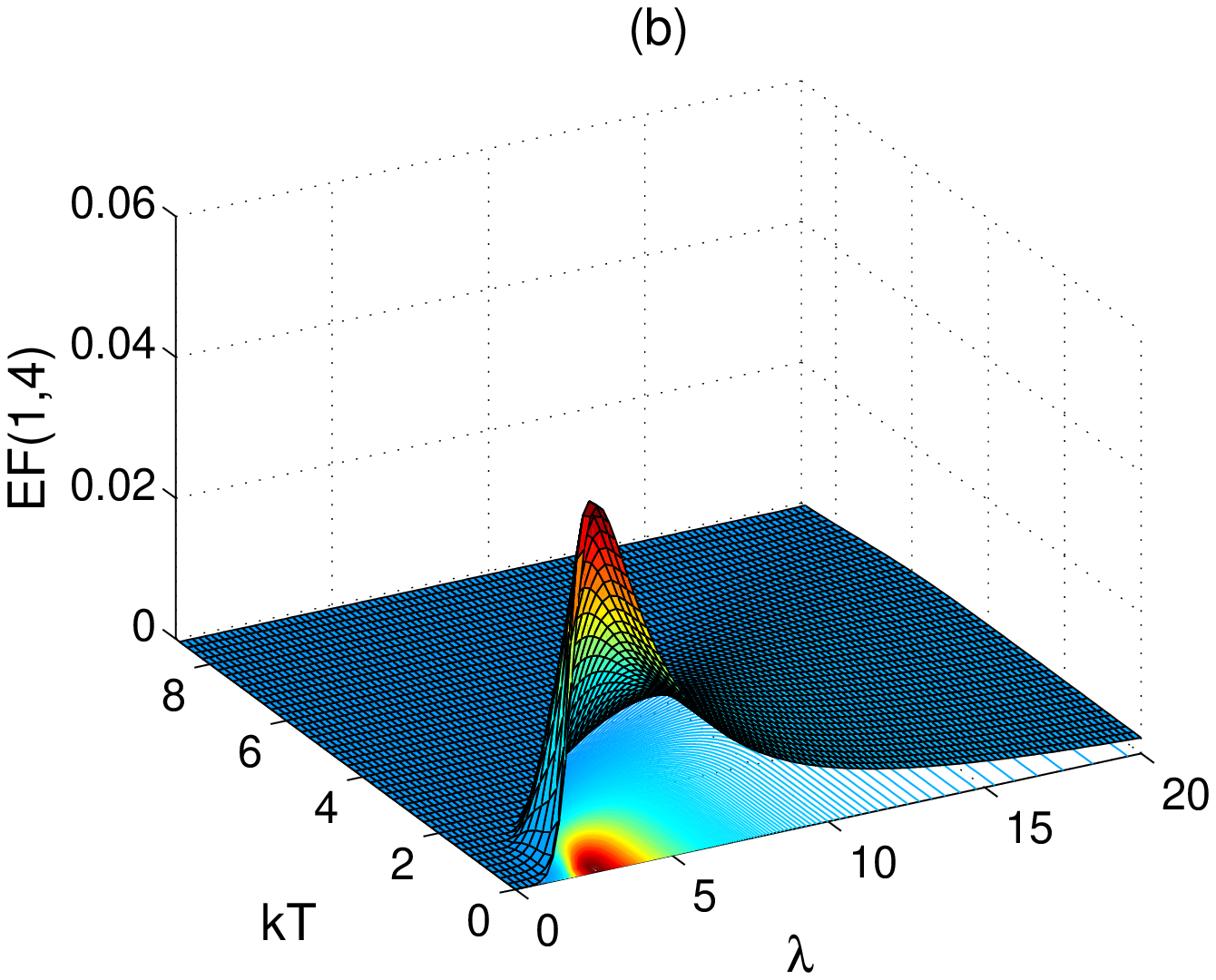}}\\
   \subfigure{\label{fig:C_G_1_c}\includegraphics[width=7.5 cm]{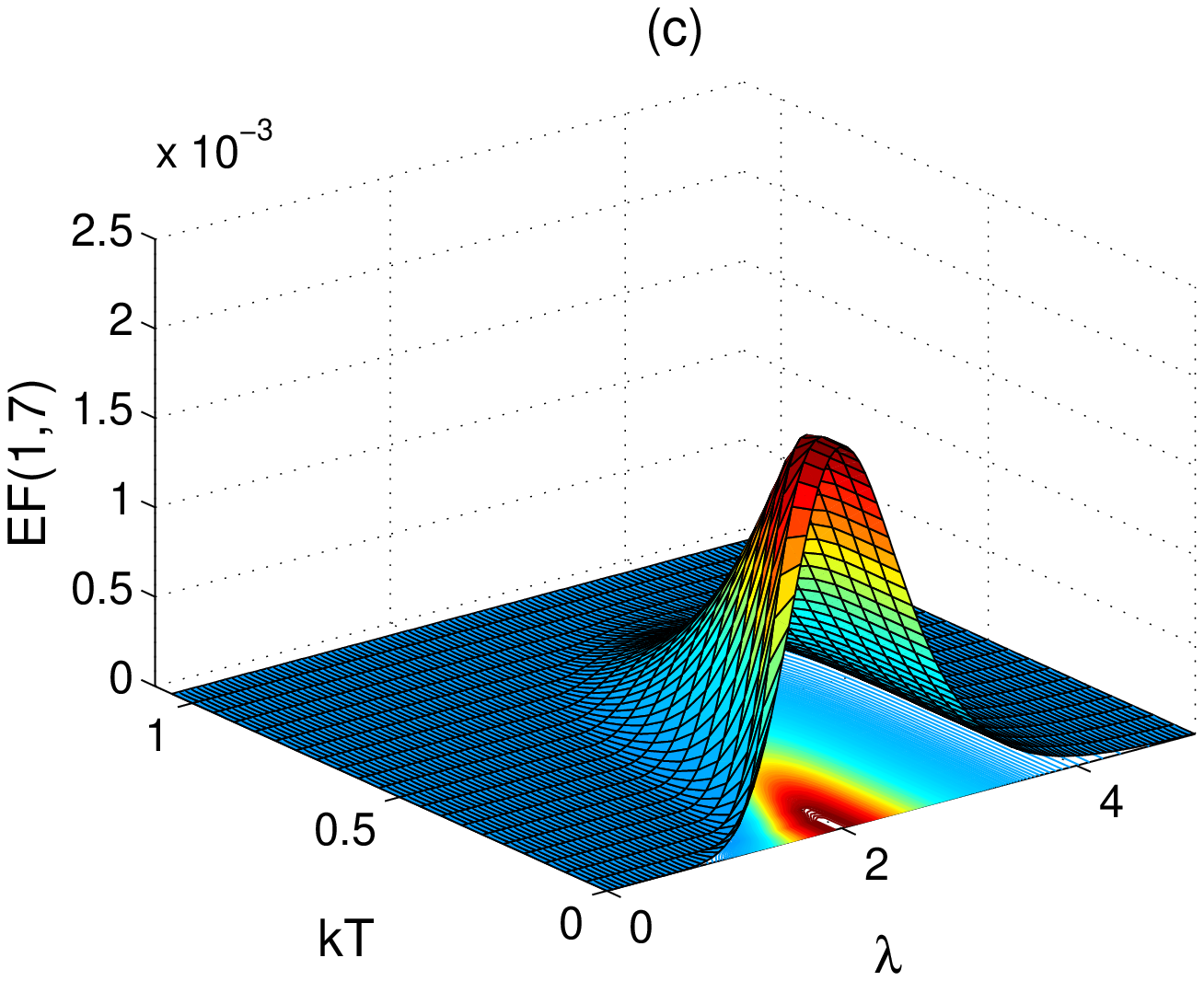}}\quad
   \caption{{\protect\footnotesize (Color online) The BE $EF(1,2)$, $EF(1,4)$ and $EF(1,7)$ of the 2D Ising system ($\gamma=1$) versus $\lambda$ and $kT$(in units of $J$).}}
 \label{C_G_1}
 \end{minipage}
\end{figure}
%fig_2
%%%%%%%%%%%%%%%%%%%%%%%%%%%%%%%%%%%%%%%%%%%%%%%%%%%%%%%%%%%%%%%%%%%%%%%%%%%%%%%%%%%%%
%%%%%%%%%%%%%%%%%%%%%%%%%%%%%%%%%%%%%%%%%%%%%%%%%%%%%%%%%%%%%%%%%%%%%%%%%%%%%%%%%%%
\begin{figure}[htbp]
\begin{minipage}[c]{\textwidth}
 \centering
   \subfigure{\label{fig:Cc_G_1_a}\includegraphics[width=7.5 cm]{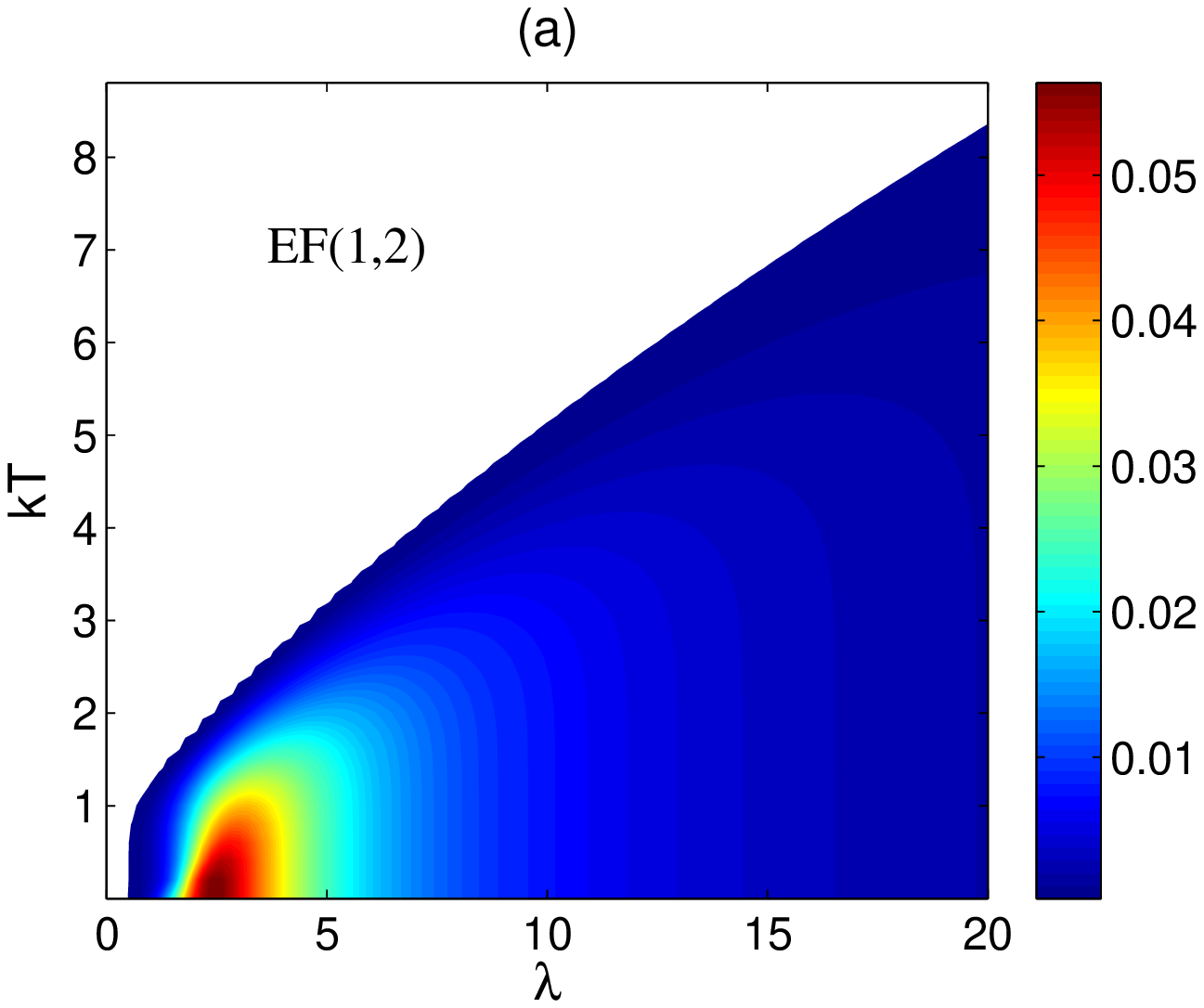}}\quad
   \subfigure{\label{fig:Cc_G_1_b}\includegraphics[width=7.5 cm]{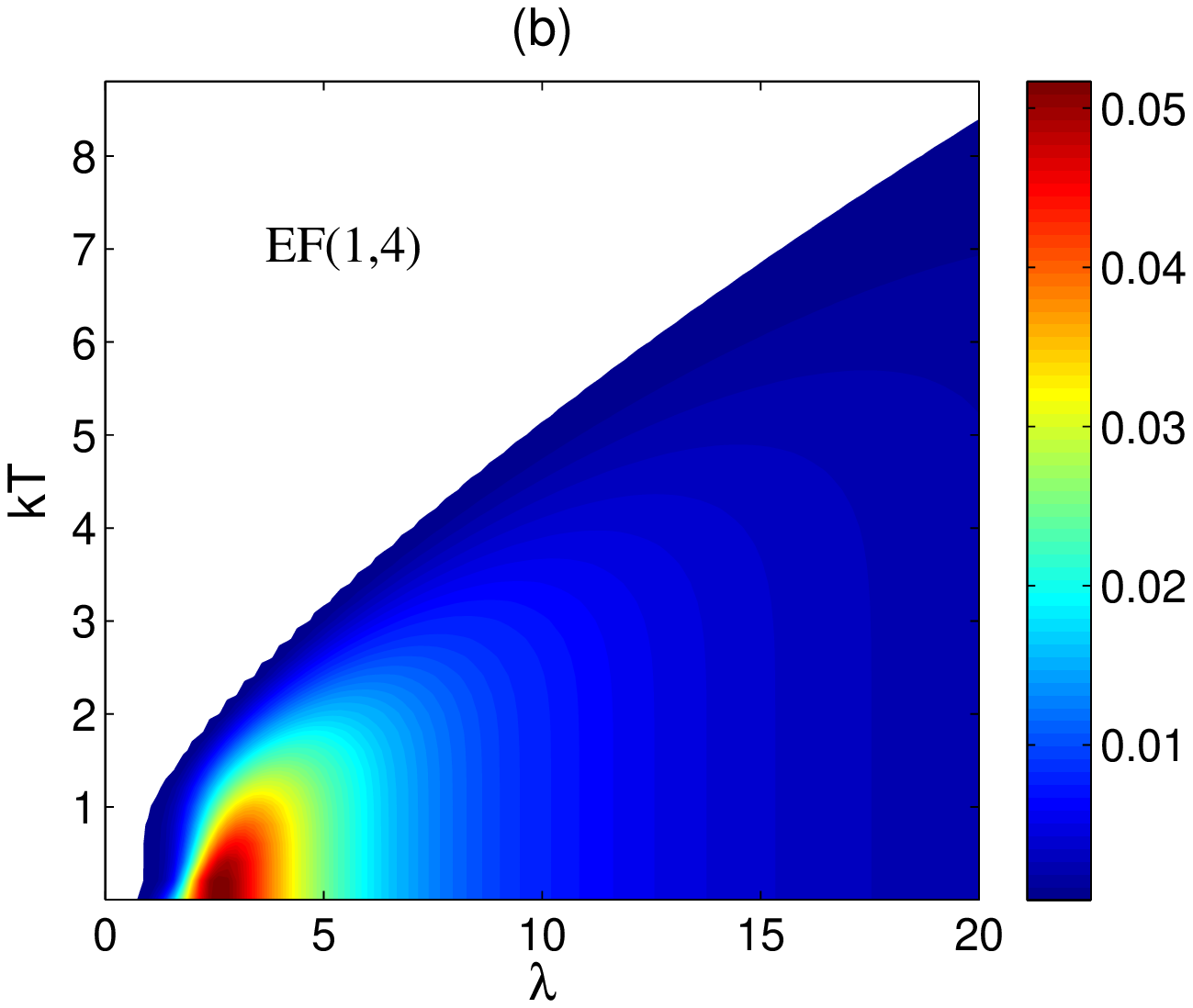}}\\
   \subfigure{\label{fig:Cc_G_1_c}\includegraphics[width=7.5 cm]{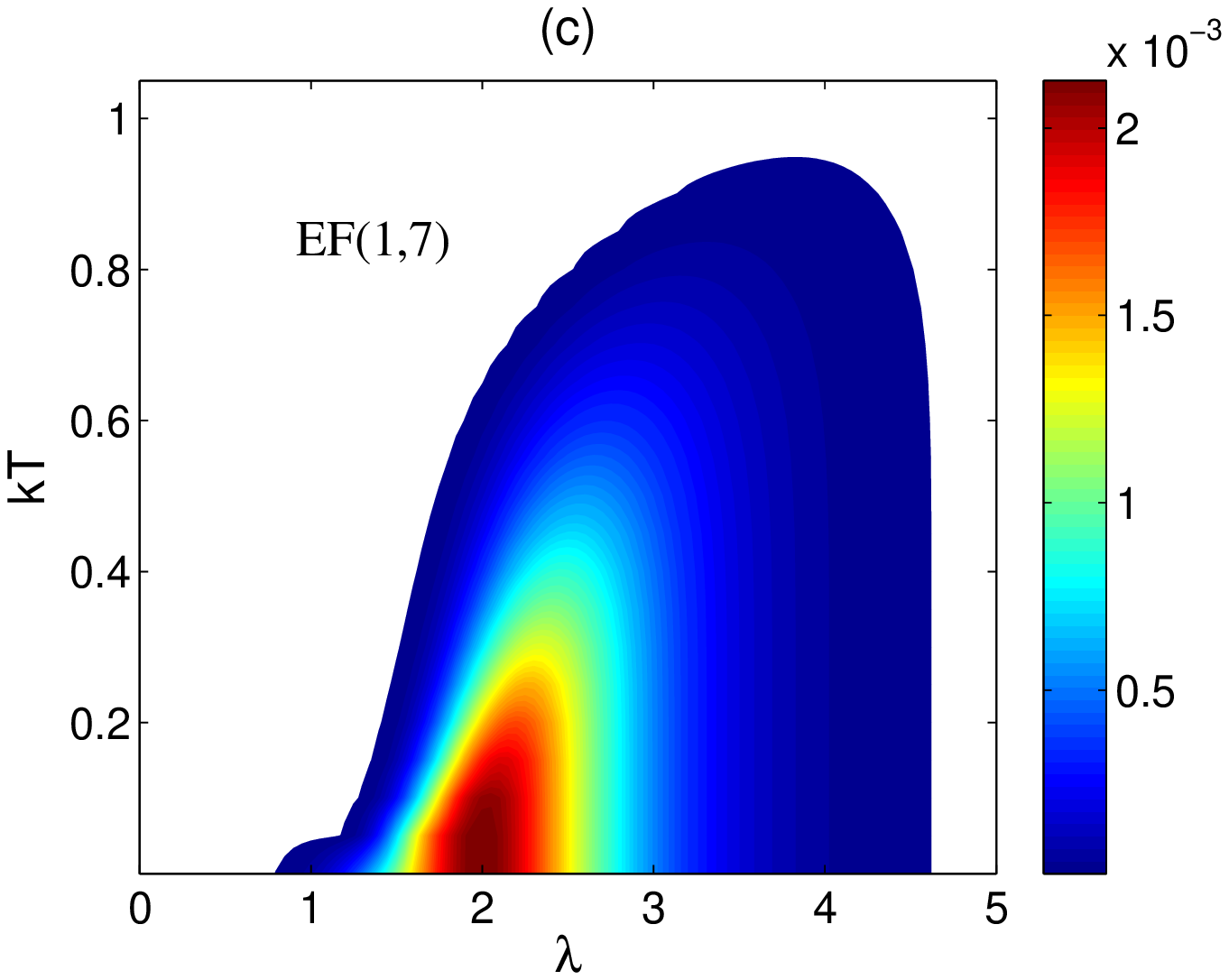}}\quad
   \caption{{\protect\footnotesize (Color online) The contour plot of the BE $EF(1,2)$, $EF(1,4)$ and $EF(1,7)$ of the 2D Ising system ($\gamma=1$) versus $\lambda$ and $kT$(in units of $J$).}}
 \label{Cc_G_1}
 \end{minipage}
\end{figure}
%fig_3
%%%%%%%%%%%%%%%%%%%%%%%%%%%%%%%%%%%%%%%%%%%%%%%%%%%%%%%%%%%%%%%%%%%%%
%%%%%%%%%%%%%%%%%%%%%%%%%%%%%%%%%%%%%%%%%%%%%%%%%%%%%%%%%%%%%%%%%%%%%
where the $\epsilon_i$'s are the eigenvalues of the Hermitian matrix
$R\equiv\sqrt{\sqrt{\rho}\tilde{\rho}\sqrt{\rho}}$ with
$\tilde{\rho}=(\sigma^y \otimes
\sigma^y)\rho^*(\sigma^y\otimes\sigma^y)$ and $\sigma^y$ is the
Pauli matrix of the spin in y direction.
For a pair of qubits the entanglement of formation is defined as,
%%%%%%%%%%%%%%%%%%%%%%%%%%%%%%%%%%%%%%%%%%%%%%%%%%%%%%%%%%%%%%%%%%%%%%
\begin{equation}
\label{entanglement}
E(\rho_{i, j})=\epsilon(C(\rho_{i, j})),
\end{equation}
where $\epsilon$ is a function of $C$
\begin{equation}
\epsilon(C)=h\left(\frac{1-\sqrt{1-C^2}}{2}\right),
\end{equation}
where $h$ is the binary entropy function
\begin{equation}
h(x)=-x \log_{2}(x) - (1-x) \log_{2}(1-x).
\end{equation}
%%%%%%%%%%%%%%%%%%%%%%%%%%%%%%%%%%%%%%%%%%%%%%%%%%%%%%%%%%%%%%%%%%%%%%%%%%
In our calculations we use the entanglement of formation $EF$ as a measure of the bipartite entanglement.
%%%%%%%%%%%%%%%%%%%%%%%%%%%%%%%%%%%%%%%%%%%%%%%%%%%%%%%%%%%%%%%%%%%%%%%%%%%%%%%%%%%
\begin{figure}[htbp]
\begin{minipage}[c]{\textwidth}
 \centering
   \subfigure{\label{fig:Cc_G_05_a}\includegraphics[width=7.5 cm]{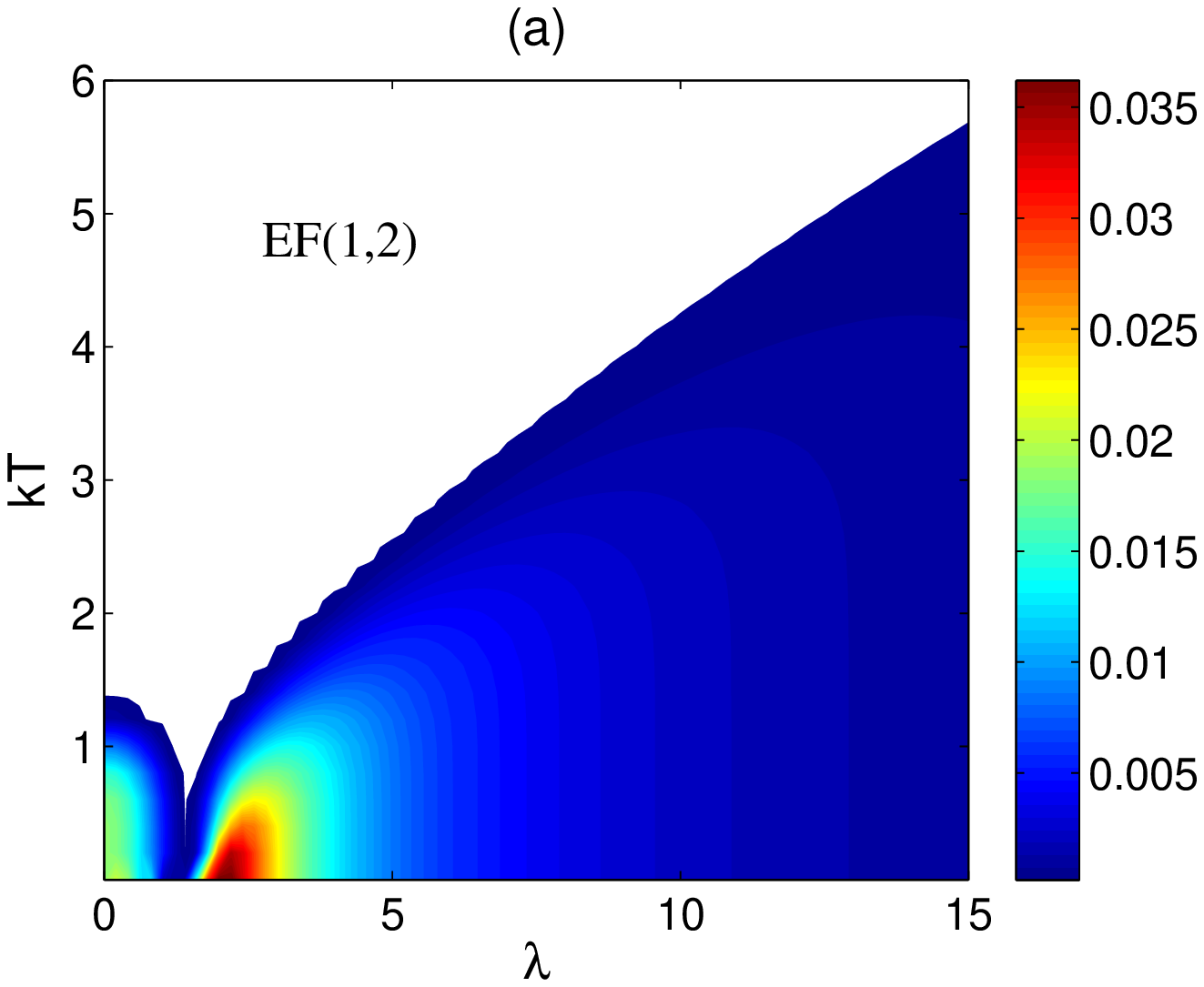}}\quad
   \subfigure{\label{fig:Cc_G_05_b}\includegraphics[width=7.5 cm]{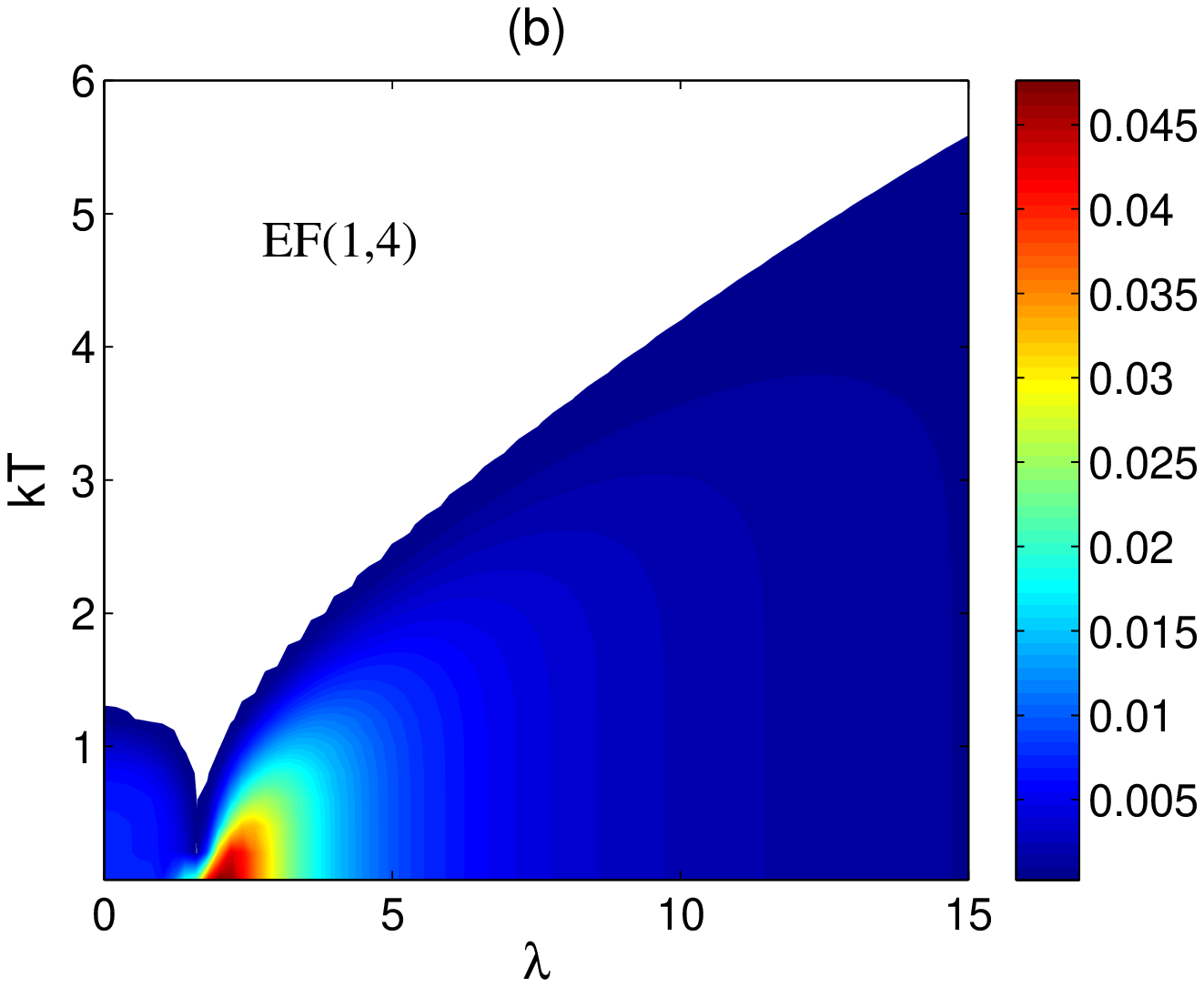}}\\
   \subfigure{\label{fig:Cc_G_05_c}\includegraphics[width=7.5 cm]{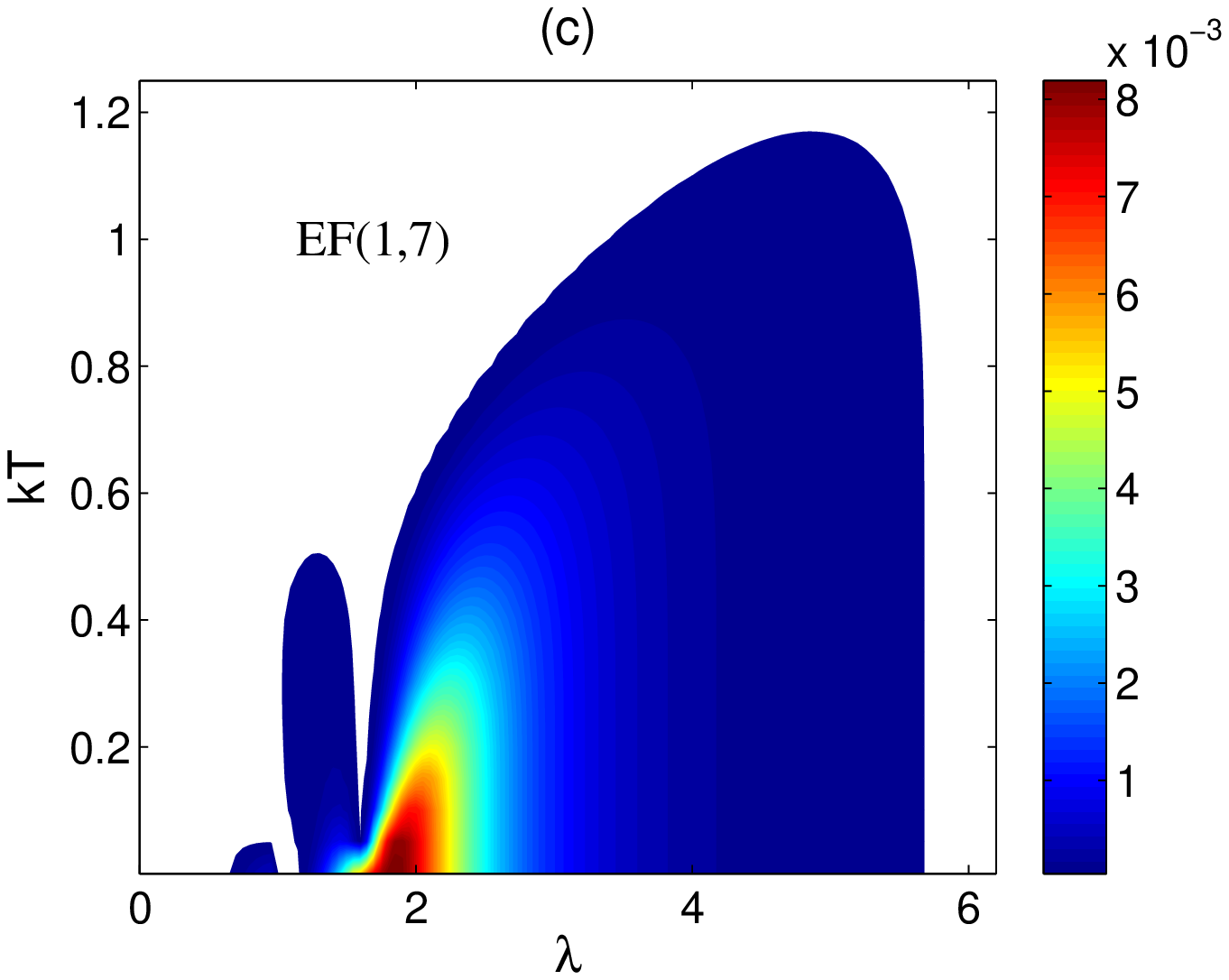}}
   \caption{{\protect\footnotesize (Color online) The contour plot of the BE $EF(1,2)$, $EF(1,4)$ and $EF(1,7)$ of the 2D partially anisotropic system ($\gamma=0.5$) versus $\lambda$ and $kT$(in units of $J$).}}
 \label{Cc_G_05}
 \end{minipage}
\end{figure}
%fig_4
%%%%%%%%%%%%%%%%%%%%%%%%%%%%%%%%%%%%%%%%%%%%%%%%%%%%%%%%%%%%%%%%%%%%%%%%%%%%%%%%%%%%%
Using the mixed density matrix $\rho_{T}$ defined in (\ref{mixed_dens_matrix}), one can evaluate the bipartite entanglement between any pair of spins in the system. 
In this section we focus on studying the bipartite entanglement only in the two-dimensional spin system sketched in fig.~\ref{Model}(a). In fig.~\ref{C_G_1} we have explored the behaviour of the entanglements of the nearest neighbors $EF(1,2)$; $EF(1,4)$ and the next-to-next-to-nearest neighbor (nnnn) $EF(1,7)$ versus $\lambda$ and the temperature $KT$ for the anisotropic Ising system ($\gamma=1$). In fact the next-to-nearest neighbor (nnn) $EF(1,5)$ is very close to $EF(1,7)$, as we will show below, but $EF(1,7)$ shows sharper changes, which makes us focus on it.
As can be noticed, the nearest neighbor bipartite entanglements between two border sites $EF(1,2)$ and between a border site and the central one $EF(1,4)$ are strongest for very small magnetic field but very fragile away from the zero temperature. On the other hand, as the magnetic field is increased the entanglement maintains a small value which is more resistant to higher temperatures. Interestingly, the threshold temperature $T_{th}$ at which the entanglement vanishes increases monotonically as the magnetic field increases. The (nnnn) entanglement EF(1,7) sustains only for very small values of magnetic field and in the vicinity of the zero temperature and its value is much smaller than the nearest neighbor entanglements. In order to further investigate the thermal robustness of the entanglement state and determine the magnitude of the entanglement precisely at high temperatures, we show the contour plot of the entanglements $EF(1,2)$, $EF(1,4)$ and $EF(1,7)$ in fig.~\ref{Cc_G_1}. As can be noticed, we can reach, for $EF(1,2)$ and $EF(1,4)$, a threshold temperature $kT=8$ and higher by applying a magnetic filed $h=20$ and higher though the entanglement magnitude is very small. EF(1,7) is very fragile to temperature regardless of the strength of the applied magnetic field as shown in fig.~\ref{Cc_G_1}(c).
%%%%%%%%%%%%%%%%%%%%%%%%%%%%%%%%%%%%%%%%%%%%%%%%%%%%%%%%%%%%%%%%%%%%%%%%%%
\begin{figure}[htbp]
\begin{minipage}[c]{\textwidth}
 \centering
   \subfigure{\label{fig:Cc_G_0_a}\includegraphics[width=7.5 cm]{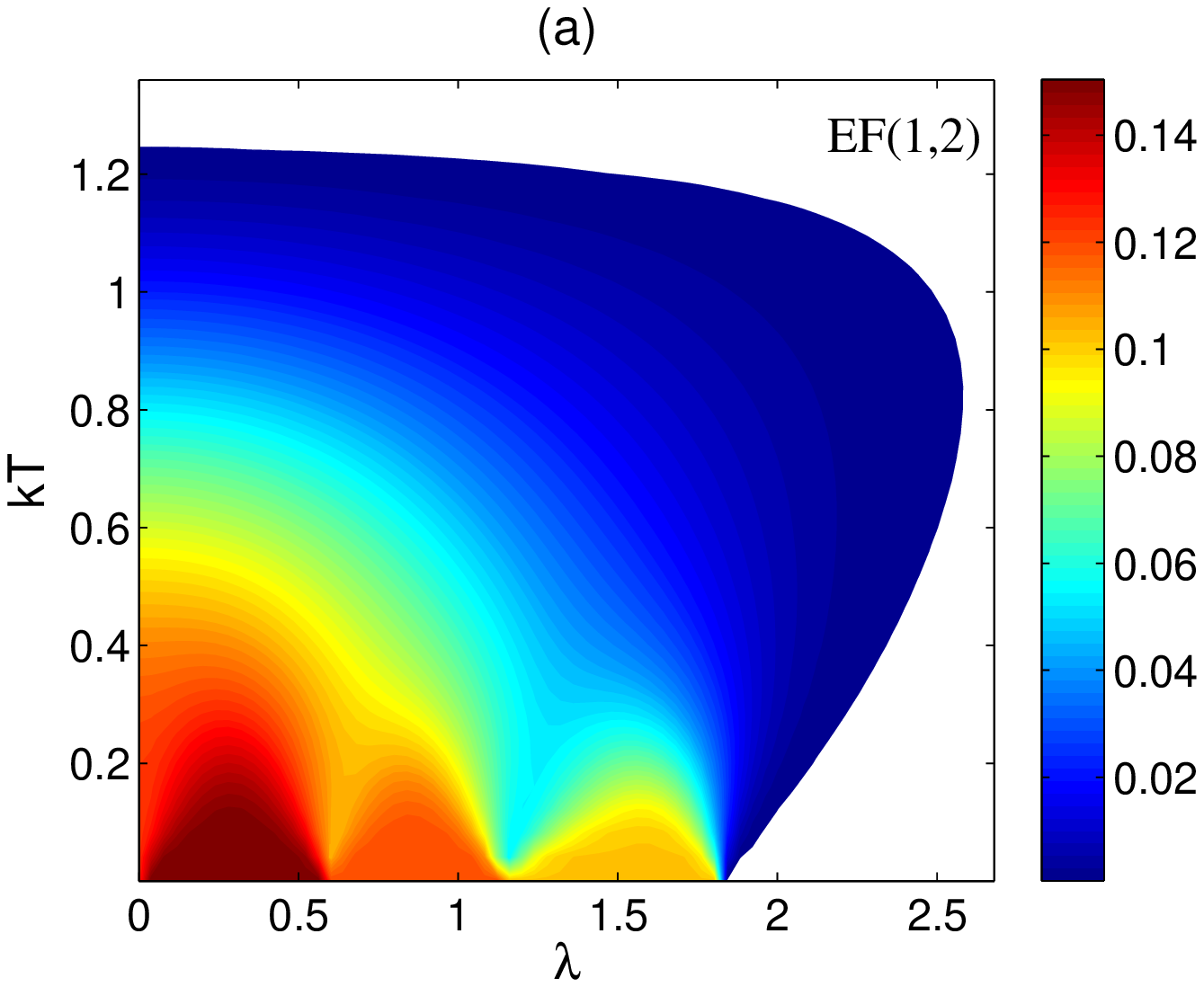}}\quad
   \subfigure{\label{fig:Cc_G_0_b}\includegraphics[width=7.5 cm]{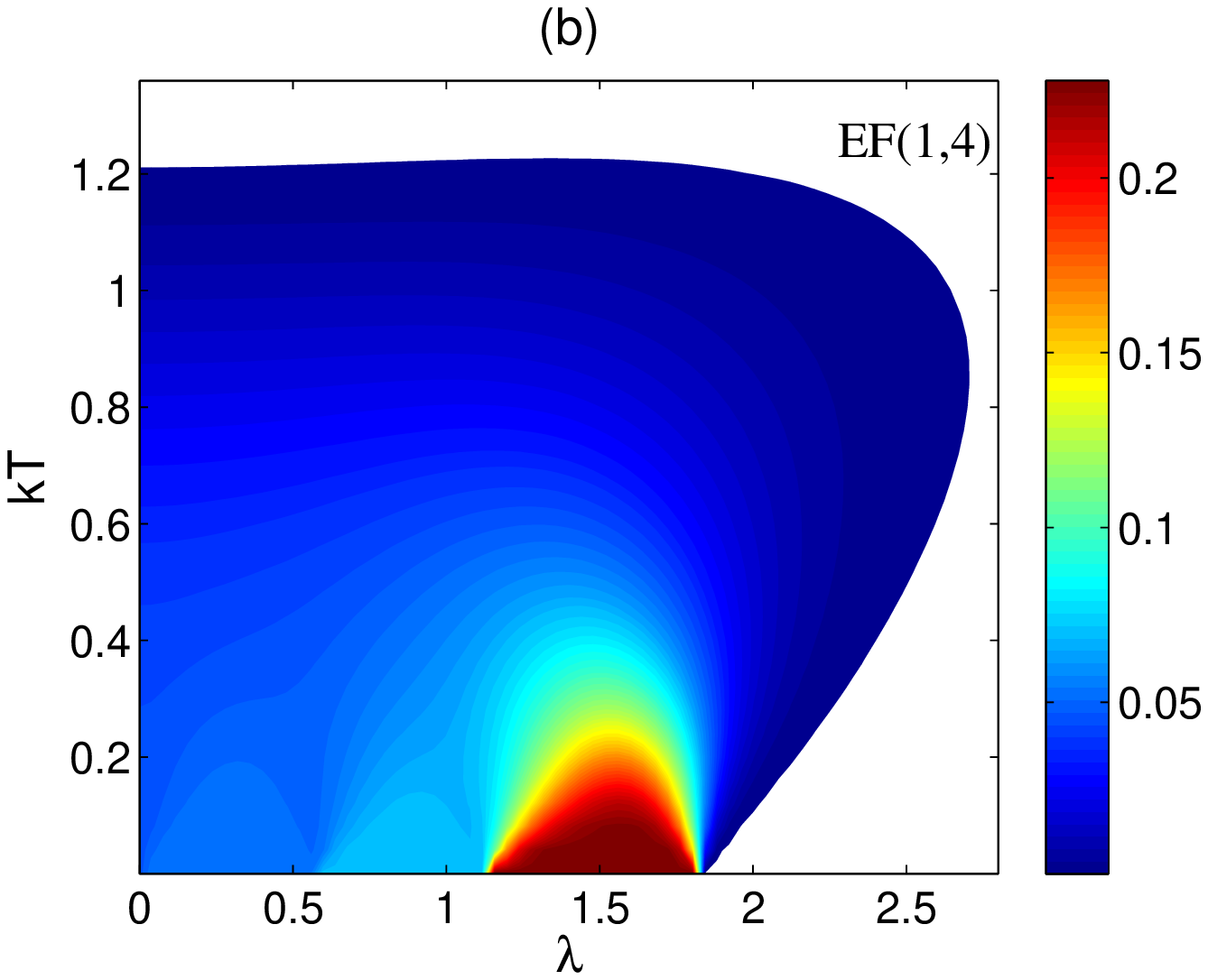}}\\
   \subfigure{\label{fig:Cc_G_0_c}\includegraphics[width=7.5 cm]{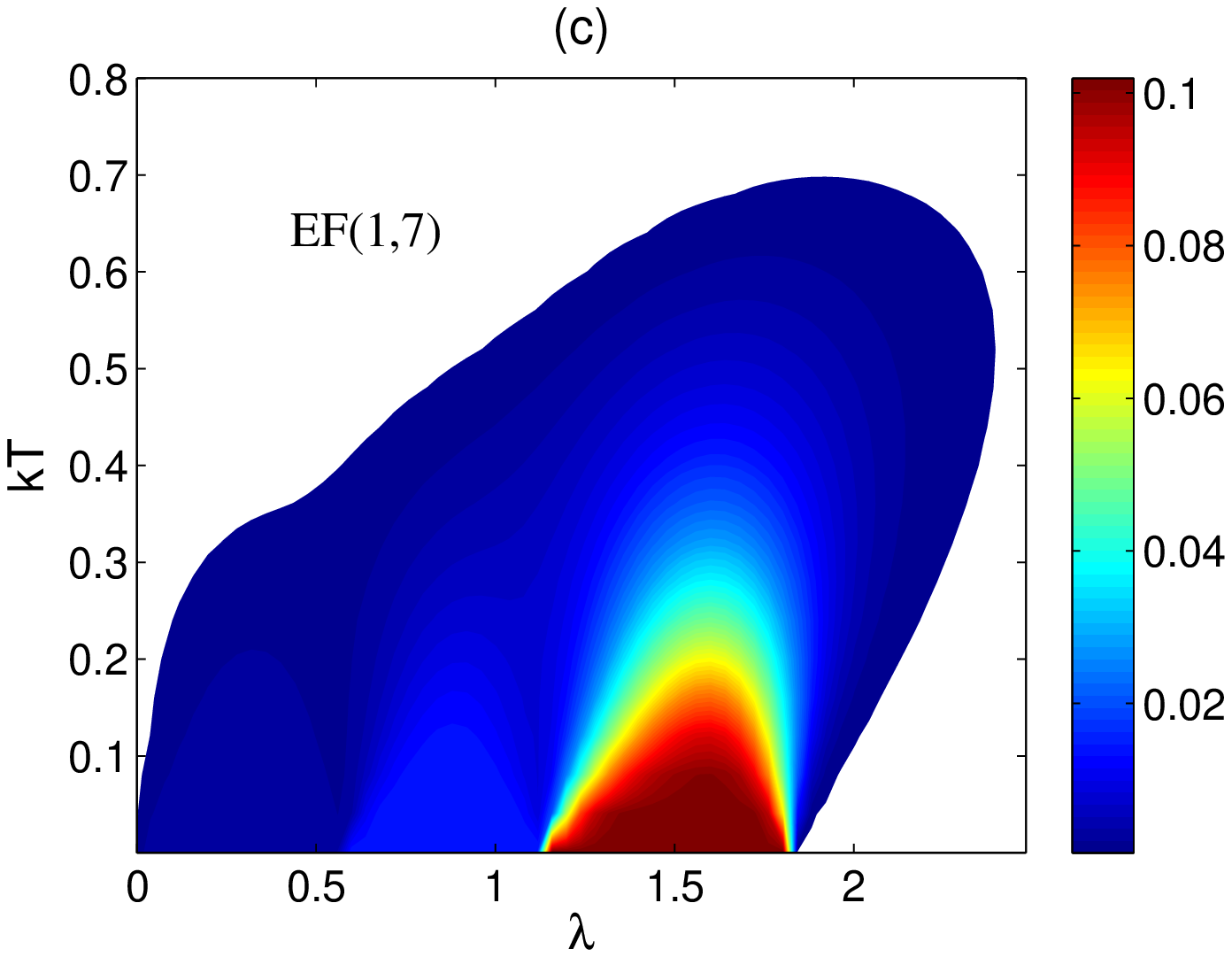}}\quad
   \caption{{\protect\footnotesize (Color online) The contour plot the BE $EF(1,2)$, $EF(1,4)$ and $EF(1,7)$ of the 2D isotropic system ($\gamma=0.$) versus $\lambda$ and $kT$(in units of $J$).}}
 \label{Cc_G_0}
 \end{minipage}
\end{figure}
%fig_5
%%%%%%%%%%%%%%%%%%%%%%%%%%%%%%%%%%%%%%%%%%%%%%%%%%%%%%%%%%%%%%%%%%%%%%%%%%%%%%%%%%%%%
%%%%%%%%%%%%%%%%%%%%%%%%%%%%%%%%%%%%%%%%%%%%%%%%%%%%%%%%%%%%%%%%%%%%%%%%%
%%%%%%%%%%%%%%%%%%%%%%%%%%%%%%%%%%%%%%%%%%%%%%%%%%%%%%%%%%%%%%%%%%%%%%%%%
\begin{figure}[htbp]
\begin{minipage}[c]{\textwidth}
 \centering
   \subfigure{\label{fig:C12_g_1_L_500_a}\includegraphics[width=7.5 cm]{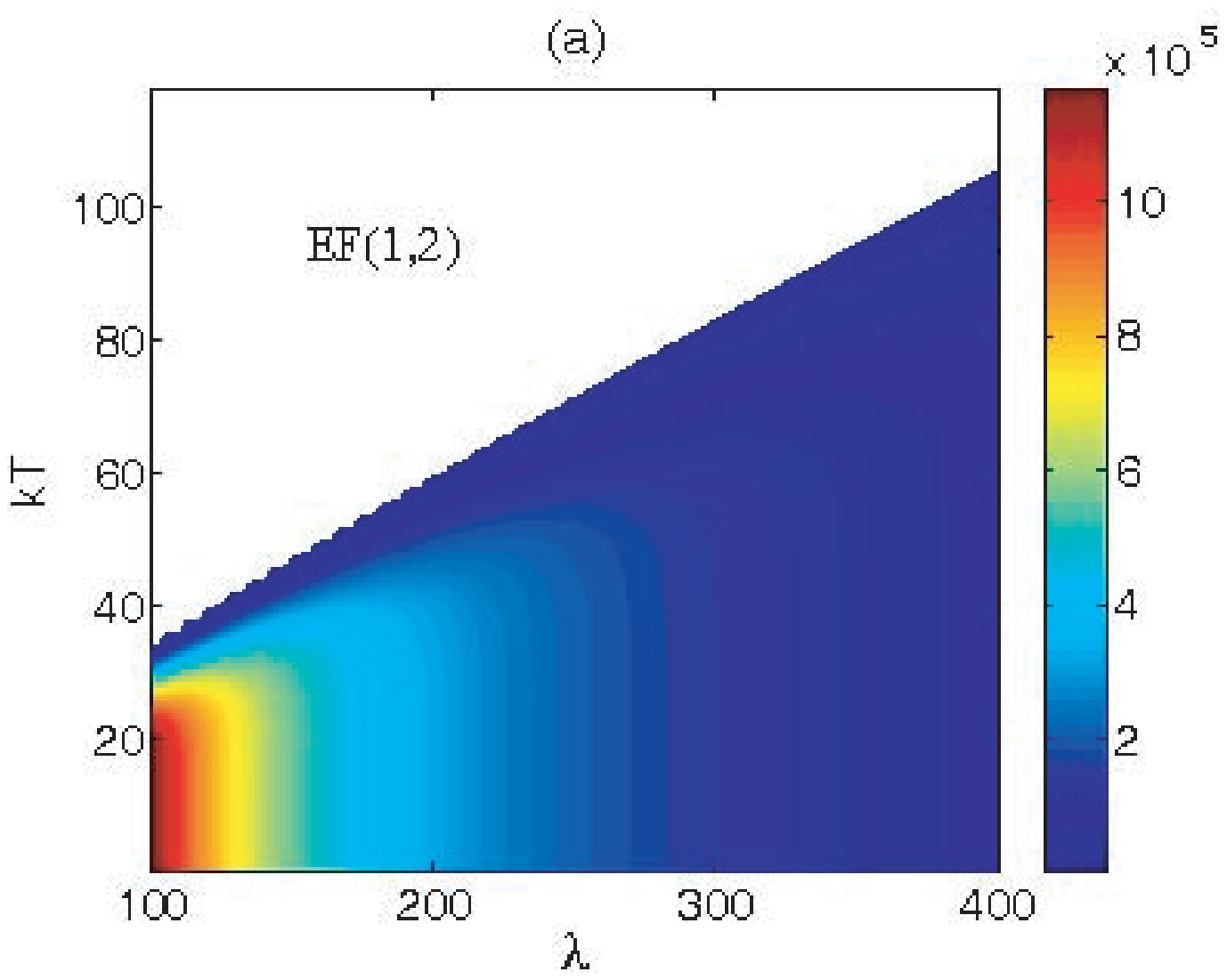}}\quad
   \subfigure{\label{fig:C14_g_05_L_500_b}\includegraphics[width=7.5 cm]{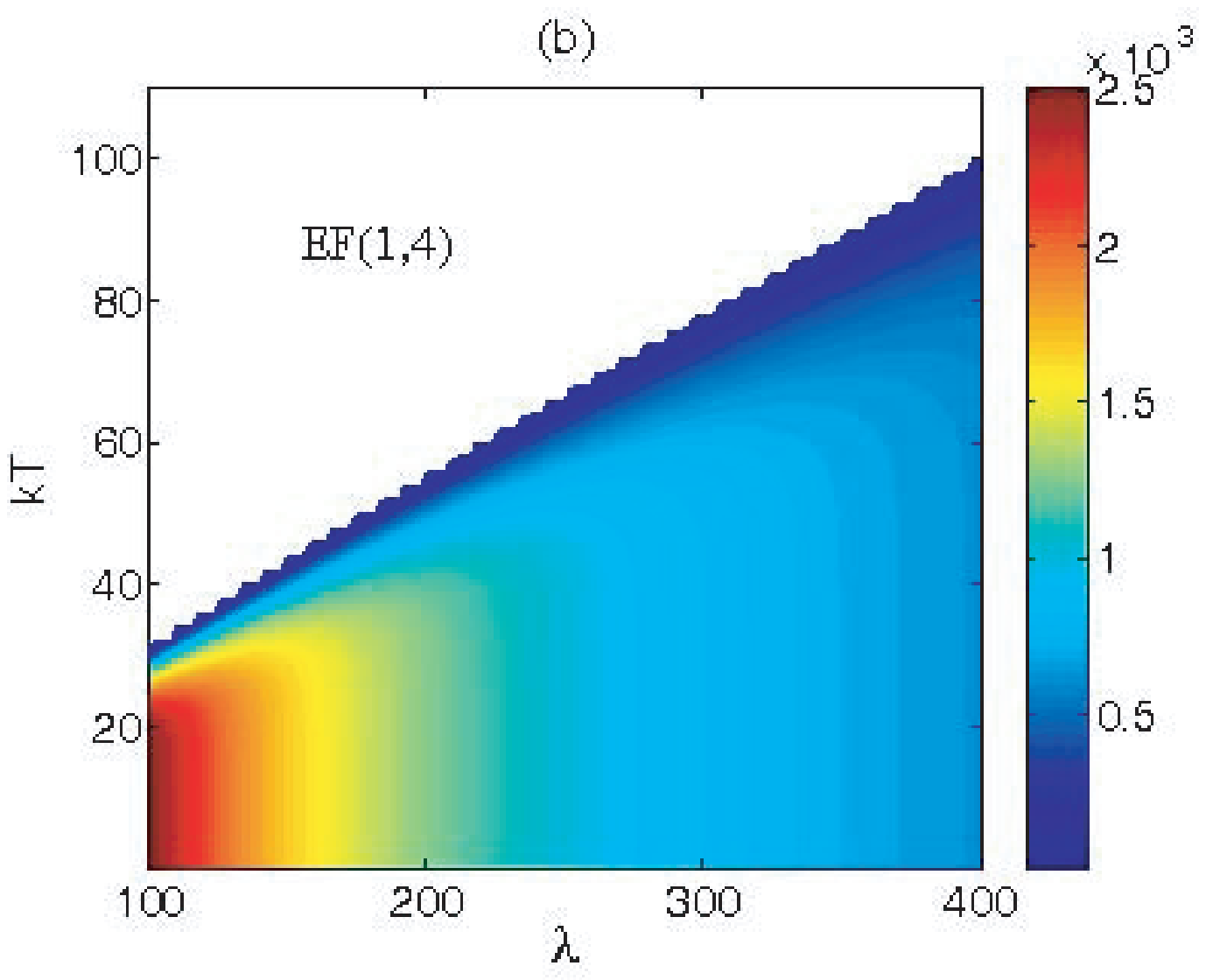}}\quad
   \subfigure{\label{fig:C14_g_0_L_500_c}\includegraphics[width=7.5 cm]{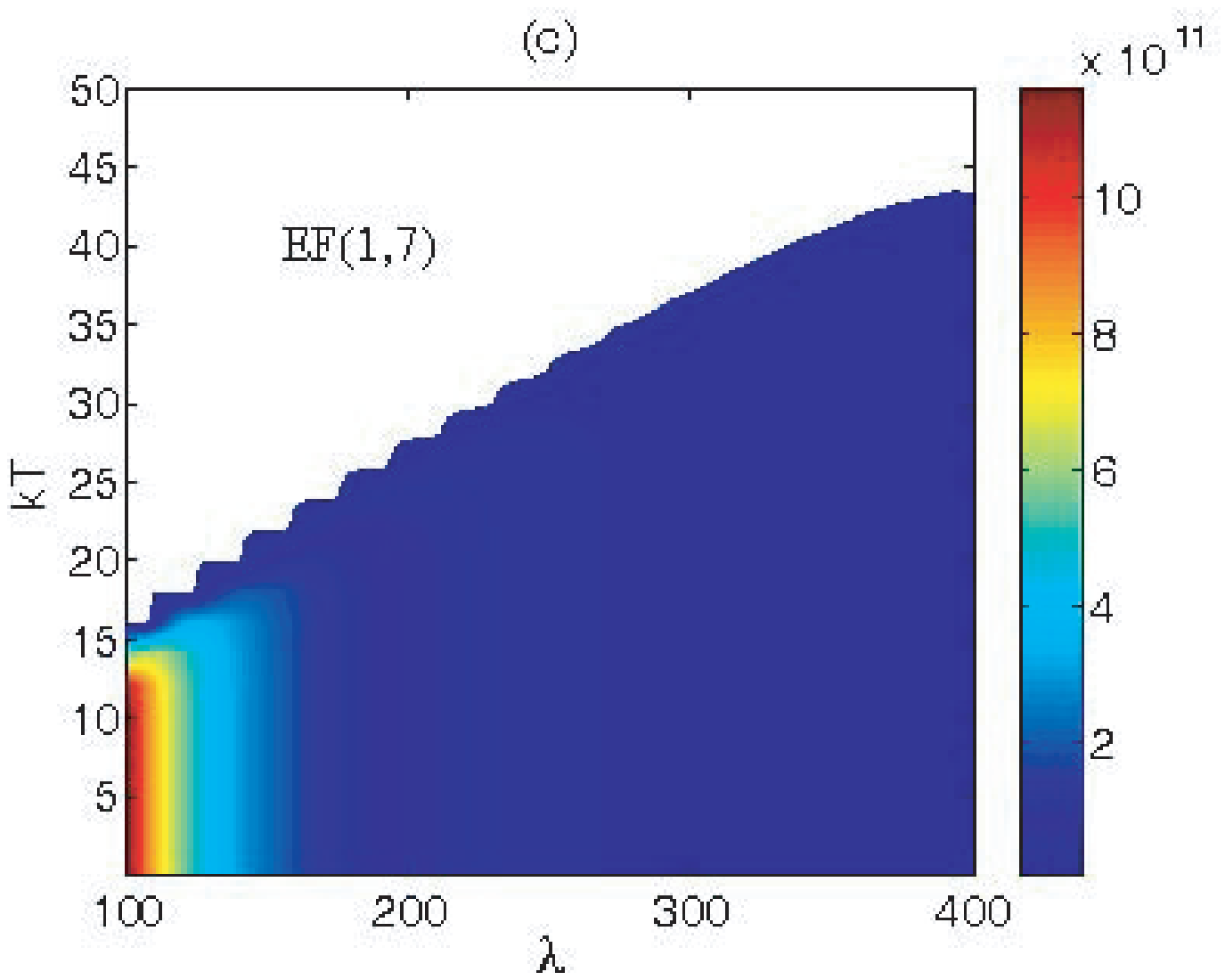}}
   \caption{{\protect\footnotesize (Color online) The contour plot of the BE (a) $EF(1,2)$ (for $\gamma=1$); (b) $EF(1,4)$ (for $\gamma=0.5$) and (c) $EF(1,7)$ (for $\gamma=0.5$) of the 2D spin system versus $\lambda$ and $kT$ (in units of $J$) for a range of $\lambda$ from 100 to 400.}}
 \label{Cc_L_500}
 \end{minipage}
\end{figure}
%fig_6
%%%%%%%%%%%%%%%%%%%%%%%%%%%%%%%%%%%%%%%%%%%%%%%%%%%%%%%%%%%%%%%%%%%%%%%%%
The partially anisotropic system with $\gamma=0.5$ was found to exhibit a close behavior to the $\gamma=1$ case as depicted in fig.~\ref{Cc_G_05}. The peak of the entanglements, at small magnetic field and in the neighborhood of zero temperature, is not single and this is due to the profile of the system energy gap for $\gamma=0.5$ as will be discussed in more details below. It is clear that the value of the threshold temperature corresponding to the different magnetic filed values is smaller compared to that of the $\gamma=1$ case as can be concluded from figs.~\ref{Cc_G_1} and ~\ref{Cc_G_05}. 
Interestingly, the completely isotropic spin system with $\gamma=0$ behaves in a completely different way compared to $\gamma=0$ and $0.5$ as shown in fig.~\ref{Cc_G_0}. As can be noticed from the figures, not only $EF(1,7)$ but also $EF(1,2)$ and $EF(1,4)$ vanish at very small temperatures, about few $kT$. Clearly the thermal fluctuations is very devastating to the isotropic system where the bipartite entanglements over the whole lattice vanishes at very small temperature. The peak of the entanglement $EF(1,2)$ is higher than that of $EF(1,4)$ but $EF(1,7)$ is much lower than both. In fact this behavior of the isotropic system ($\gamma=0$) should not be very surprising, as it is known that the systems described by the Hamiltonian Eq. (\ref{Hamiltonian}) at the thermodynamic limit belongs to different universality classes based on the value of $\gamma$. The isotropic system is characterized by a separable state at a small value of the magnetic field, and for the considered system at $\gamma=0$, the ground state is separable for $\lambda \approx 1.85$ and higher.

In order to investigate the robustness of entanglement at much higher temperatures, we depict, as an example, the contour of the entanglements $EF(1,2)$ (at $\gamma=1$), $EF(1,4)$ (at $\gamma=0.5$), at very high magnetic field in fig.~\ref{Cc_L_500}(a) and (b) respectively, which confirms the survival of entanglement, though very low in magnitude, at high temperature. Interestingly when we considered $EF(1,7)$ in the Ising system and the partially anisotropic system ($\gamma = 0.5$) at high magnetic fields and temperatures, we found that it is not exactly zero (only for $\gamma=0.5$) though its extremely small as shown in fig.~\ref{Cc_L_500}(c).
%%%%%%%%%%%%%%%%%%%%%%%%%%%%%%%%%%%%%%%%%%%%%%%%%%%%%%%%%%%%%%%%%%%%%%%%%%%%%%%%%%%%%
\begin{figure}[htbp]
\begin{minipage}[c]{\textwidth}
 \centering
   \subfigure{\label{fig:dE_g_1_c_L=10}\includegraphics[width=7.5 cm]{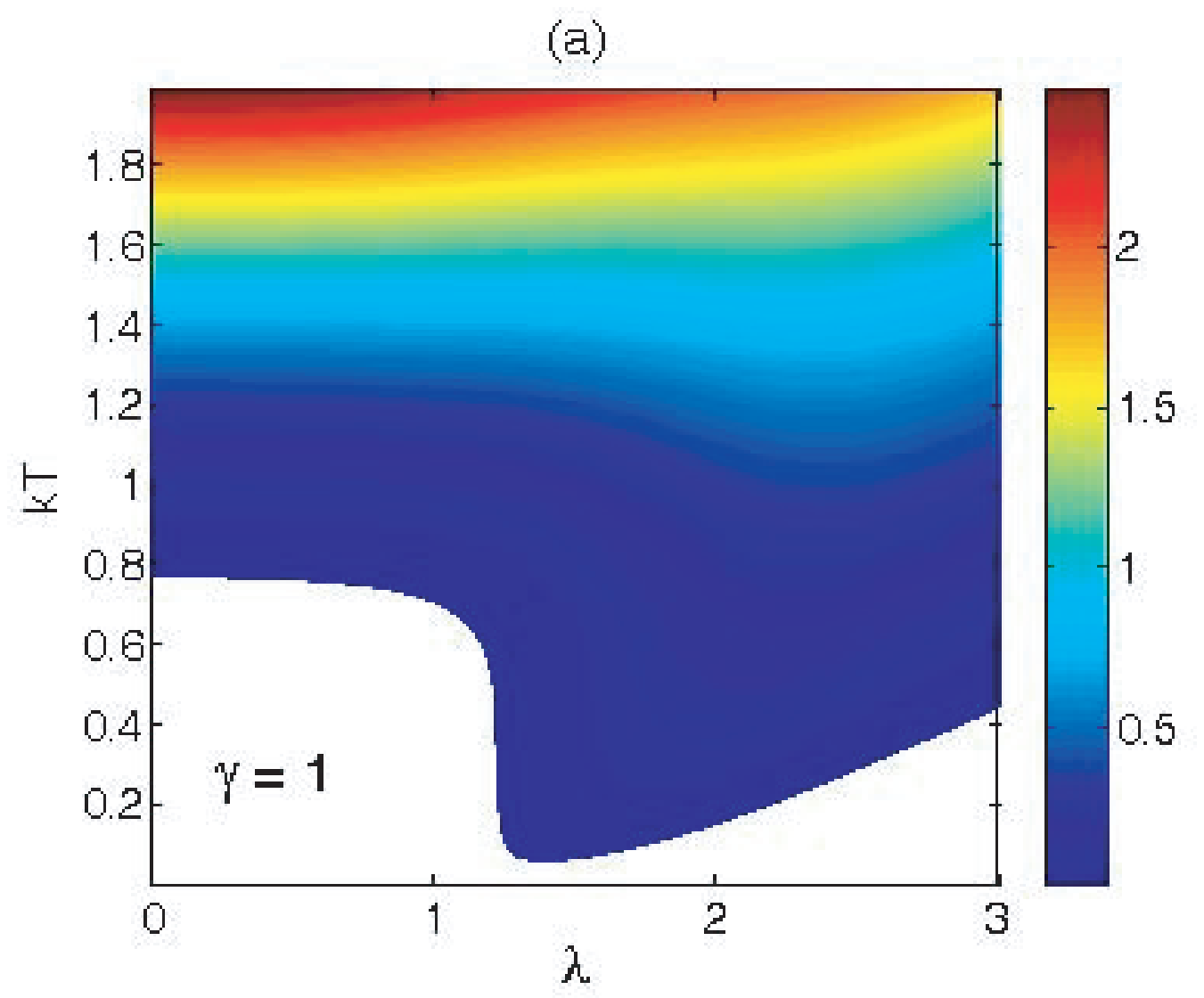}}\quad
   \subfigure{\label{fig:dE_g_05_c_L=10}\includegraphics[width=7.5 cm]{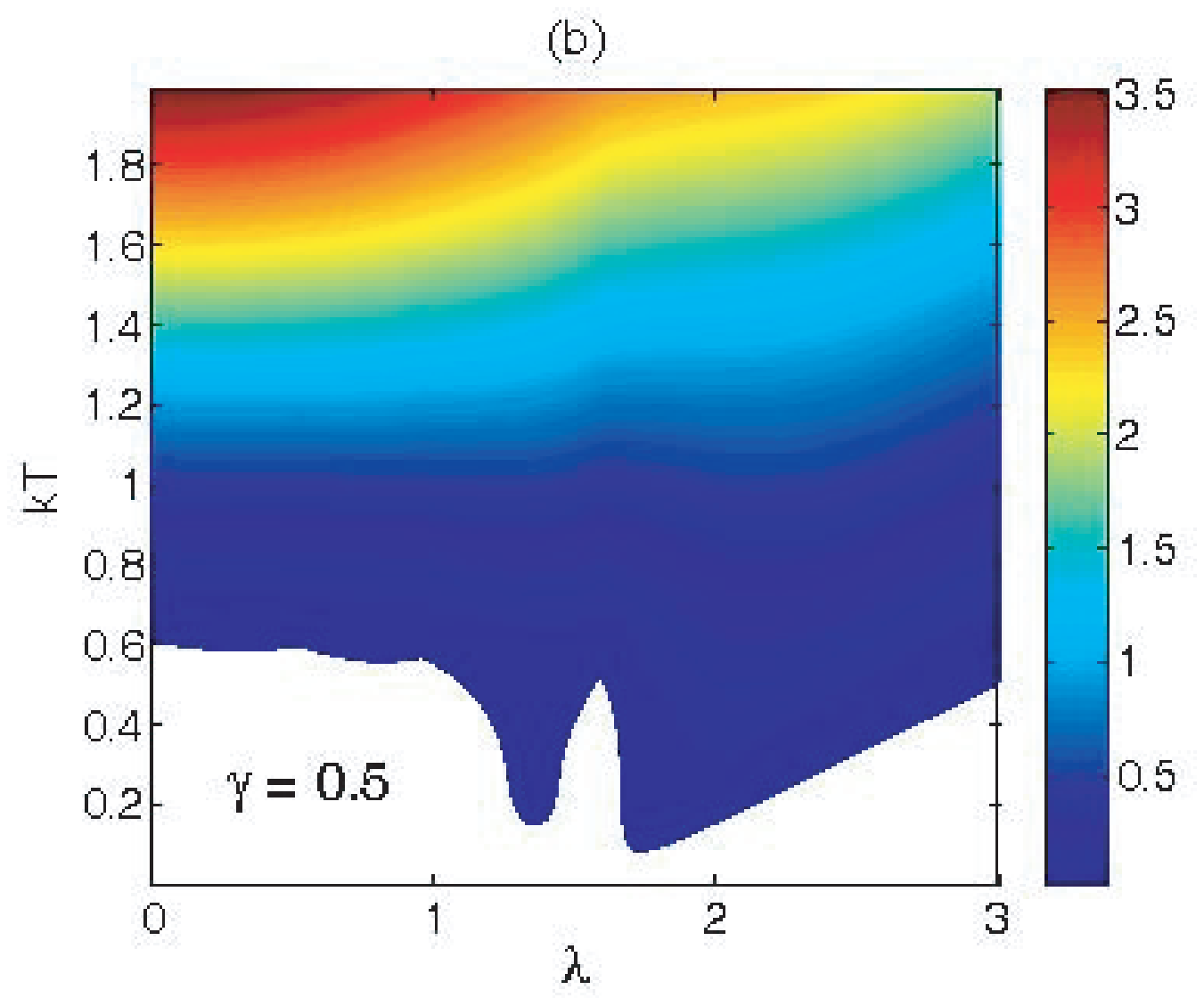}}\\
   \subfigure{\label{fig:dE_g_0_c_L=10}\includegraphics[width=7.5 cm]{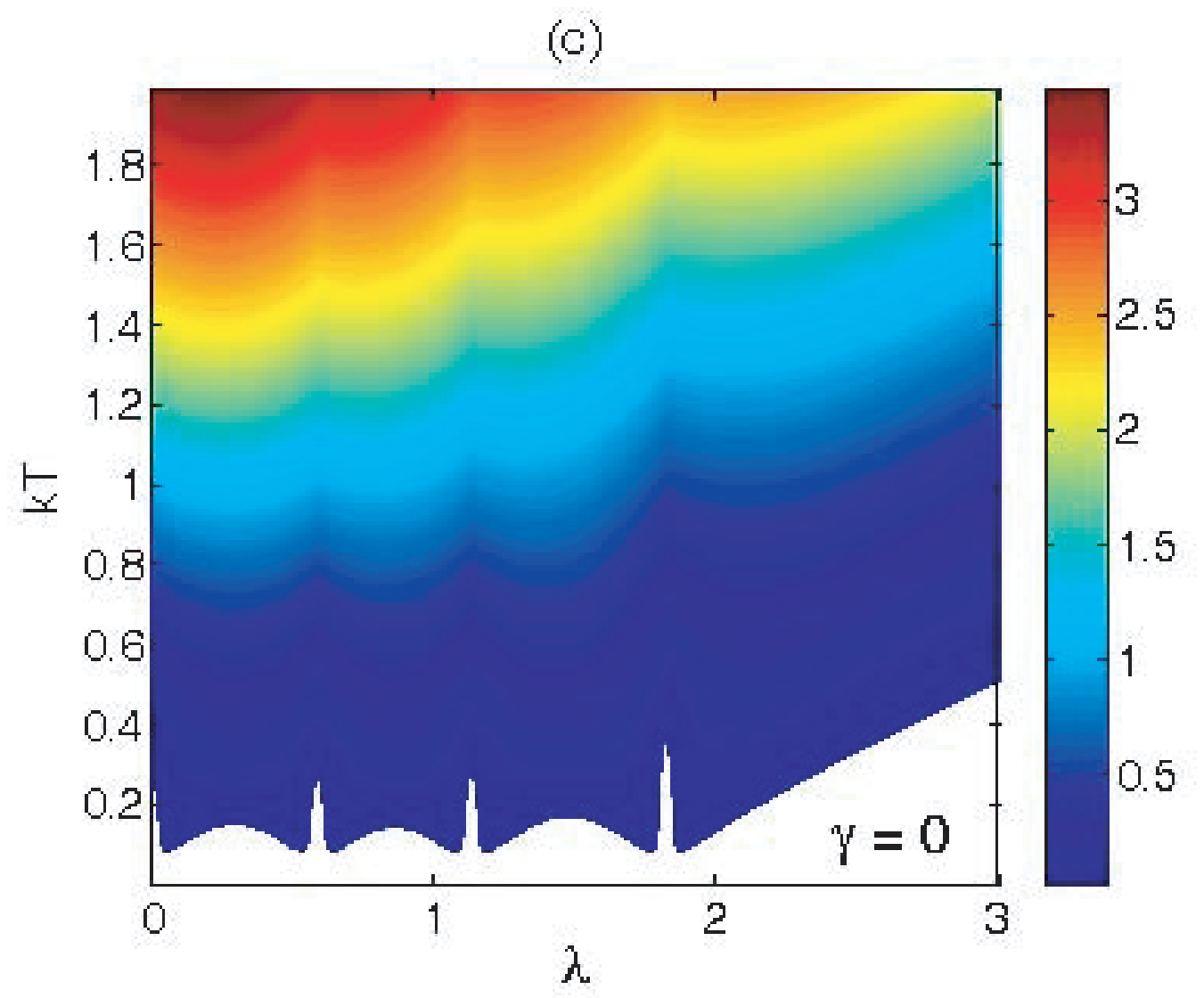}}\quad
   \subfigure{\label{fig:dE_g_0_c}\includegraphics[width=7 cm]{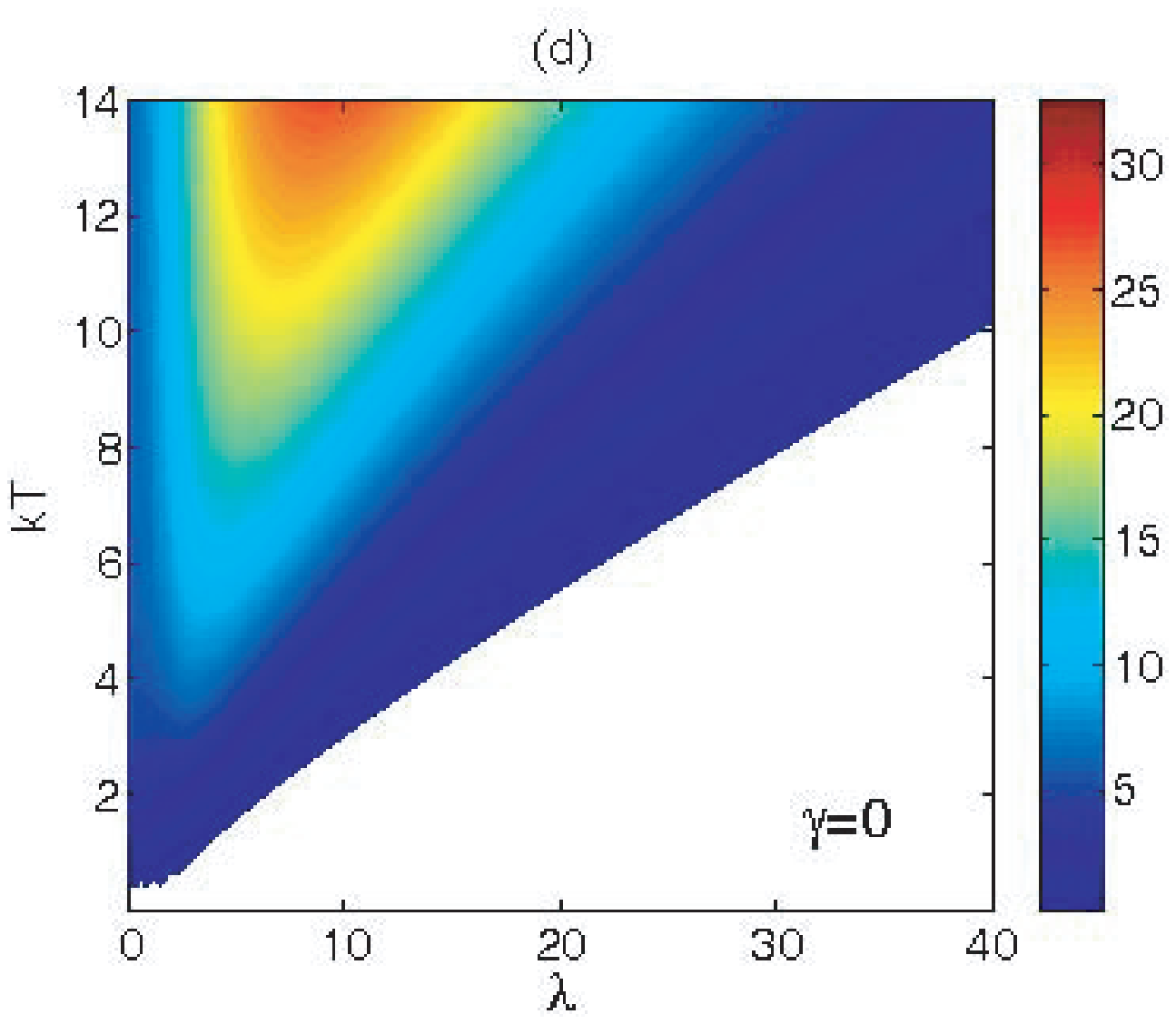}}\\   \caption{{\protect\footnotesize (Color online) The contour plot of thermal energy gap (in units of $J$) of the 2D spin system versus $\lambda$ and $kT$(in units of $J$) for (a) $\gamma=1$; (b) $\gamma=0.5$; (c) $\gamma=0$ and (d) $\gamma=0$.}}
 \label{dE_c_L=10}
 \end{minipage}
\end{figure}
%fig_7
%%%%%%%%%%%%%%%%%%%%%%%%%%%%%%%%%%%%%%%%%%%%%%%%%%%%%%%%%%%%%%%%%%%%%%%%%%%%%%%%%%%%%

It is of great interest to examine the relationship between the robustness of thermal entanglement and the corresponding thermal energy gap of the system. By the thermal energy gap, $\Delta E_{th}$, we mean the difference between the mean (ensemble average) energy of the system at temperature $T$ and the system ground state energy, i.e. $\Delta E_{th} = \langle E \rangle - E_0$. 

In fig.~\ref{dE_c_L=10} (a), (b), and (c) we present the contour plots of energy gap for the systems with $\gamma=1$, 0.5 and 0 respectively versus $\lambda$ and $kT$ for small values of $\lambda$. As can be seen, there is a clear differences between the different cases, where the energy gap in the Ising system shows one sharp minimum before increasing montonically as $\lambda$ increases which gives raise to one corresponding sharp peak of entanglement in that case at the small values of $\lambda$. The energy gaps in the partially anisotropic and isotropic systems show two and multiple minima respectively before they also increase montonically with $\lambda$, which causes the double peaks and multiple peaks, with different relative intensities, in the two systems respectively. This explains the different profiles of the entanglement peaks, at small values of the magnetic field, as the degree of anisotropy changes as demonstrated in figs.~\ref{Cc_G_1}, ~\ref{Cc_G_05} and ~\ref{Cc_G_0}. On the other hand, the thermal energy gap at the different anisotropic values looks asymptotically (at high magnetic field) the same, as shown in fig.~\ref{dE_c_L=10}(d) for the case $\gamma=0$ for instance. 

Remarkably, one can see a strong correspondence between the strength and survival of entanglement, particularly for $\gamma=0.5$ and 1 and the value of the energy gap when comparing figs.~\ref{Cc_G_1}, \ref{Cc_G_05} and \ref{dE_c_L=10}. The energy gap is either zero (the white regions of the contour plot) or quit small in the domains of non-zero entanglement. 
As can be noticed the thermal energy gap increases monotonically as the magnetic field intensity increases which explains the survival of the entanglement, despite its small magnitude, at relatively high temperatures and its strong resistance against thermal excitations compared to the high magnitude entanglement at the small values of the magnetic field which is very fragile to temperature. It is important to emphasis here that though the energy gap profile looks asymptotically almost the same for the three anisotropic parameter values, nevertheless the entanglement in the isotropic system ($\gamma=0$) in contrary to the other two cases vanishes at very low temperature regardless of the energy gap value and this is due to the fact that this system ground state, as we mentioned before, is disentangled for $\lambda \approx 1.85$ and higher.

%%%%%%%%%%%%%%%%%%%%%%%%%%%%%%%%%%%%%%%%%%%%%%%%%%%%%%%%%%%%%%%%%%%%%%%%%%%%%%%%%%%%%
\section{Robustness of thermal entanglement in Two-dimensional Spin System}
%%%%%%%%%%%%%%%%%%%%%%%%%%%%%%%%%%%%%%%%%%%%%%%%%%%%%%%%%%%%%%%%%%%%%%%%%%%%%%%%%%%%%
%%%%%%%%%%%%%%%%%%%%%%%%%%%%%%%%%%%%%%%%%%%%%%%%%%%%%%%%%%%%%%%%%%%%%%%%%%%%%%%%%%%%%
%%%%%%%%%%%%%%%%%%%%%%%%%%%%%%%%%%%%%%%%%%%%%%%%%%%%%%%%%%%%%%%%%%%%%%%%%%%%%%%%%%%%%
In order to study the multipartite entanglement of the entire lattice, a distance-like  measure of entanglement, namely the global robustness of entanglement $R(\rho)$ \cite{Vidal1999,Markham2008} is commonly used, which is defined for a general state $\rho$ as the minimum amount of noise t needed to destroy the entanglement content of $\rho$ and is given by
\begin{equation}
%R(\rho) := \stackrel{\textstyle{min}}{\omega} t  
R(\rho) := \mbox{min}_{\;\omega} \; \;  t \; ,
\end{equation}
where $\omega$ is the state when added to $\rho$ converts it to a separable state $\phi$ such that
\begin{equation}
\phi(\omega,t) := \left\lbrace \frac{1}{1+t}(\rho + t \; \omega) \right\rbrace \in \it{S} \; ,
\end{equation}
where $\it{S}$ is the set of all separable states. The resultant separable state $\phi$ can be regarded as a mixture of two states $\rho$ and $\omega$ with relative populations $1/(1+t)$ and $t/(1+t)$ respectively.    
This general approach can be applied, in particular, to a system in contact with a heat reservoir to determine the threshold temperature, $T_{th}$, below it the system is guaranteed to be entangled \cite{Markham2008}. Therefore, for instance in the spin system, if the state $\rho$ is identified as the ground state $\psi_0$ and the rest of the states $\{\psi_i\}$, which get mixed with $\psi_0$ as the temperature is raised, as $\omega$, then the population of the state $\rho$ is given by $1/(1+t)=e^{-E_0/kT}/Z$. As a result, the condition for the system to be guaranteed entanglement at a temperature $T$ will read  
\begin{equation}
\label{ineq_equ}
\frac{e^{-E_0/kT}}{Z} \; > \;  \frac{1}{1+R(\psi_0)} \; ,
\end{equation}
where $R(\psi_0)$ is the global robustness of the ground state $\psi_0$, which has an energy eigenvalue $E_0$. To obtain the threshold temperature one has to turn the inequality in Eq.(\ref{ineq_equ}) into an equality and we get
\begin{equation}
\frac{e^{-E_0/kT_{th}}}{Z} \; = \; \frac{1}{1+R(\psi_0)} \; .
\label{Threshold_temp}
\end{equation}
%%%%%%%%%%%%%%%%%%%%%%%%%%%%%%%%%%%%%%%%%%%%%%%%%%%%%%%%%%%%%%%%%%%%%%%%%%%%%%%%%%%%%%%%%%%%%
\begin{figure}[htbp]
\begin{minipage}[c]{\textwidth}
 \centering
   \subfigure{\label{fig:2D_E_G_1}\includegraphics[width=7.5 cm]{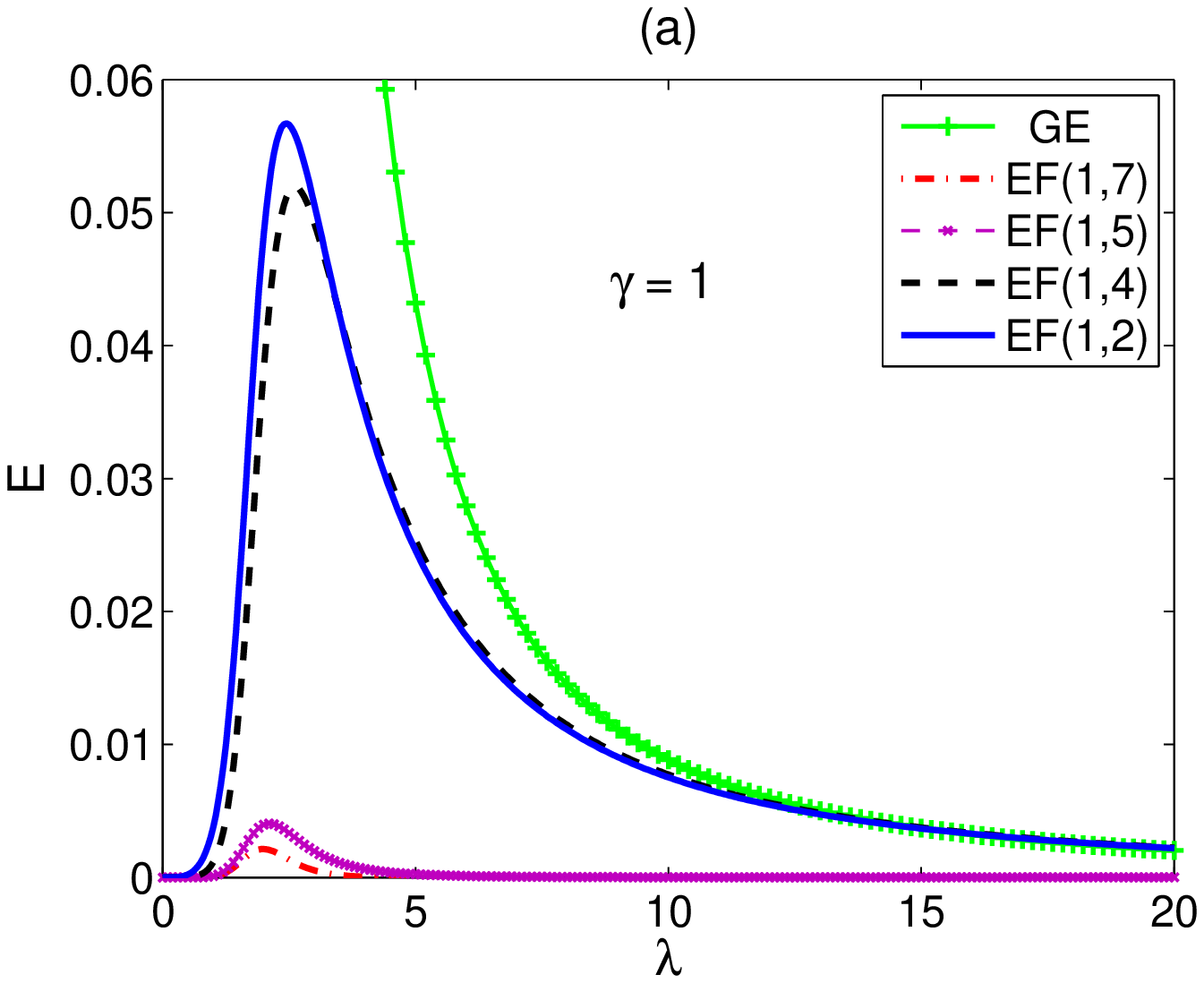}}\quad
   \subfigure{\label{fig:2D_E_G_05_close}\includegraphics[width=7.5 cm]{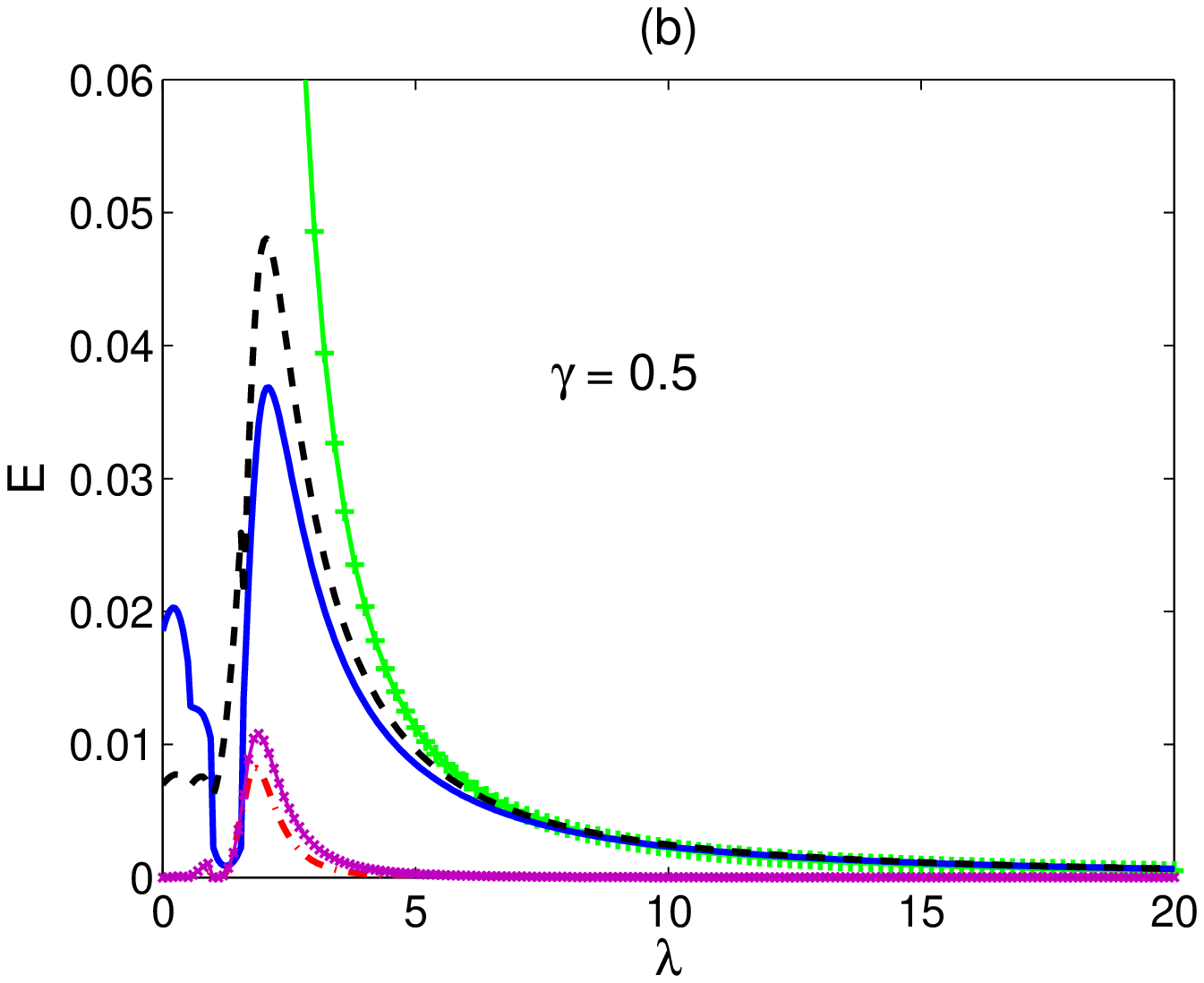}}\\
      \subfigure{\label{fig:2D_E_G_0}\includegraphics[width=7.5 cm]{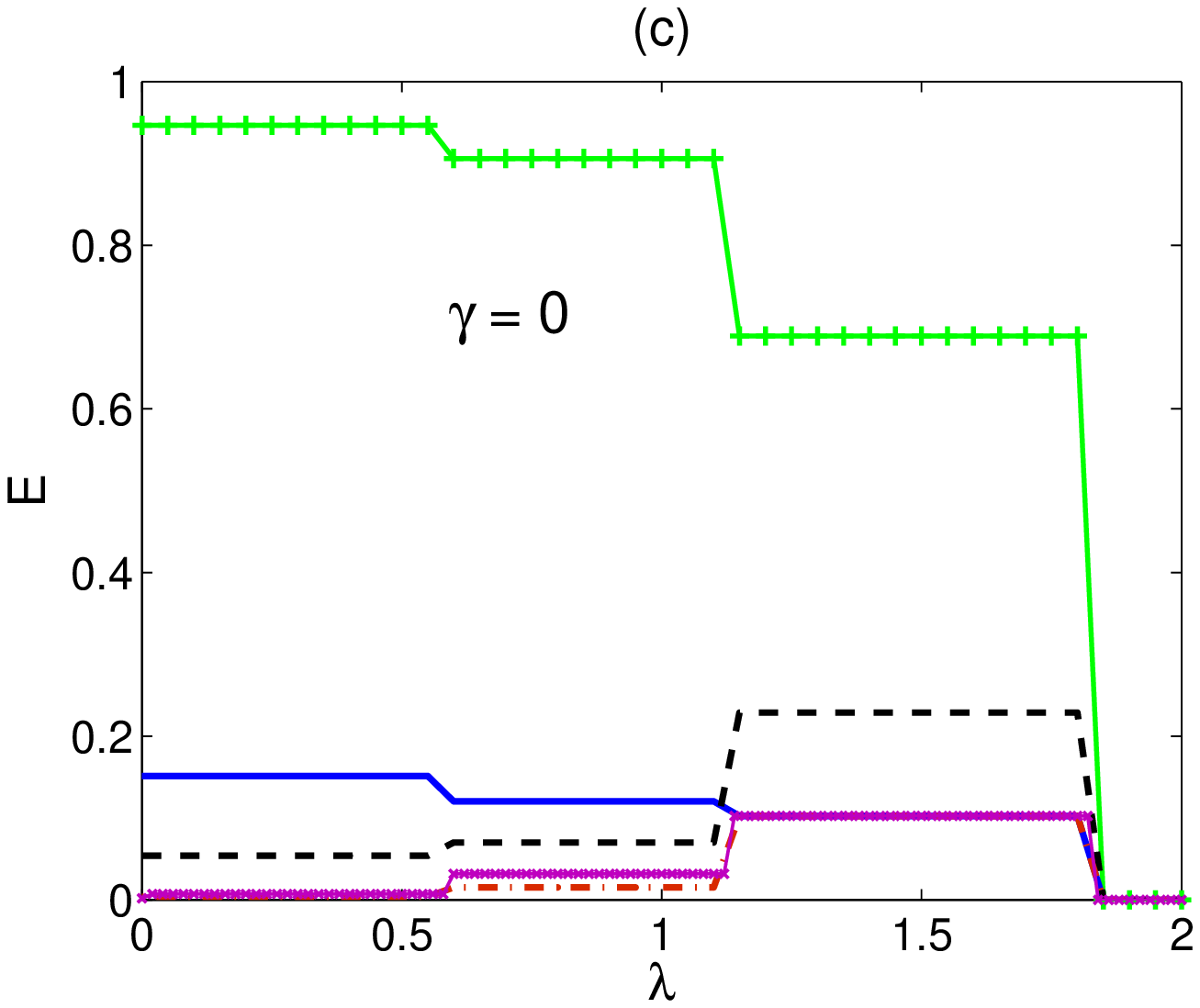}}
   \caption{{\protect\footnotesize (Color online) The entanglements $EF(1,2)$, $EF(1,4)$, $EF(1,5)$, $EF(1,7)$ and geometric entanglement of the 2D spin system versus $\lambda$ for $\gamma=0, \; 0.5$ and 1 at zero temperature. The legends are as shown in subfigure (a).}}
 \label{2D_E}
 \end{minipage}
\end{figure}
%fig_8
%%%%%%%%%%%%%%%%%%%%%%%%%%%%%%%%%%%%%%%%%%%%%%%%%%%%%%%%%%%%%%%%%%%%%%%%%%%%%%%%%%%%%%%%%
%%%%%%%%%%%%%%%%%%%%%%%%%%%%%%%%%%%%%%%%%%%%%%%%%%%%%%%%%%%%%%%%%%%%%%%%%%%%%%%%%%%%%%%%%%%%
%%%%%%%%%%%%%%%%%%%%%%%%%%%%%%%%%%%%%%%%%%%%%%%%%%%%%%%%%%%%%%%%%%%%%%%%%%%%%%%%%%%%%
\begin{figure}[htbp]
\begin{minipage}[c]{\textwidth}
 \centering
   \subfigure{\label{fig:2D_T_G_1}\includegraphics[width=7.5 cm]{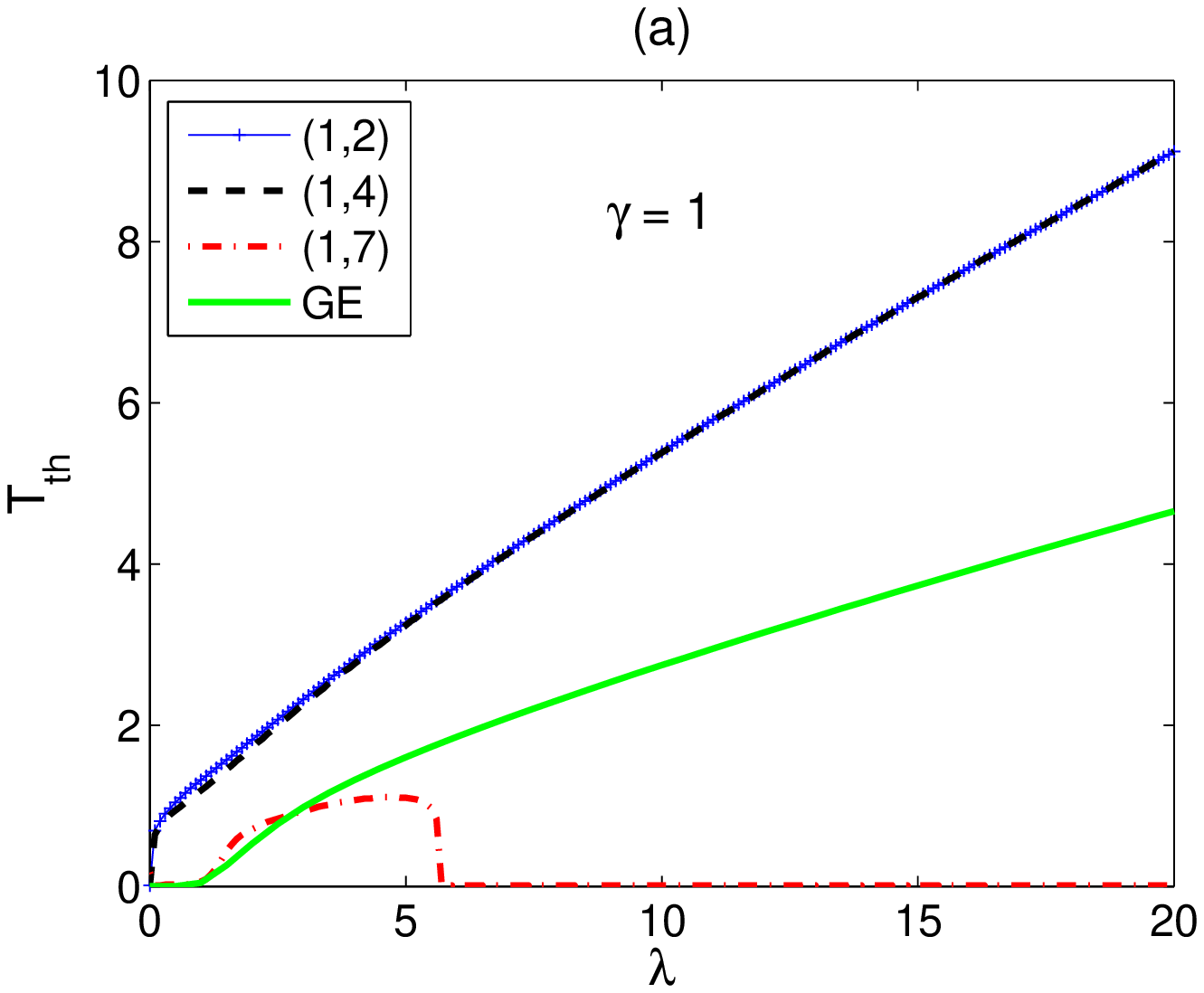}}\quad
   \subfigure{\label{fig:2D_T_G_05}\includegraphics[width=7.5 cm]{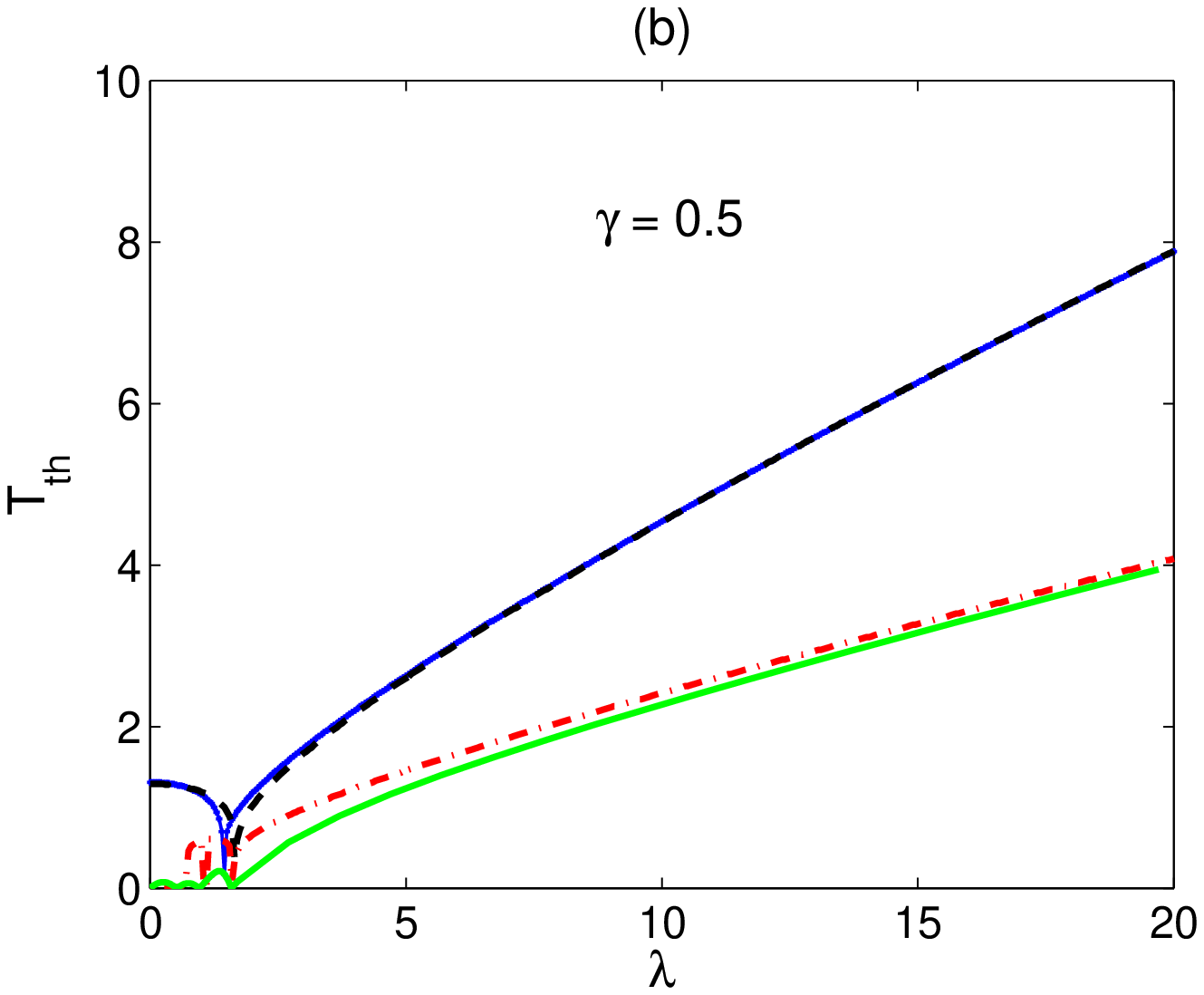}}\\
   \subfigure{\label{fig:2D_T_G_05_close}\includegraphics[width=7.5 cm]{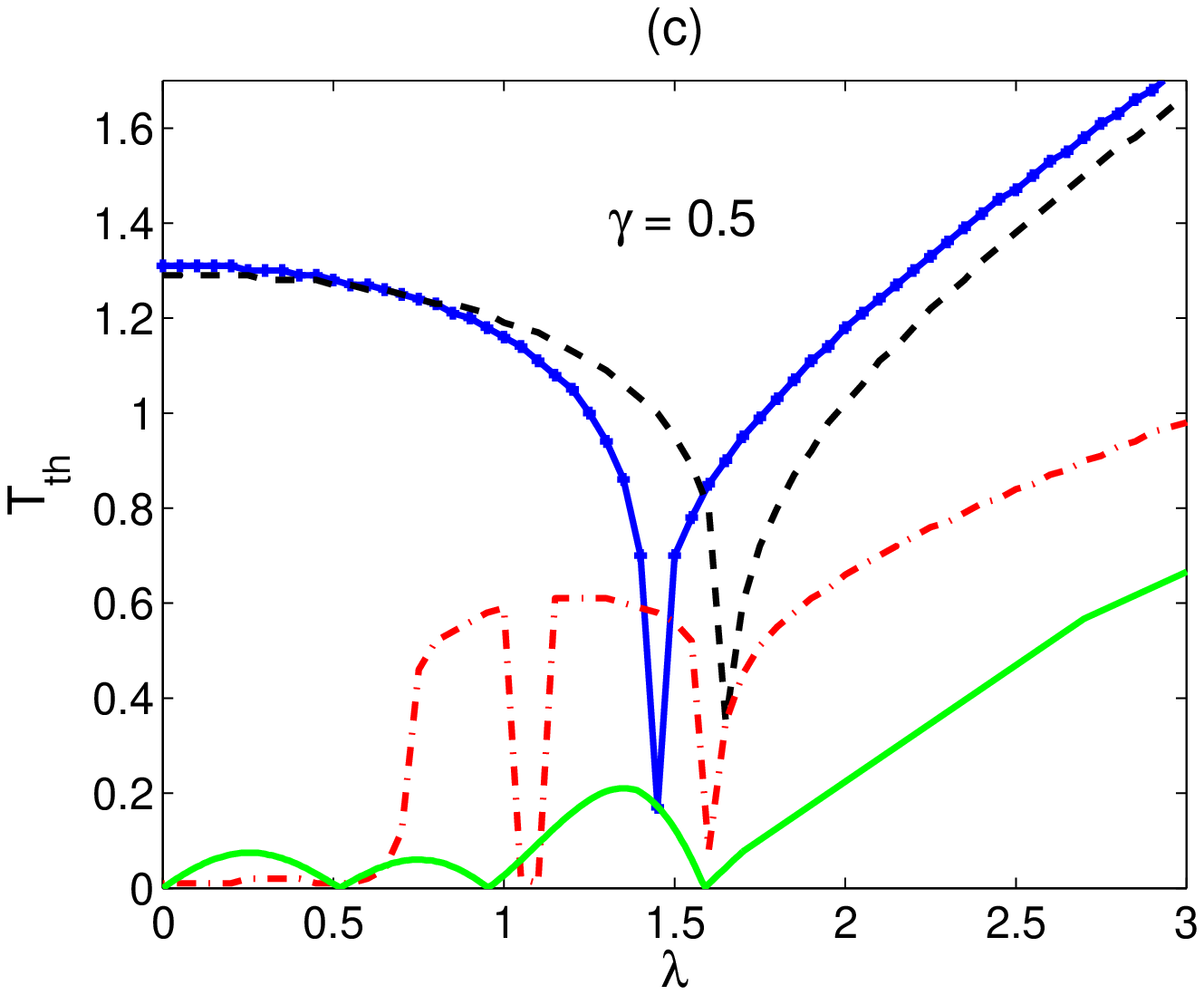}}\quad
      \subfigure{\label{fig:2D_T_G_0}\includegraphics[width=7.5 cm]{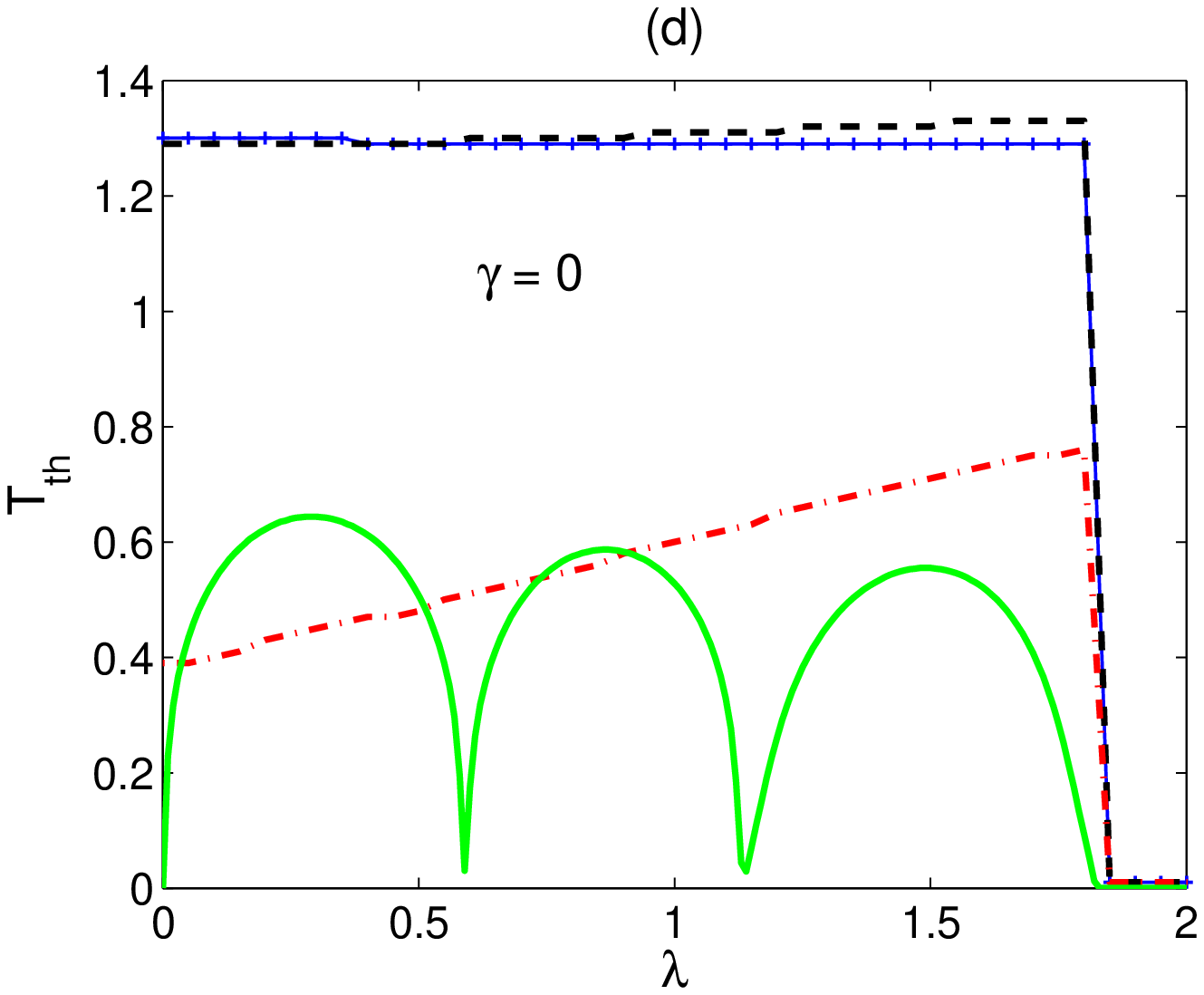}}\\
   \caption{{\protect\footnotesize (Color online) The threshold temperatures (in units of $J$) corresponding to the entanglements EF(1,2), EF(1,4), EF(1,7) and geometric entanglement of the 2D spin system versus $\lambda$ for $\gamma=0, \; 0.5$ and 1. The legends are as shown in subfigure (a).}}
 \label{2D_T}
 \end{minipage}
\end{figure}
%fig_9
%%%%%%%%%%%%%%%%%%%%%%%%%%%%%%%%%%%%%%%%%%%%%%%%%%%%%%%%%%%%%%%%%%%%%%%%%%%%%%%%%%%%%%%%%
%%%%%%%%%%%%%%%%%%%%%%%%%%%%%%%%%%%%%%%%%%%%%%%%%%%%%%%%%%%%%%%%%%%%%%%%%%%%%%%%%%%%%%%%%%%%
%%%%%%%%%%%%%%%%%%%%%%%%%%%%%%%%%%%%%%%%%%%%%%%%%%%%%%%%%%%%%%%%%%%%%%%%%%%%%%%%%%%%%
\begin{figure}[htbp]
\begin{minipage}[c]{\textwidth}
 \centering
   \subfigure{\label{fig:deltaE_g=0}\includegraphics[width=7.5 cm]{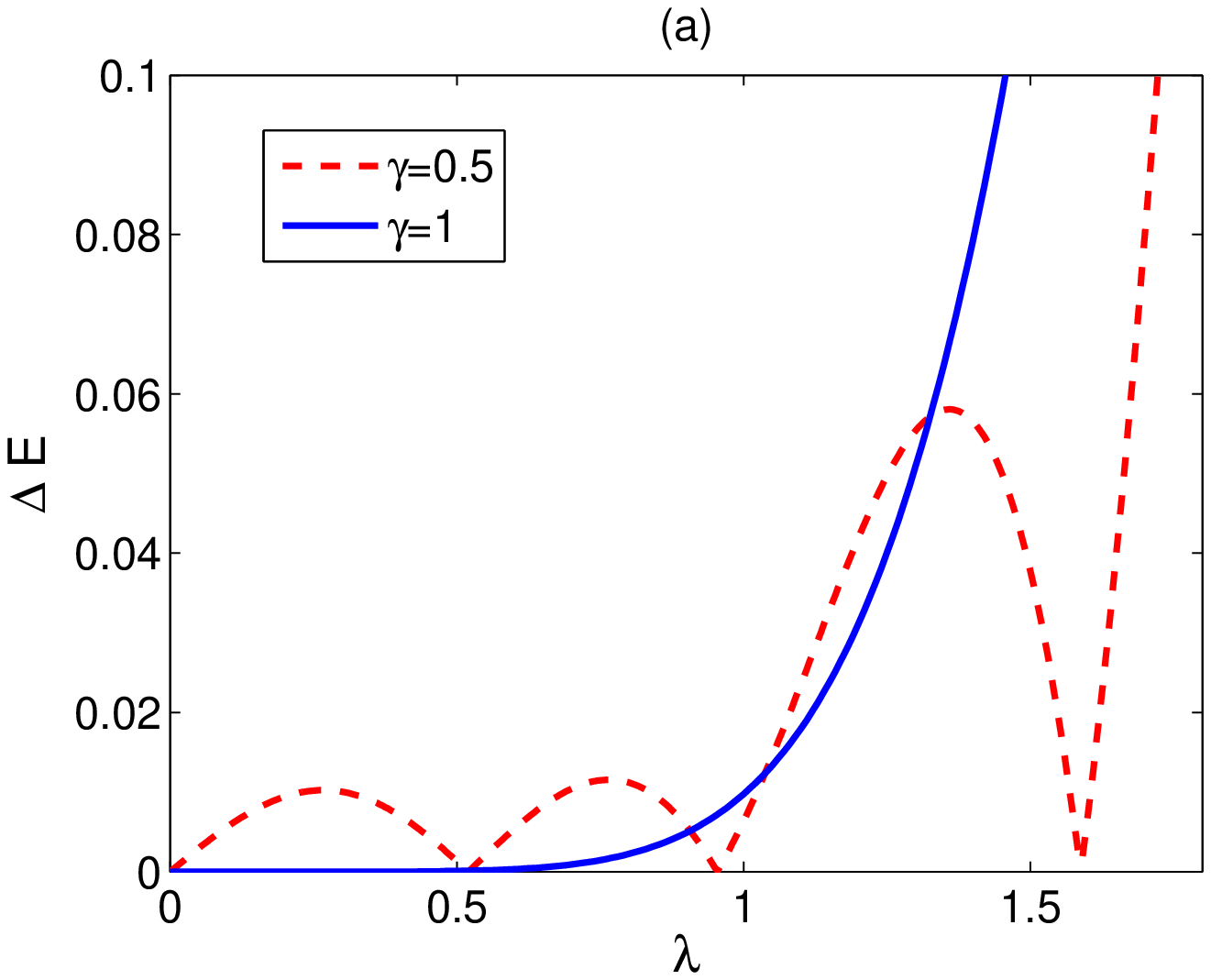}}\quad
   \subfigure{\label{fig:deltaE_g=05}\includegraphics[width=7.5 cm]{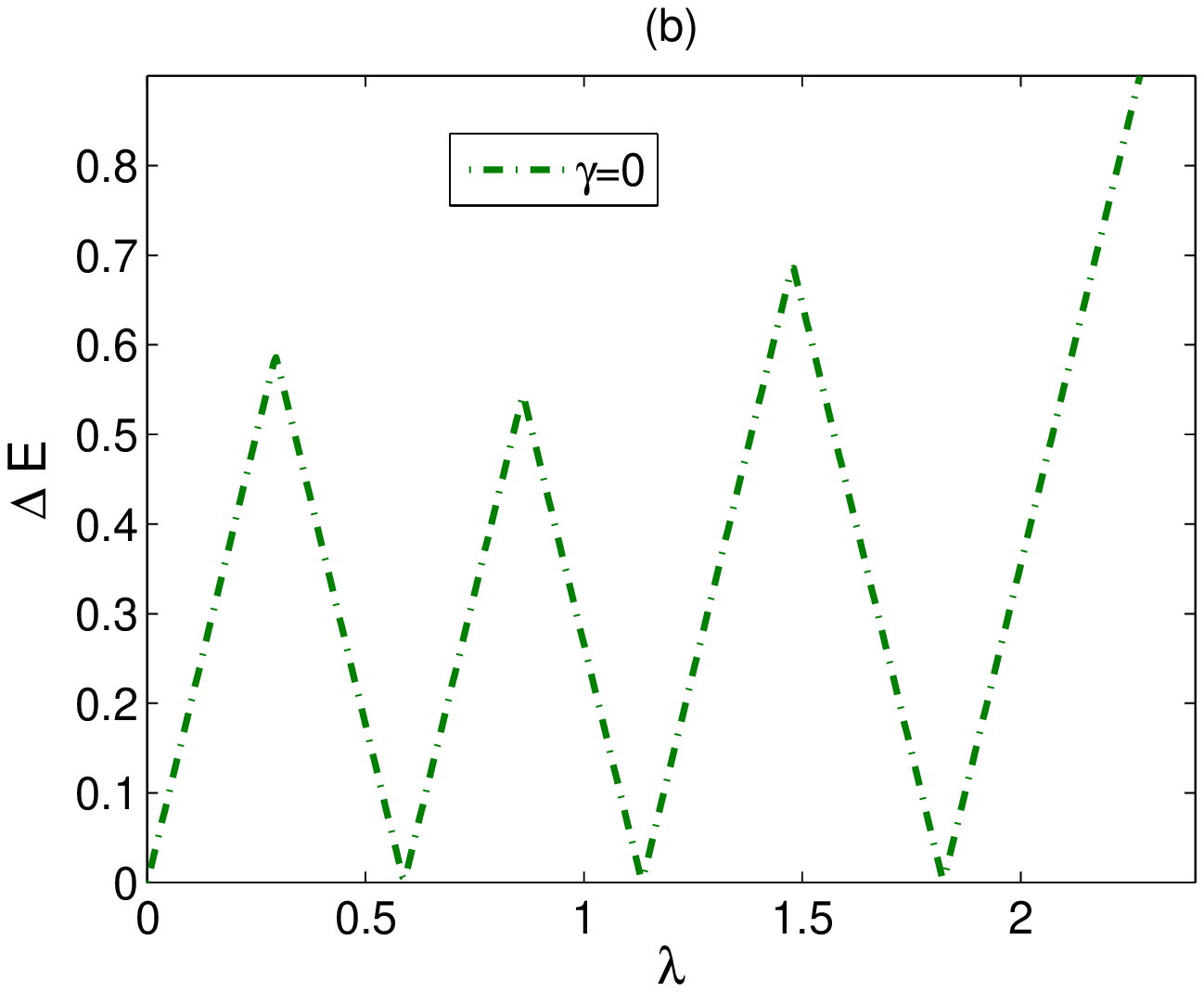}}\\
 \subfigure{\label{fig:deltaE_g=1}\includegraphics[width=7.5 cm]{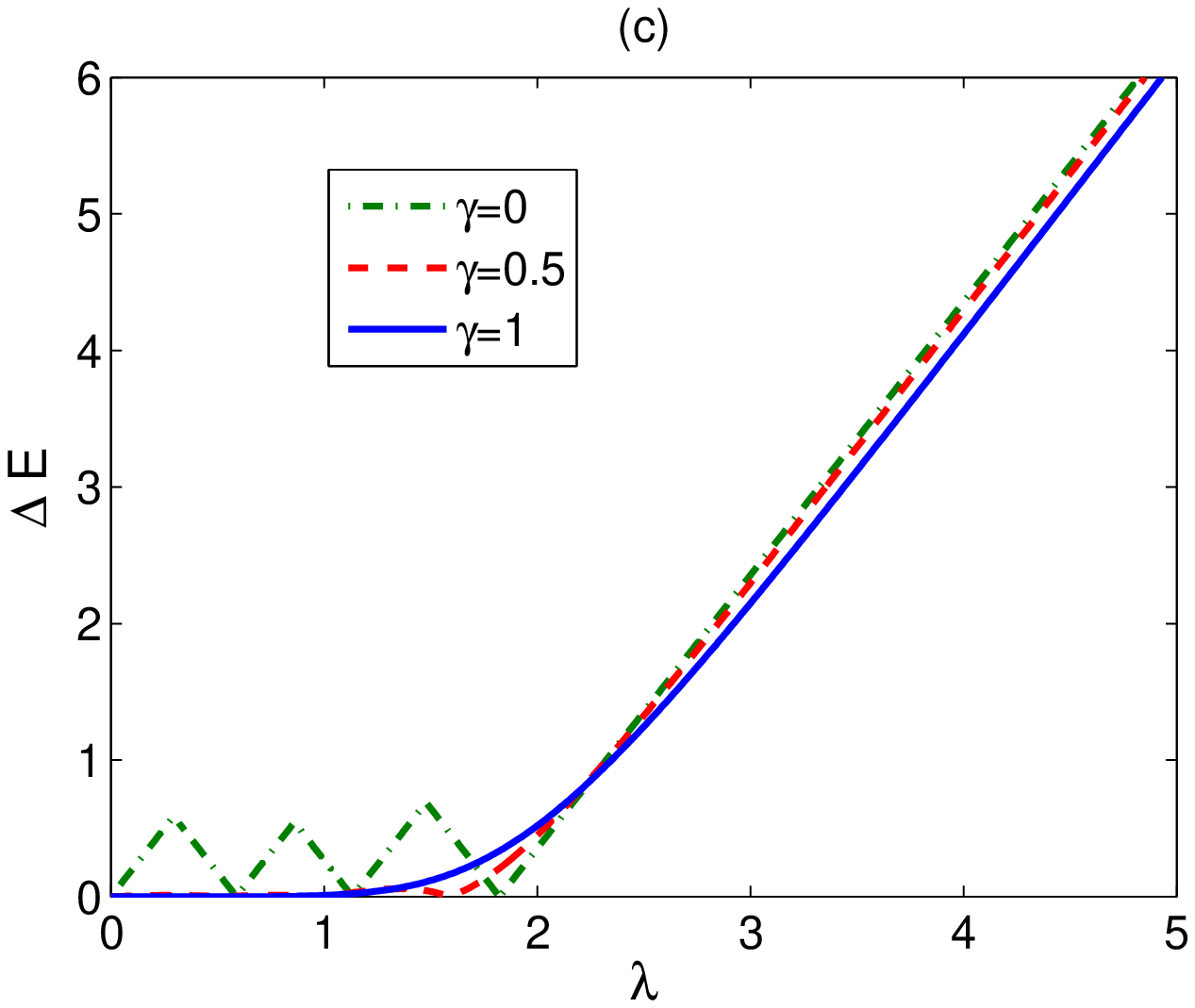}}\quad
   \caption{{\protect\footnotesize (Color online) The energy gap (in units of $J$) of the 2D spin system versus $\lambda$ for $\gamma=0, \; 0.5$ and 1.}}
 \label{2D_dE}
 \end{minipage}
\end{figure}
%fig_10
%%%%%%%%%%%%%%%%%%%%%%%%%%%%%%%%%%%%%%%%%%%%%%%%%%%%%%%%%%%%%%%%%%%%%%%%%%%%%%%%%%%%%
%%%%%%%%%%%%%%%%%%%%%%%%%%%%%%%%%%%%%%%%%%%%%%%%%%%%%%%%%%%%%%%%%%%%%%%%%%%%%%%%%%%%%
\begin{figure}[htbp]
\begin{minipage}[c]{\textwidth}
 \centering
   \subfigure{\label{fig:compare_a}\includegraphics[width=7.5 cm]{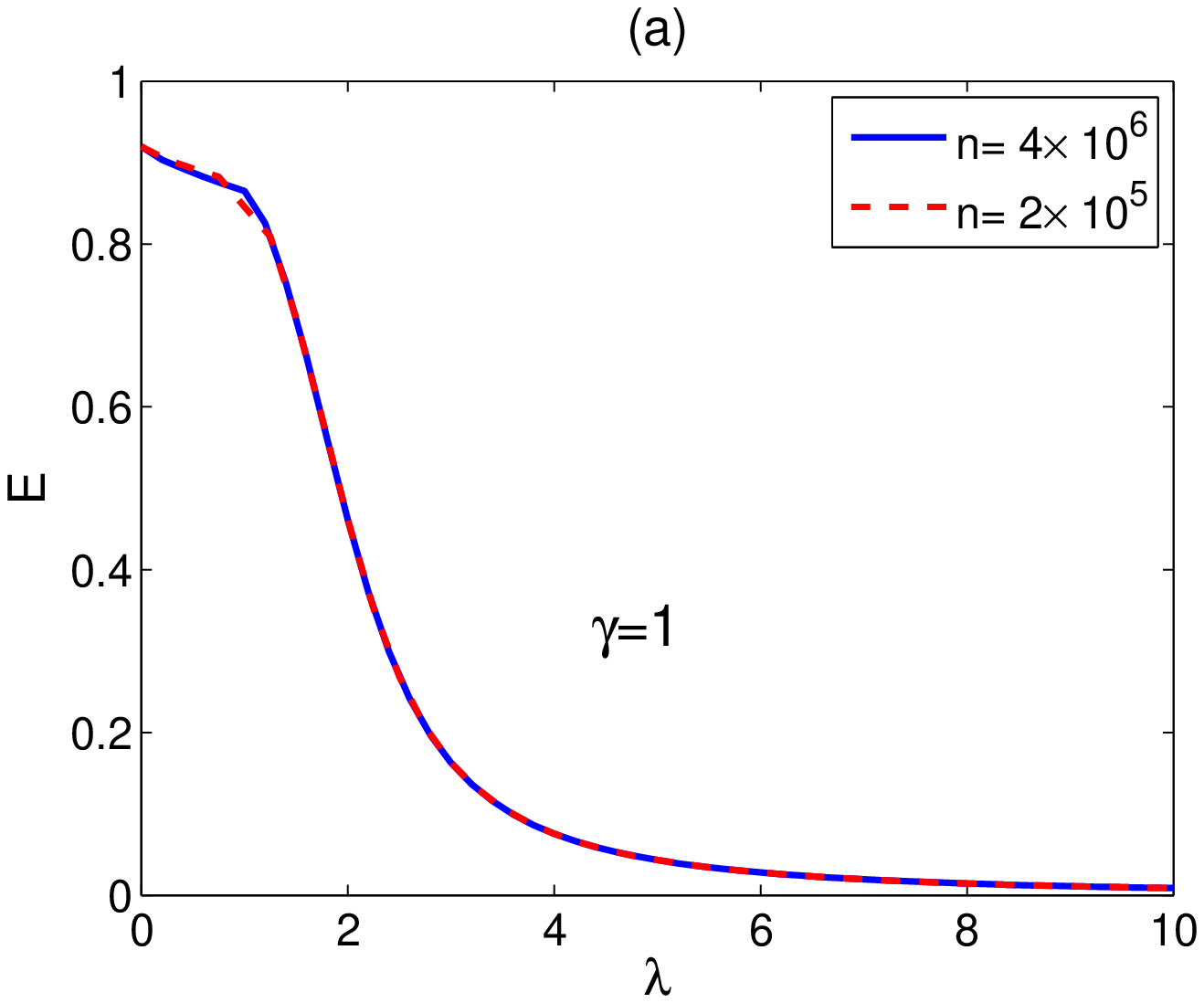}}\quad
   \subfigure{\label{fig:compare_b}\includegraphics[width=7.5 cm]{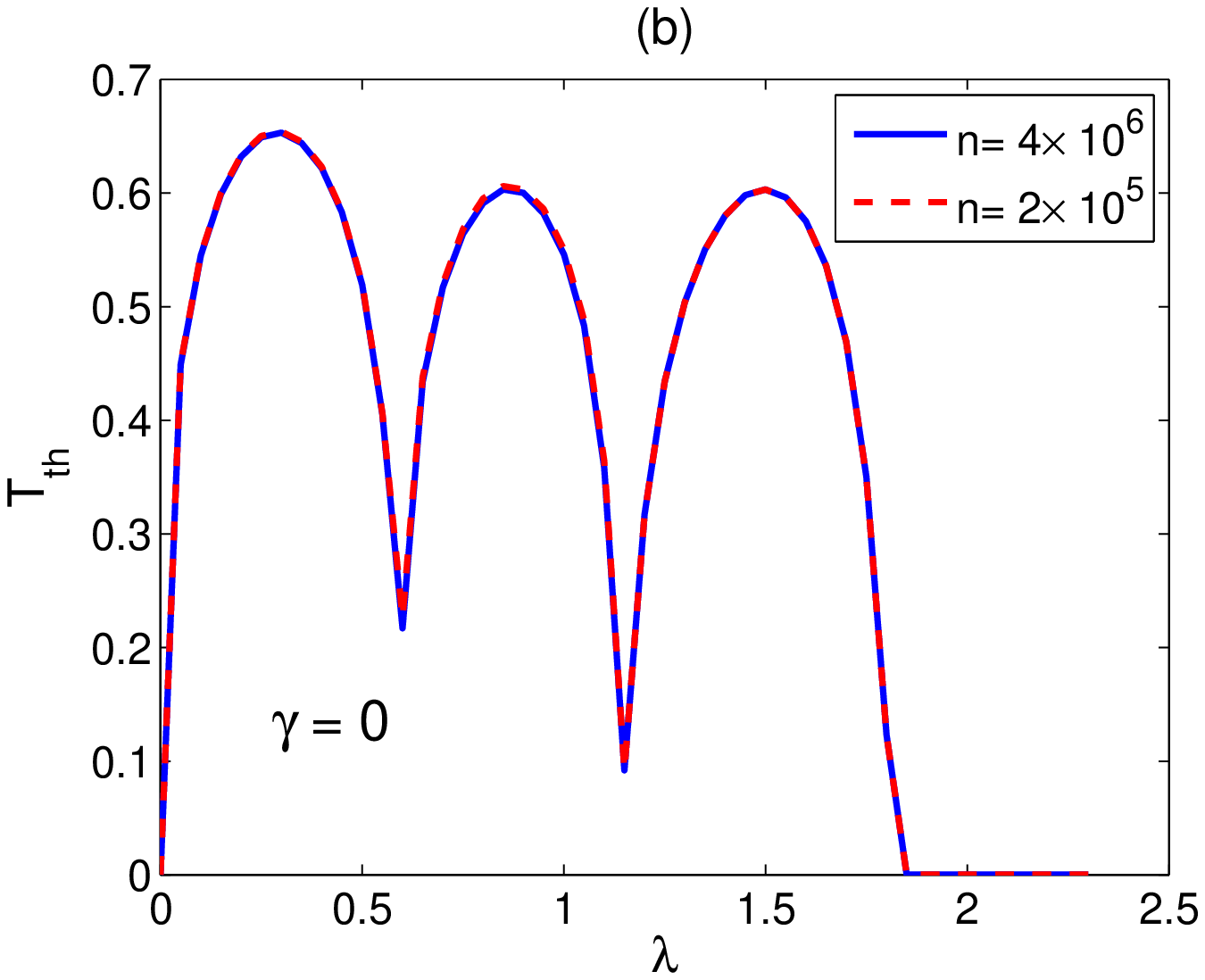}}\\
   \caption{{\protect\footnotesize (Color online) (a) The geometric entanglement of the 2D Ising system ($\gamma=1$) and (b) the threshold temperature (in units of $J$) of the 2D isotropic system ($\gamma=0$), for two different numbers of the set of parameters $\{P_i\}$, $n=2\times 10^5$ and $4\times 10^6$.}}
 \label{compare}
 \end{minipage}
\end{figure}
%fig_11
%%%%%%%%%%%%%%%%%%%%%%%%%%%%%%%%%%%%%%%%%%%%%%%%%%%%%%%%%%%%%%%%%%%%%%%%%%%%%%%%%%%%%%%
To determine the threshold temperature, $T_{th}$, one has to evaluate the robustness of entanglement of the ground state $R(\psi_0)$ which is very difficult task, where the entire system Hilbert state has to be searched for the noise mixed state $\omega$. However, it was found that a lower bound for the robustness of entanglement can be obtained \cite{Markham2008} by evaluating the geometric entanglement $G(\psi_0)$ instead \cite{Wei2003}, which is easier to evaluate, where in general for any pure state $\psi$
\begin{equation}
\frac{1}{1+R(\psi)} \; \leq  \; \frac {1}{2^{\; G(\psi)}} \;,
\label{Robustness_GE}
\end{equation}
which would enable us calculating a lower bound for the threshold temperature, where below it the system is guaranteed to be entangled. In all the figures we simply denote this temperature as $T_{th}$.

The geometric measure of multipartite entanglement utilizes the geometric properties of the Hilbert space to find the distance (or angle) between a pure state, $\psi$, representing the system and the closest pure separable state, $\phi$ to it, i.e. $ ||\; |\psi \rangle - |\phi \rangle \;||$. The square $sine$ of the angle between the two states $\psi$ and $\phi$ represents a good measure of the global geometric entanglement, where the smallest value of the square $\sin$ specifies the closet separable state to the pure state $\psi$ and is defined by
\begin{equation}
G(\psi) := 1 - [\; \mbox{max}_{\;\phi} \; || \langle \psi |\phi \rangle || \;]^2 \; ,
\label{GE_def}
\end{equation} 
where $|| \langle \psi |\phi \rangle ||$ represents cosine the angle between the two states $\psi$ and $\phi$. By evaluating the the geometric entanglement (GE), and using Eqs. (\ref{Threshold_temp}) and (\ref{Robustness_GE}) we can find the lower limit of the threshold temperature $T_{th}$.

In order to find the closet separable state to the state $\psi$, we assume an arbitrary separable state $\phi$ as a product of the single spin states of the 7 spins which takes the form
\begin{equation}
|\phi \rangle = \prod_{i=1}^{i=7} \{ P_i |0\rangle + \sqrt{1-P_{i}^2} e^{i \delta} |1\rangle \} \; .
\end{equation}
Utilizing the reality of the wavefunction, where the eigenstates of this class of Hamiltonians are real, we set the azimuthal angle $\delta=0$ \cite{Wei2003,Wei2005}. In addition, we have also examined numerically the independence of the results on the azimuthal angle.
The set of parameters $\{P_i, i=1,2,\cdots 7 \}$ has to be varied over its entire range to cover the whole system Hilbert space searching for the closet distance between $\psi$ and $\phi$ where $0 \leq P_i \leq 1$. Searching the entire Hilbert space is a computationally difficult task and therefore we have tested around $4 \times 10^6$ different distinct set of values for the P's parameters, uniformally distributed over the system Hilbert space, for the calculations of the geometric entanglement.

In this section we investigate only the two dimensional spin system.
In fig.~\ref{2D_E}, we compare the bipartite entanglements $EF(1,2)$, $EF(1,4)$, $EF(1,5)$ and $EF(1,7)$ with the multipartite geometric entanglement $GE$ versus the parameter $\lambda$ at zero temperature for different degrees of anisotropy. As can be noticed, the behavior of the nnn entanglement $EF(1,5)$ is very close to the nnnn entanglement $EF(1,7)$ for all degrees of anisotropy. In fig.~\ref{2D_E}(a), the nearest neighbour bipartite entanglements $EF(1,2)$ and $EF(1,4)$ of the Ising system after reaching its peak value at around $\lambda=2.5$, they decay as $\lambda$ increases, whereas the nnn and nnnn bipartite entanglements $EF(1,5)$ and $EF(1,7)$ reach exactly zero magnitude at small values of the magnetic field.
On the other hand, the multipartite entanglement $GE$ starts with large magnitude $\approx 0.92$ at $\lambda=0$ then decays abruptly as $\lambda$ increases before it asymptotically approaches the nearest neighbor entanglements, where all sustain, with quite small magnitudes up to large values of the magnetic field. In fact, examining the nearest neighbor bipartite and geometric entanglements at very large magnetic field strength shows that they reach quite small values that are of the same order of magnitude, for instance at $\lambda=300$ they are of the order of $10^{-5}$, though the magnitude of $GE$ is always less than that of $EF$.
A similar behavior of the bipartite and multipartite entanglements is observed again in the partially anisotropic system as shown in fig.~\ref{2D_E}(b), but in this case, $EF(1,5)$ and $EF(1,7)$ sustain to larger value of $\lambda$ reaching very small magnitudes compared to $EF(1,2)$ and $EF(1,4)$.
The behavior of the entanglement in the isotropic system is depicted in fig. ~\ref{2D_E}(c) where all types of entanglement vanish at $\lambda \approx 1.85$ and as we mentioned before this stems from the fact that the ground state of the system is separable at this value and higher. In fig.~\ref{2D_T} we compare the threshold temperature of the bipartite entanglements, which is the temperature at which the bipartite entanglement vanish as was demonstrated in figs. \ref{Cc_G_1}, \ref{Cc_G_05} and \ref{Cc_G_0}, to the threshold temperature of the multipartite geometric entanglement, as defined before versus the parameter $\lambda$. In the Ising model, explored in fig.~\ref{2D_T}(a), the threshold temperatures of the nearest neighbour bipartite entanglements $EF(1,2)$ and $EF(1,4)$ are very close and increase monotonically as the magnetic field increases. On the other hand, $T_{th}$ for $EF(1,7)$ is very close to that of the geometric entanglement at small values of the magnetic field where it increases monotonically but suddenly drops to zero around $\lambda=6$, whereas $T_{th}$ for the geometric entanglement maintains its monotonic behavior but is much smaller than $T_{th}$ for $EF(1,2)$ and $EF(1,4)$. In fig.~\ref{2D_T}(b), the threshold temperatures of the partially anisotropic system, $\gamma=0.5$, behave in a similar way to the isotropic case where the temperatures for the nearest neighbor bipartite entanglements are very close but what is even more interesting is that the threshold temperatures for the nnnn bipartite sustains as the magnetic field increases and asymptotically becomes very close to that of the geometric entanglement.

This means that the multipartite entanglement over the lattice along with the bipartite entanglement between the far spins (such as $EF(1,7))$ are more fragile to temperature than the bipartite entanglement between the nearest neighbor spins, such as $EF(1,2)$ and $EF(1,4)$, which manifests higher resistance and assumes higher magnitude compared to GE and $EF(1,7)$ at the same temperature.
A closer look at the behavior of the threshold temperatures of the partially anisotropic system at small values of the magnetic field is given in fig.~\ref{2D_T}(c). One can see sharp changes in the threshold temperatures specially for the nnnn entanglement and the geometric entanglement, which can be explained in terms of the energy gap of the system as will be discussed shortly. The threshold temperatures of the completely isotropic system, $\gamma=0$, is explored in fig.~\ref{2D_T}(d), where again the threshold temperatures of the nearest neighbor entanglements $EF(1,2)$ and $EF(1,4)$ are very close and maintain an almost constant value before suddenly dropping to zero at $\lambda \approx 1.85$. The threshold temperature of $EF(1,7)$, which is considerably lower than that of the nearest neighbors, increases linearly before suddenly dropping to zero also at the same value $\lambda \approx 1.85$. The threshold temperature of the multipartite entanglement exhibits an oscillatory behavior with an average value within that of the nnnn bipartite value.

Figure~\ref{2D_dE} explains the sharp changes in the threshold temperatures at the small range of values of the magnetic field. As can be noticed in fig.~\ref{2D_dE}(a), (b) and (c), the energy gap of the Ising system increases monotonically over the entire $\lambda$ range with no sharp changes, which explains the smooth monotonic increase in the threshold temperatures shown in fig.~\ref{2D_T}(a). In contrary, the energy gaps in the partially anisotropic and isotropic systems exhibit oscillating and sharp oscillating changes respectively at the small values of the parameter $\lambda \leq 1.85$ before coinciding with the anisotropic curve and increasing monotonically as shown in fig.~\ref{2D_dE}. Interestingly, by comparing the behavior of the threshold temperatures, particularly of the multipartite entanglement, in fig.~\ref{2D_T} to that of the energy gaps in fig.~\ref{2D_dE} one can notice the strict correspondence between them; the minima (and maxima) in the threshold temperatures coincide with that of the energy gaps on the parameter $\lambda$ scale and when the energy gap increases monotonically, the threshold temperature follows that behavior too. The impact of the energy gap on the multipartite entanglement is stronger compared to the bipartite entanglement due to the fact that the energy gap is calculated for the entire (multipartite) system.
The effect of the number of distinctive set of parameters $\{P_i, i=1,2,\cdots 7 \}$ on the accuracy of the results is examined in fig.~\ref{compare} where two different numbers, $2 \times 10^5$ and $4 \times 10^6$ are compared in plotting the multipartite entanglement for $\gamma=1$ and the threshold temperature for $\gamma=0.5$ , which shows a very strong coincidence.
%%%%%%%%%%%%%%%%%%%%%%%%%%%%%%%%%%%%%%%%%%%%%%%%%%%%%%%%%%%%%%%%%%%%%%%%%%%%%%%%%%%%%%%%%
%%%%%%%%%%%%%%%%%%%%%%%%%%%%%%%%%%%%%%%%%%%%%%%%%%%%%%%%%%%%%%%%%%%%%%%%%%%%%%%%%%%%%%%%%
\begin{figure}[htbp]
\begin{minipage}[c]{\textwidth}
 \centering
   \subfigure{\label{Imp_g_1_a}\includegraphics[width=7.5 cm]{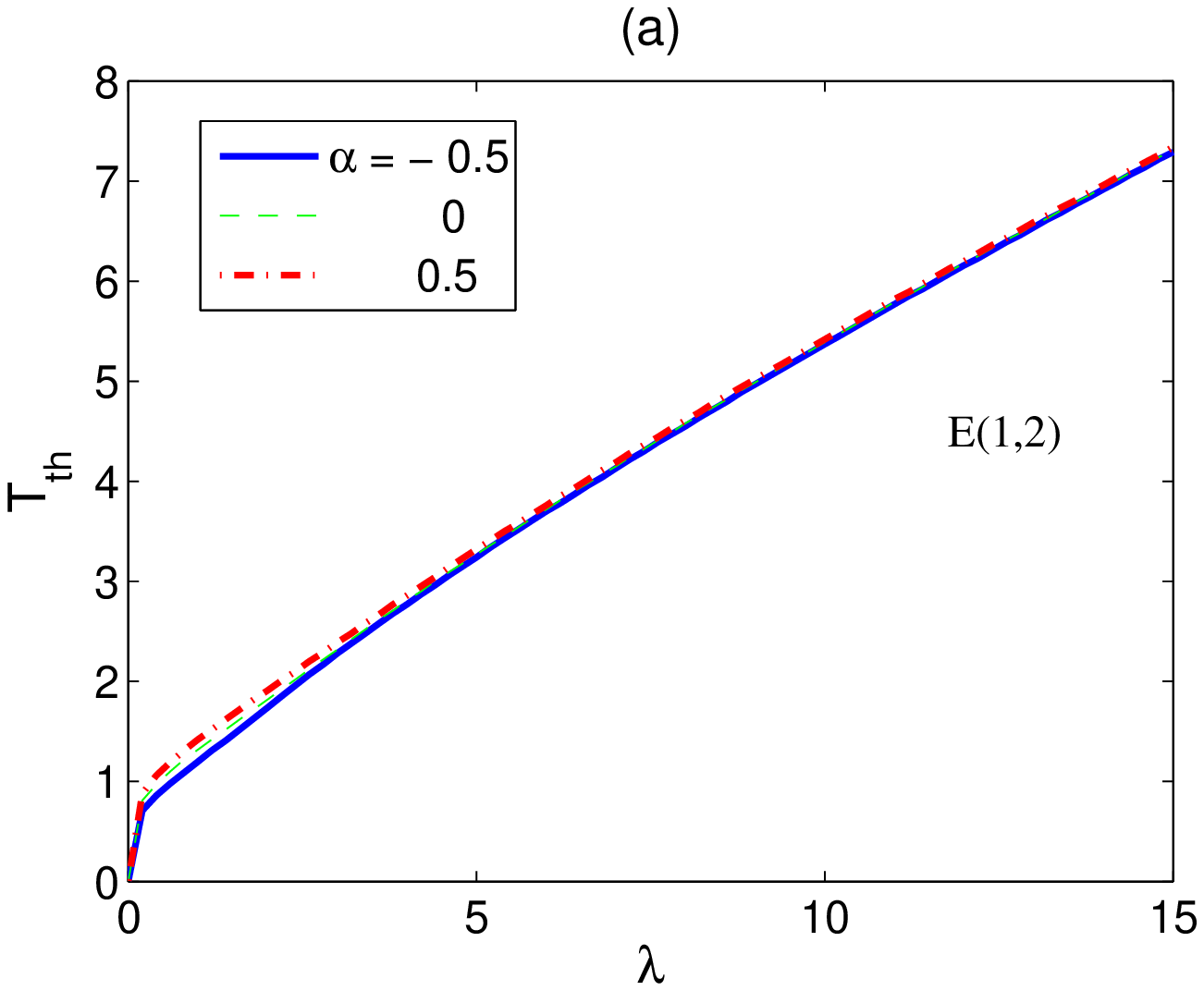}}\quad
   \subfigure{\label{Imp_g_1_b}\includegraphics[width=7.5 cm]{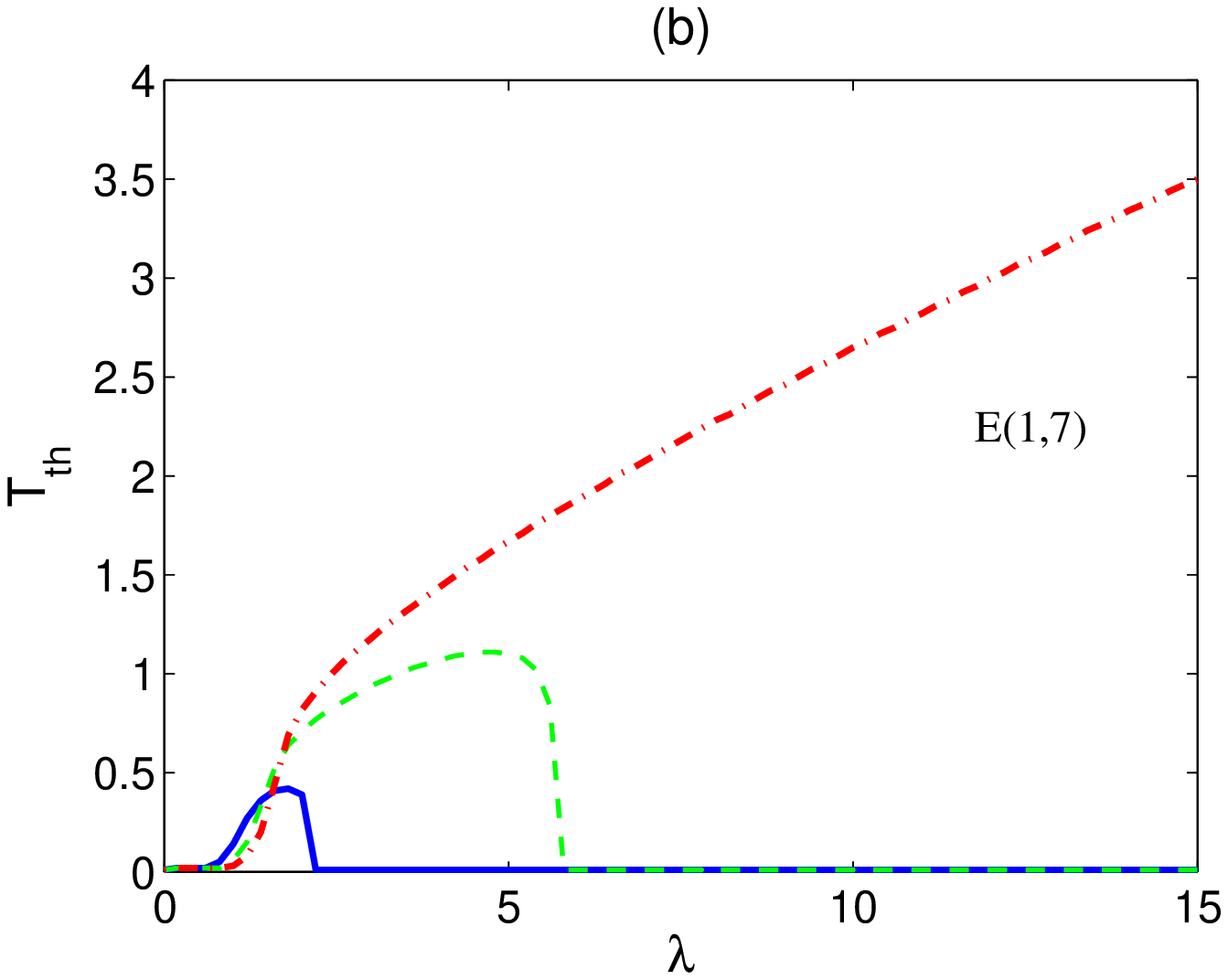}}\\
   \subfigure{\label{Imp_g_1_c}\includegraphics[width=7.5 cm]{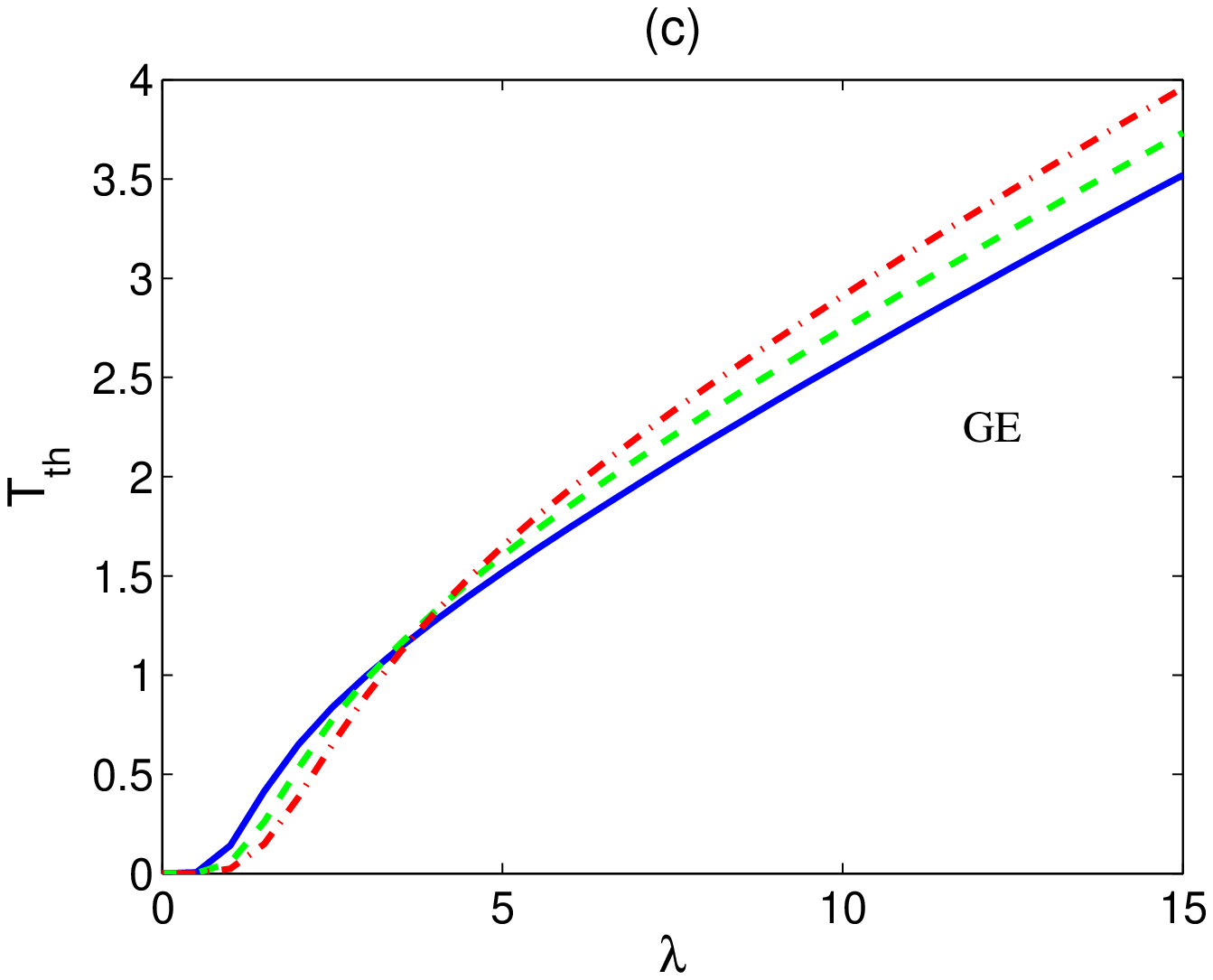}}\\
   \caption{{\protect\footnotesize (Color online) The threshold temperature of the entanglements EF(1,2), EF(1,7) and GE in the Ising 2D spin system with a central impurity versus $\lambda$ at different impurity strengths. The legends are as shown in subfigure (a).}}
 \label{Imp_g_1}
 \end{minipage}
\end{figure}
%fig_12
%%%%%%%%%%%%%%%%%%%%%%%%%%%%%%%%%%%%%%%%%%%%%%%%%%%%%%%%%%%%%%%%%%%%%%%%%%%%%%%%%%%%%%
%%%%%%%%%%%%%%%%%%%%%%%%%%%%%%%%%%%%%%%%%%%%%%%%%%%%%%%%%%%%%%%%%%%%%%%%%%%%%%%%%%%%%%
%%%%%%%%%%%%%%%%%%%%%%%%%%%%%%%%%%%%%%%%%%%%%%%%%%%%%%%%%%%%%%%%%%%%%%%%%%%%%%%%%%%%%%%%%
\begin{figure}[htbp]
\begin{minipage}[c]{\textwidth}
 \centering
   \subfigure{\label{Imp_g_05_a}\includegraphics[width=7.5 cm]{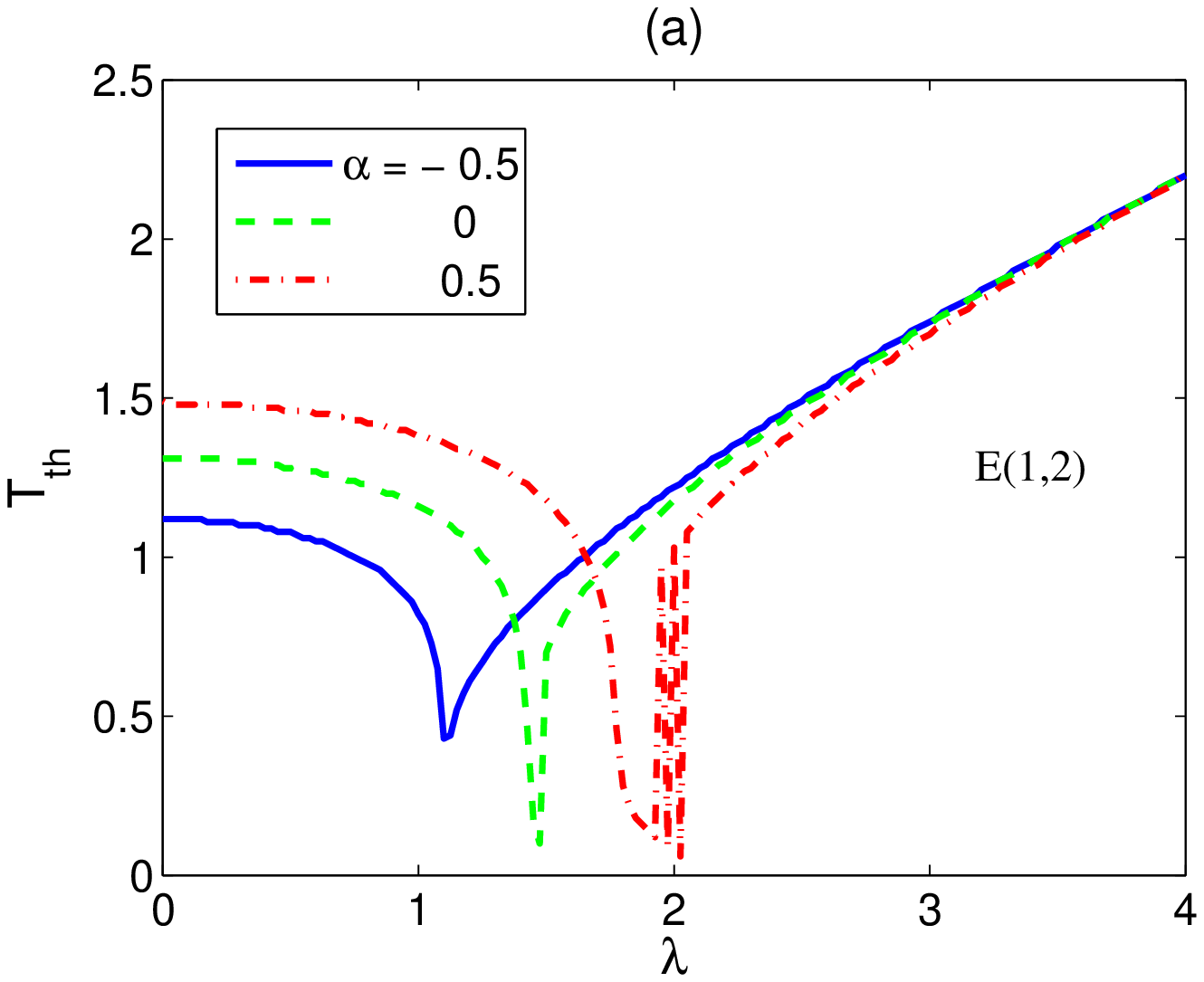}}\quad
   \subfigure{\label{Imp_g_05_b}\includegraphics[width=7.5 cm]{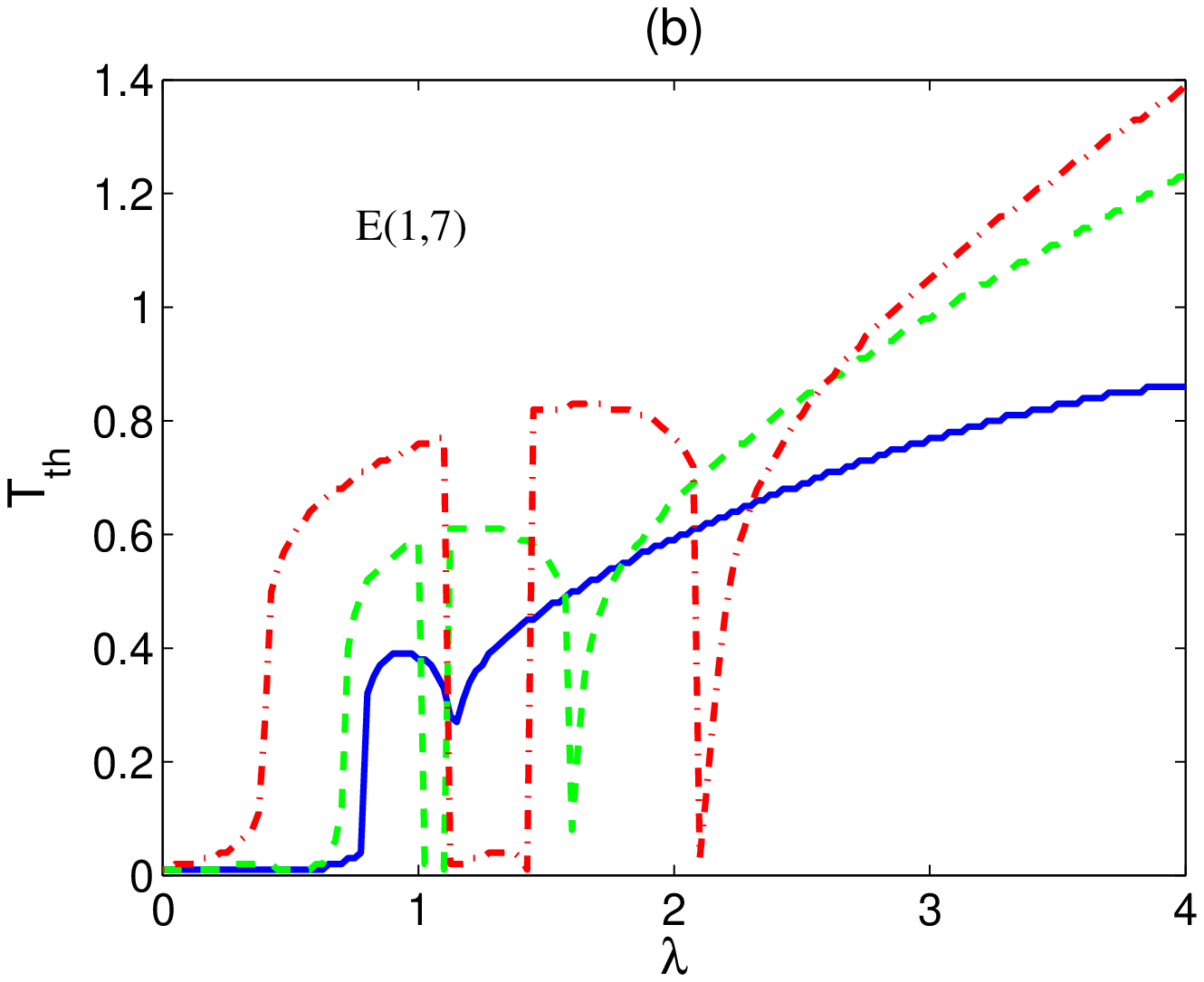}}\\
   \subfigure{\label{Imp_g_05_c}\includegraphics[width=7.5 cm]{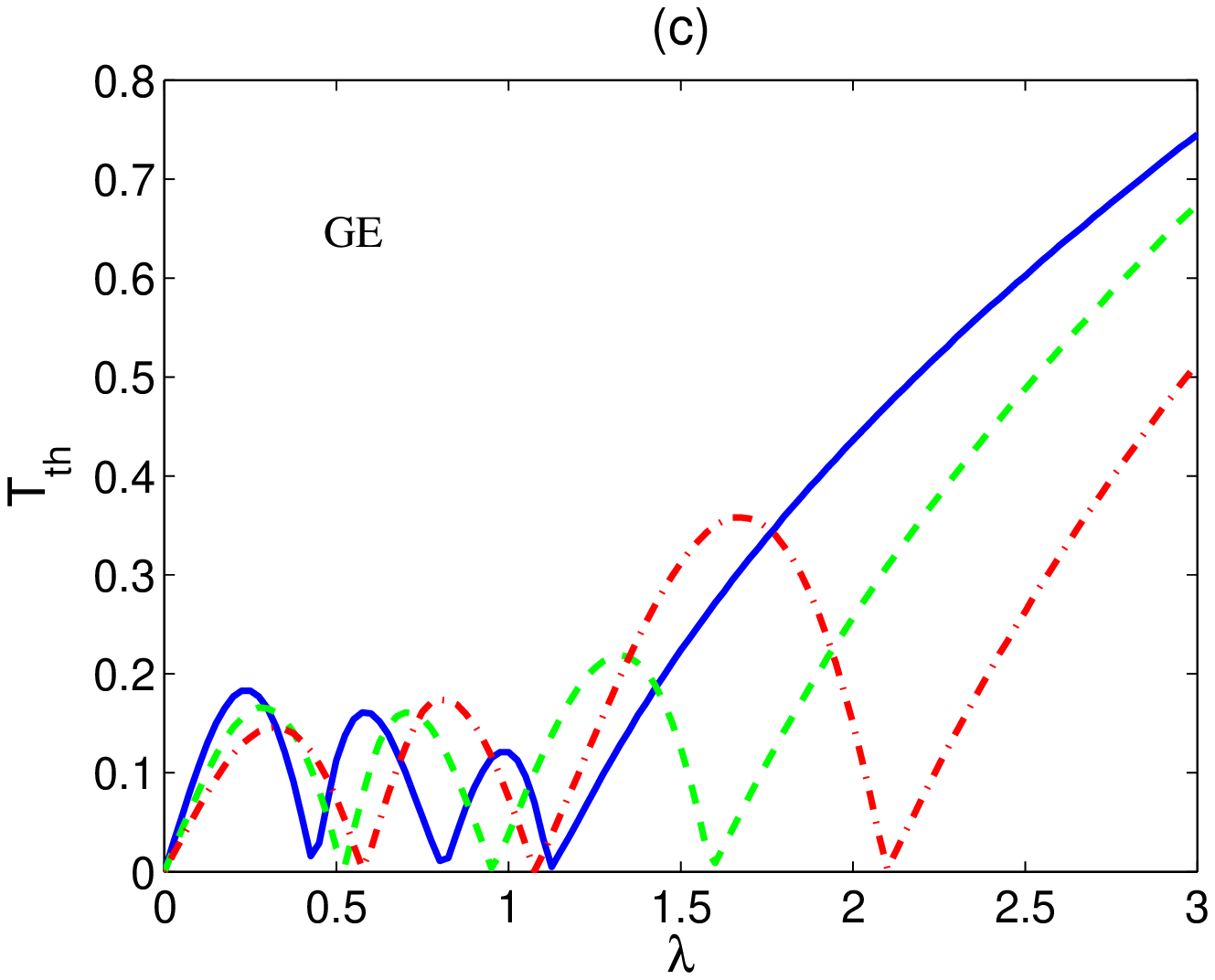}}\\
   \caption{{\protect\footnotesize (Color online) The threshold temperature of the entanglements EF(1,2), EF(1,7) and GE in the partially anisotropic 2D spin system with a central impurity versus $\lambda$ at different impurity strengths. The legends are as shown in subfigure (a).}}
 \label{Imp_g_05}
 \end{minipage}
\end{figure}
%fig_13
%%%%%%%%%%%%%%%%%%%%%%%%%%%%%%%%%%%%%%%%%%%%%%%%%%%%%%%%%%%%%%%%%%%%%%%%%%%%%%%%%%%%%%
%%%%%%%%%%%%%%%%%%%%%%%%%%%%%%%%%%%%%%%%%%%%%%%%%%%%%%%%%%%%%%%%%%%%%%%%%%%%%%%%%%%%%%%%%%%%
%%%%%%%%%%%%%%%%%%%%%%%%%%%%%%%%%%%%%%%%%%%%%%%%%%%%%%%%%%%%%%%%%%%%%%%%%%%%%%%%%%%%%%%%%
\begin{figure}[htbp]
\begin{minipage}[c]{\textwidth}
 \centering
   \subfigure{\label{Imp_g_0_a}\includegraphics[width=7.5 cm]{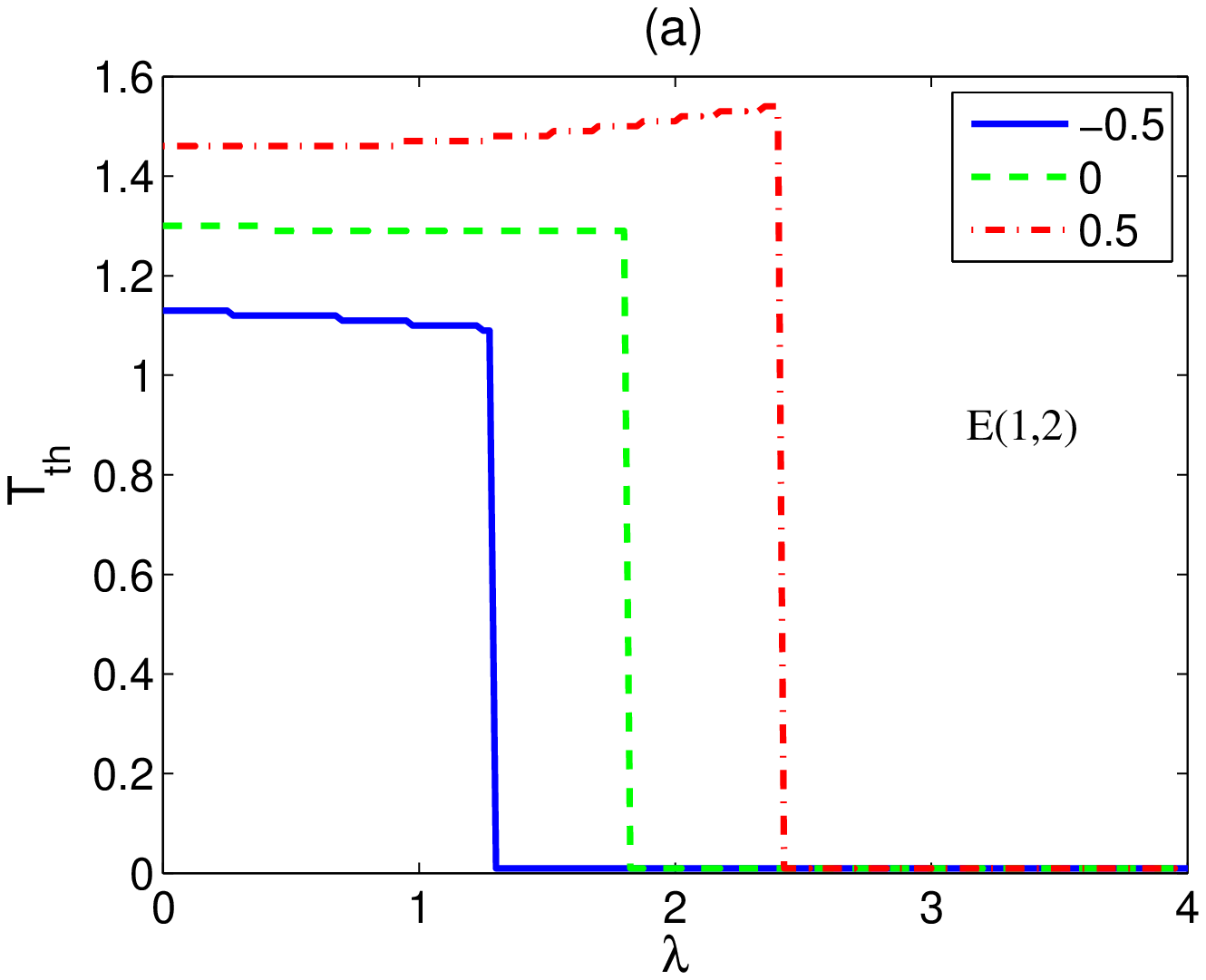}}\quad
   \subfigure{\label{Imp_g_0_b}\includegraphics[width=7.5 cm]{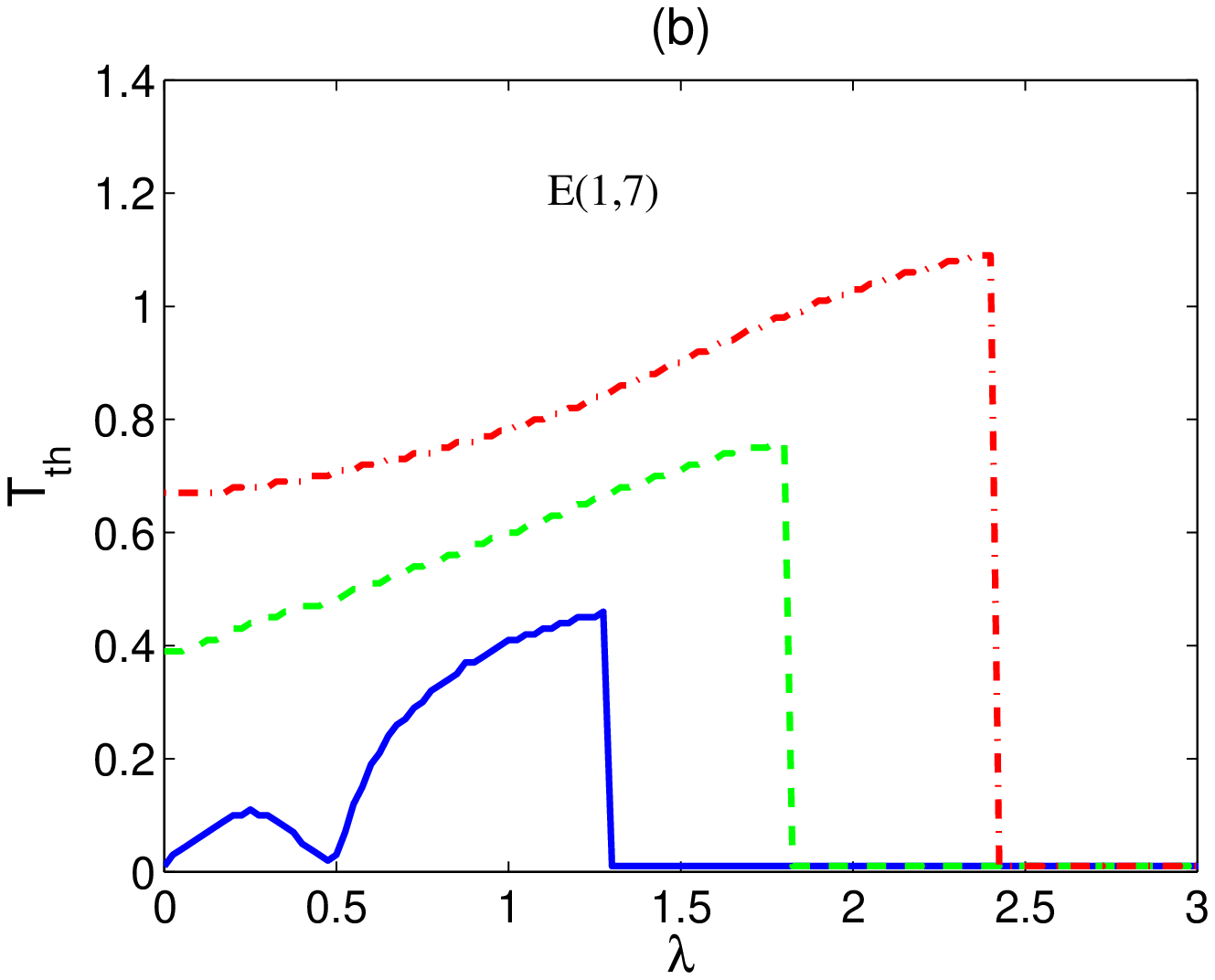}}\\
   \subfigure{\label{Imp_g_0_c}\includegraphics[width=7.5 cm]{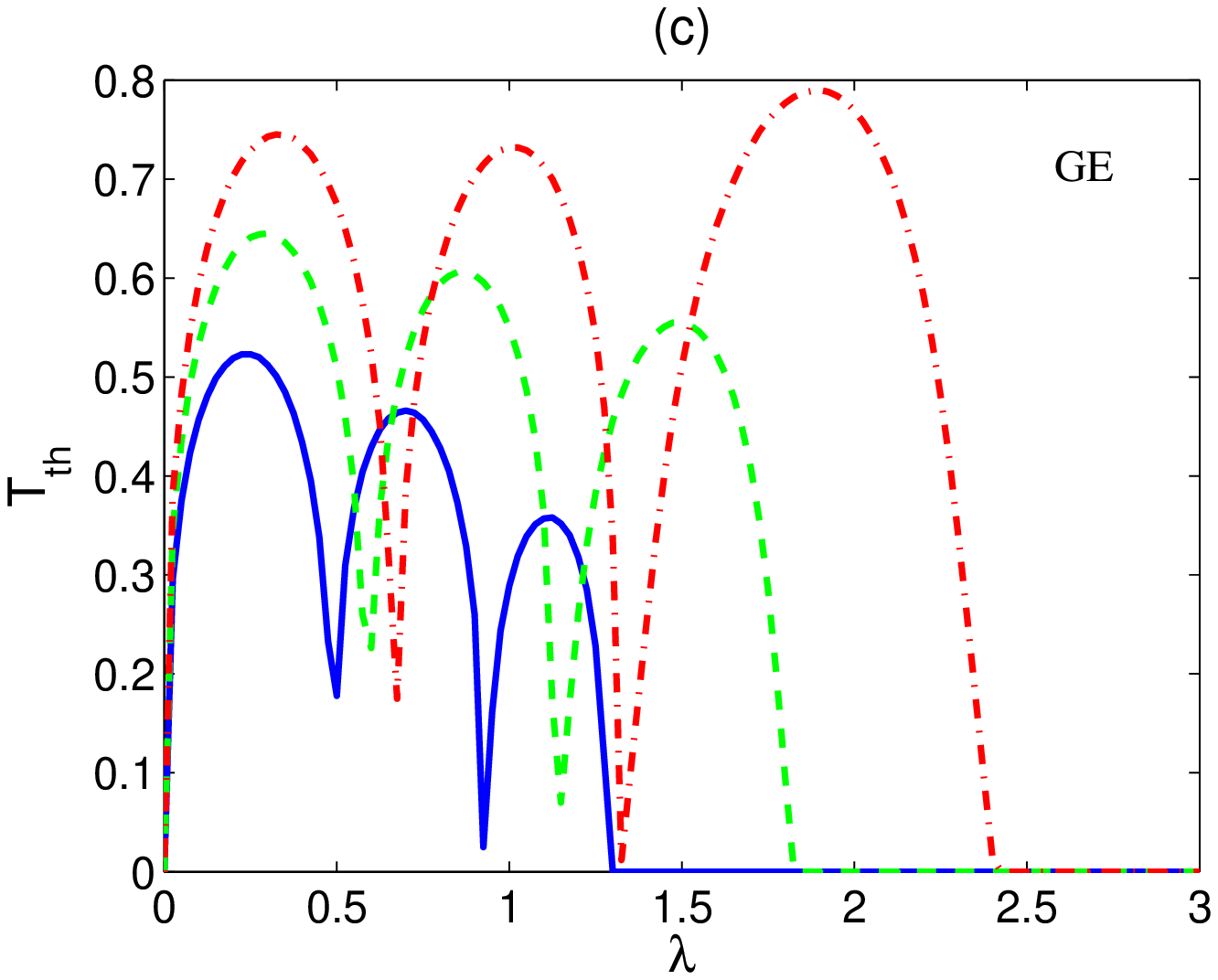}}\\
   \caption{{\protect\footnotesize (Color online) The threshold temperature of the entanglements EF(1,2), EF(1,7) and GE in the isotropic 2D spin system with a central impurity versus $\lambda$ at different impurity strengths. The legends are as shown in subfigure (a).}}
 \label{Imp_g_0}
 \end{minipage}
\end{figure}
%fig_14
%%%%%%%%%%%%%%%%%%%%%%%%%%%%%%%%%%%%%%%%%%%%%%%%%%%%%%%%%%%%%%%%%%%%%%%%%%%%%%%%%%%%%%
%%%%%%%%%%%%%%%%%%%%%%%%%%%%%%%%%%%%%%%%%%%%%%%%%%%%%%%%%%%%%%%%%%%%%%%%%%%%%%%%%%%%%%%%%
%%%%%%%%%%%%%%%%%%%%%%%%%%%%%%%%%%%%%%%%%%%%%%%%%%%%%%%%%%%%%%%%%%%%%%%%%%%%%%%%%%%%%%%%%
\section{Impurity effect}
The imperfection and disorder in real physical systems have been always a big concern when studying the different quantum properties of many body systems \cite{Latorre2009, Affleck2009}. Disorder and lack of homogeneity and isotropy cause a break of the translational symmetry and consequently the scaling of the entropy and all related quantities. An essential source of disorder is the presence of impurities in the physical system. The effect of quantum impurities in many body systems and the quantification of the entanglement in these systems have been investigated \cite{Affleck2009}. The Von Neumann entropy was used to quantify the single site impurity entanglement in the considered systems. At finite temperature, the thermodynamic impurity entropy is used to quantify entanglement especially in Kondo impurity systems \cite{Andrei1980, Vigman1980}. It was shown that the entanglement is significantly affected by the presence of the impurity even in absence of physical coupling to the impurity itself. In a previous work, it was demonstrated that the entanglement and ergodicity in two dimensional spin system can be tuned using impurities and anisotropy \cite{Sadiek2012}. The effect of impurities on the spin relaxation rate \cite{Huang2006} and dynamics of entanglement in one-dimensional spin systems have been investigated \cite{Apollaro2008}. The decay rate of the spin oscillation was found to be significantly affected by the coupling strength of the impurity spin.

In this section we study the effect of a single impurity located at the central site on the threshold temperature of the different types of entanglement in the lattice. The single impurity is a spin that is coupled to its nearest neighbors through an exchange interaction $J'$ which differs from that between the other sites. We set $J'=(\alpha+1)J$ where $\alpha$ is the impurity strength parameter. Here we consider 3 different cases for the impurity strength, $\alpha=-0.5, 0$ and $0.5$ representing weak, null and strong ones respectively. In fact, such a system of 7-spins with the central spin having a different coupling strength from the surrounding ones can be realized, for instance, in a system of coupled semiconductor quantum dots where the coupling between the valence electrons on the different dots is the exchange interaction which can be controlled by raising or lowering the potential barrier between the dots \cite{Burkard1999}.

In figs.~\ref{Imp_g_1}(a), (b) and (c) we study the effect of the central impurity, with different strengths, in the two-dimensional Ising lattice on the nn bipartite $EF(1,2)$, nnnn bipartite $EF(1,7)$ and multipartite entanglement $GE$ respectively. As can be noticed, the impurity strength has minor effect on $EF(1,2)$ where almost no change can be observed.
Interestingly, in the case of $EF(1,7)$, the critical value of the magnetic field at which the entanglement vanishes increases as the impurity strength increases. For strong impurity case, $EF(1,7)$ never vanish and increases montonically as $\lambda$ increases. This means that the impurity can be used to significantly preserve and enhance nnnn entanglement at high temperature and magnetic field in the Ising system, which is also the case for $GE$ as can be noticed in fig.~\ref{Imp_g_1}(c). 
The partially anisotropic system is explored in fig.~\ref{Imp_g_05}. The impurity has a significant effect on $EF(1,2)$ only at small values of $\lambda$, where it shifts the threshold minimum towards the right and creates an oscillation for strong impurity value. The asymptotic value of $EF(1,2)$ at large $\lambda$ is not affected by the impurity strength. On the other hand, while the impurity strength enhances the nnnn asymptotic entanglement, it reduces the global entanglement but shifts the minima of $T_{th}$ towards higher magnetic field values. The effect of the impurity on the isotropic system, with $\gamma=0$, is shown in fig.~\ref{Imp_g_0}, where increseing the impurity strength clearly enhances all types of entanglement. In addition, it also shift the magnetic field critical value at which all entanglements vanish toward higher values.
In fact, our results concerning the threshold temperatures here confirm the findings in a previous work \cite{Sadiek2012} where it was shown that the entanglement can be enhanced or quenched in the spin system depending on the degree of anisotropy and the strength of the impurities.
%%%%%%%%%%%%%%%%%%%%%%%%%%%%%%%%%%%%%%%%%%%%%%%%%%%%%%%%%%%%%%%%%%%%%%%%%%%%%%%%%%%%%%%%%
%%%%%%%%%%%%%%%%%%%%%%%%%%%%%%%%%%%%%%%%%%%%%%%%%%%%%%%%%%%%%%%%%%%%%%%%%%%%%%%%%%%%%%%%%
\section{Entanglements and threshold temperatures in one dimensional spin system}
%%%%%%%%%%%%%%%%%%%%%%%%%%%%%%%%%%%%%%%%%%%%%%%%%%%%%%%%%%%%%%%%%%%%%%%%%%%%%%%%%%%%%%%%%
%%%%%%%%%%%%%%%%%%%%%%%%%%%%%%%%%%%%%%%%%%%%%%%%%%%%%%%%%%%%%%%%%%%%%%%%%%%%%%%%%%%%%%%%%
\begin{figure}[htbp]
\begin{minipage}[c]{\textwidth}
 \centering
   \subfigure{\label{fig:1D_E_G_1}\includegraphics[width=7.5 cm]{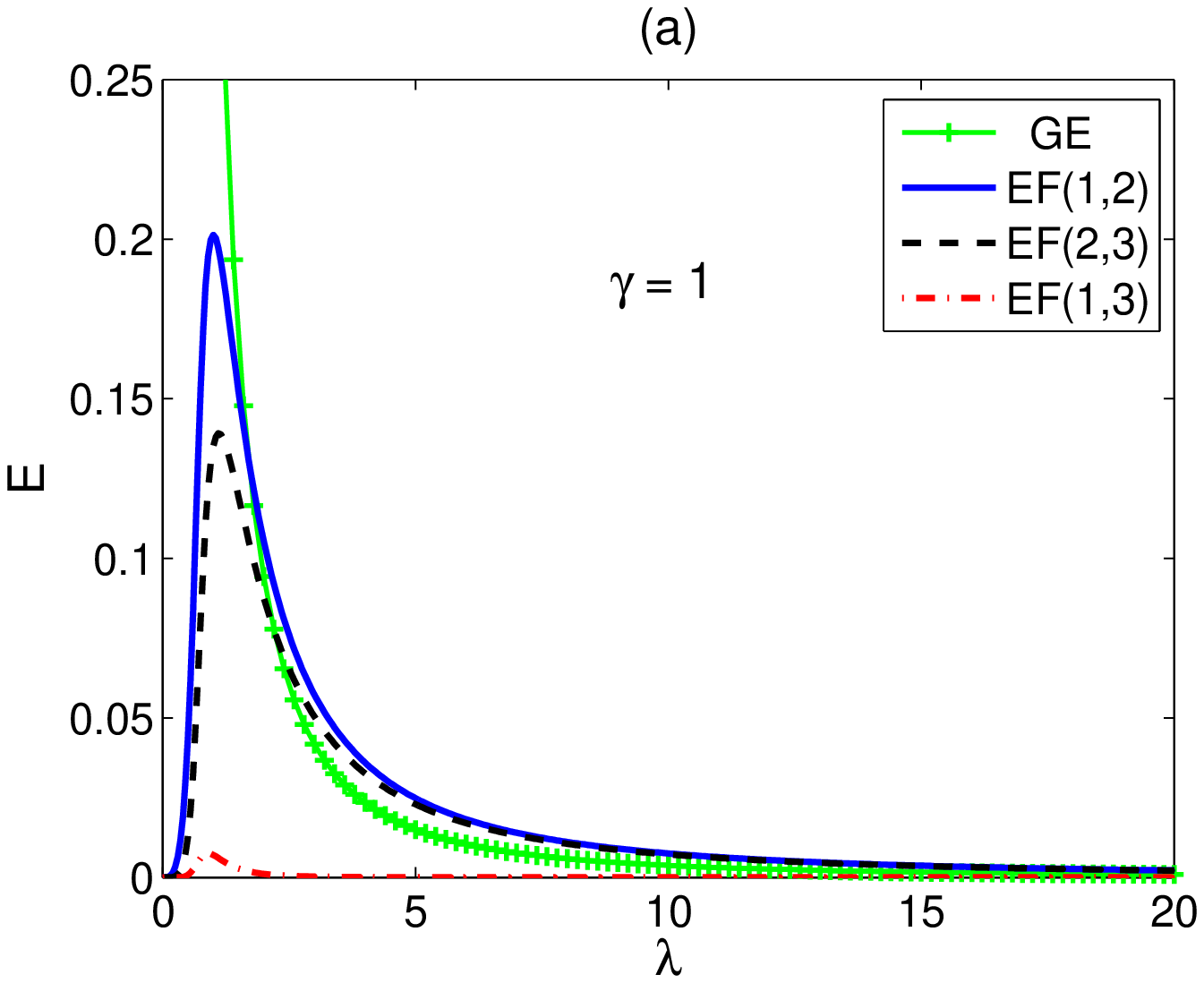}}\quad
   \subfigure{\label{fig:1D_E_G_05_close}\includegraphics[width=7.5 cm]{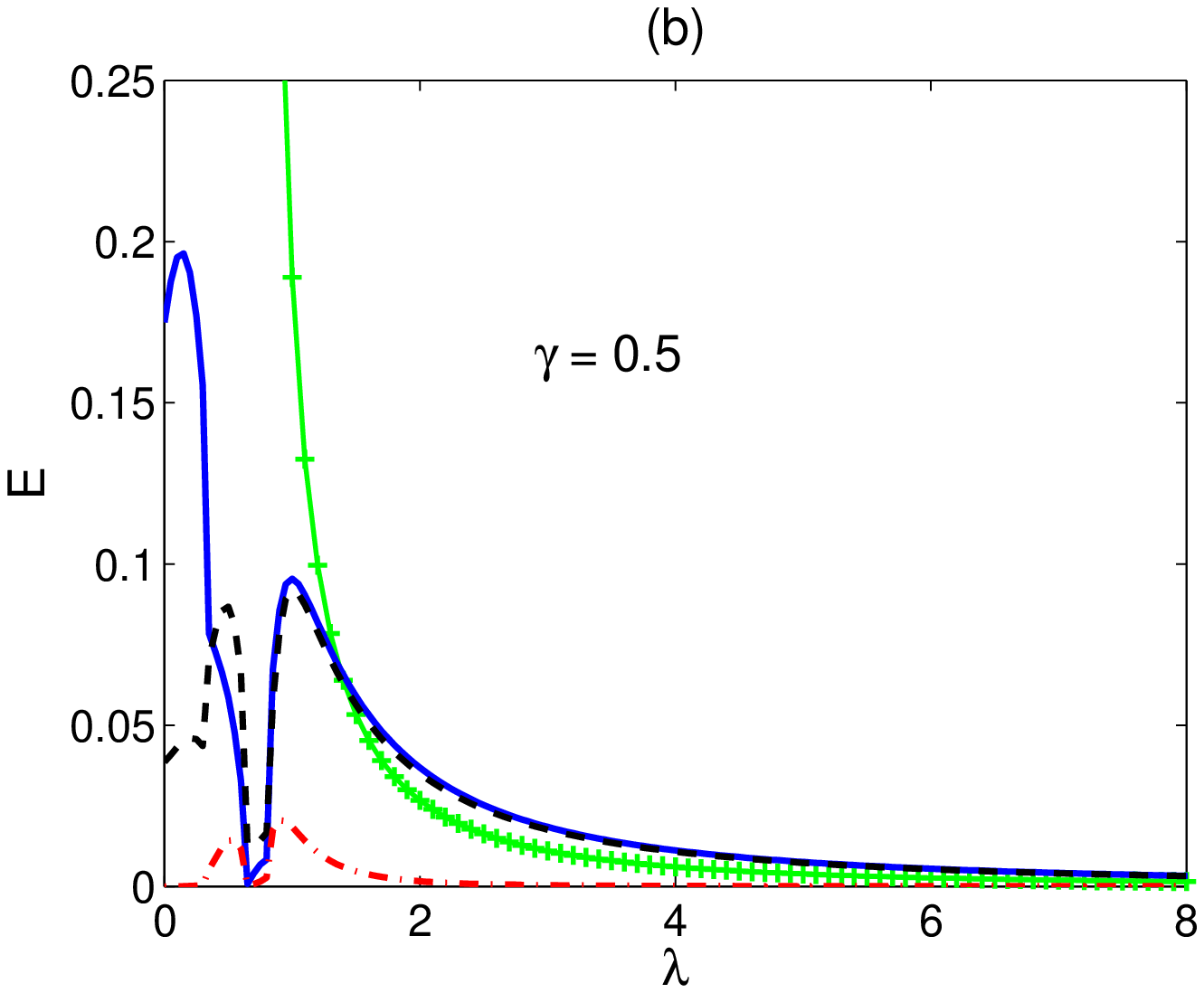}}\\
   \subfigure{\label{fig:1D_E_G_0}\includegraphics[width=7.5 cm]{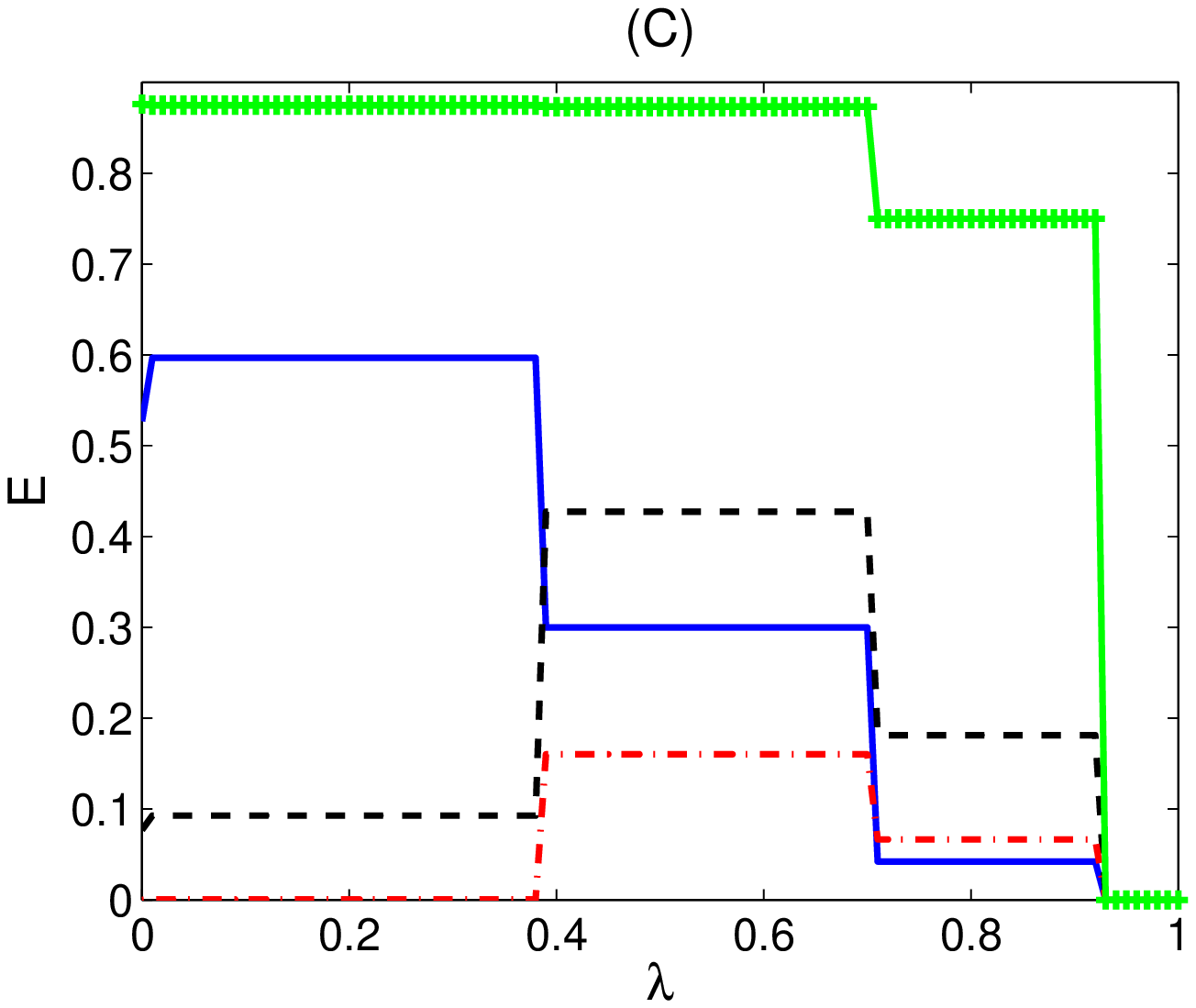}}\\
   \caption{{\protect\footnotesize (Color online) The entanglements EF(1,2), EF(2,3), EF(1,3) and geometric entanglement of the 1D spin system versus $\lambda$ for $\gamma=0, \; 0.5$ and 1 at zero temperature. The legends are as shown in subfigure (a).}}
 \label{1D_E}
52. \end{minipage}
\end{figure}
%fig_15
%%%%%%%%%%%%%%%%%%%%%%%%%%%%%%%%%%%%%%%%%%%%%%%%%%%%%%%%%%%%%%%%%%%%%%%%%%%%%%%%%%%%
%%%%%%%%%%%%%%%%%%%%%%%%%%%%%%%%%%%%%%%%%%%%%%%%%%%%%%%%%%%%%%%%%%%%%%%%%%%%%%%%%%%%
\begin{figure}[htbp]
\begin{minipage}[c]{\textwidth}
 \centering
   \subfigure{\label{fig:1D_T_G_1}\includegraphics[width=7.5 cm]{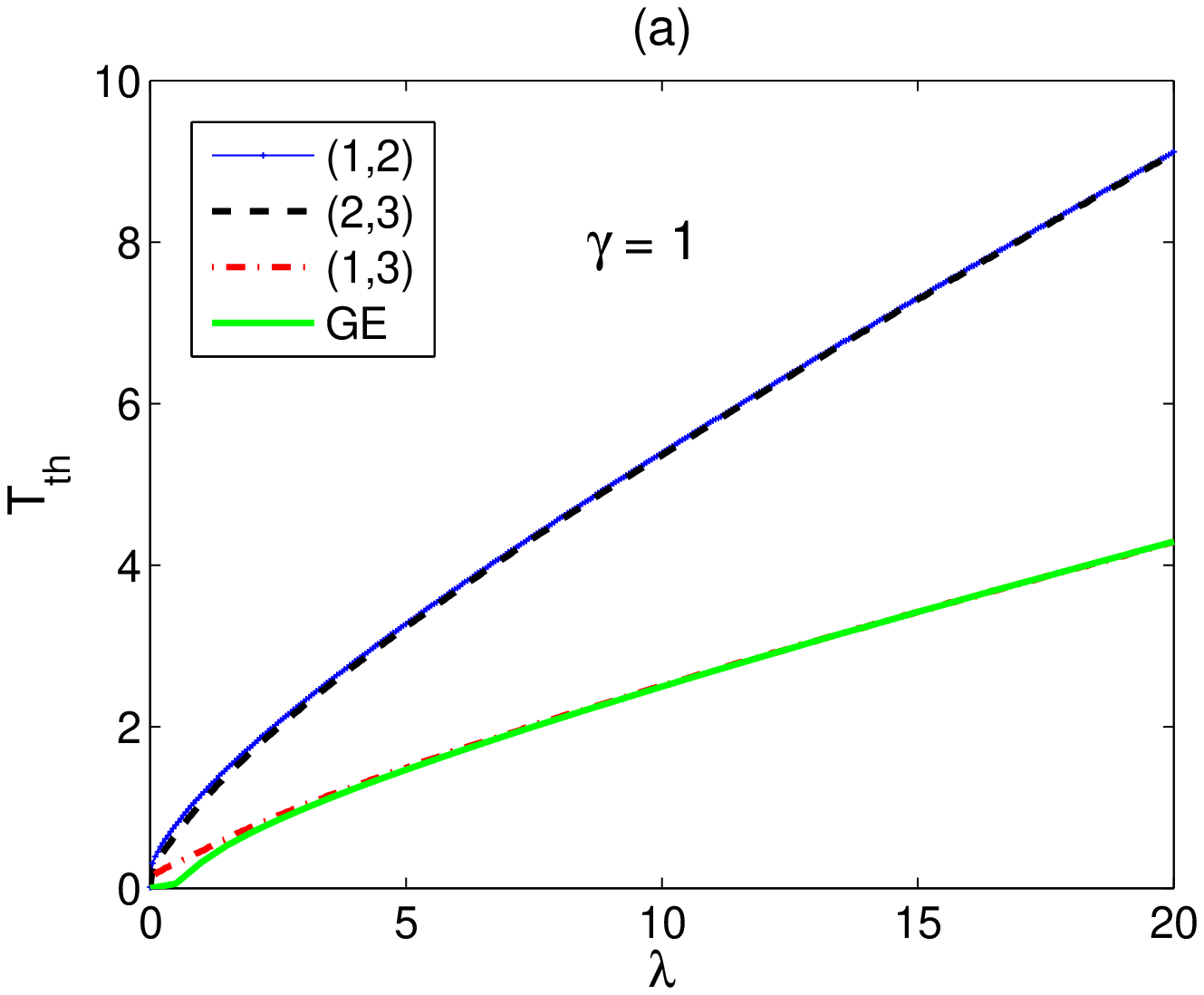}}\quad
   \subfigure{\label{fig:1D_T_G_05}\includegraphics[width=7.5 cm]{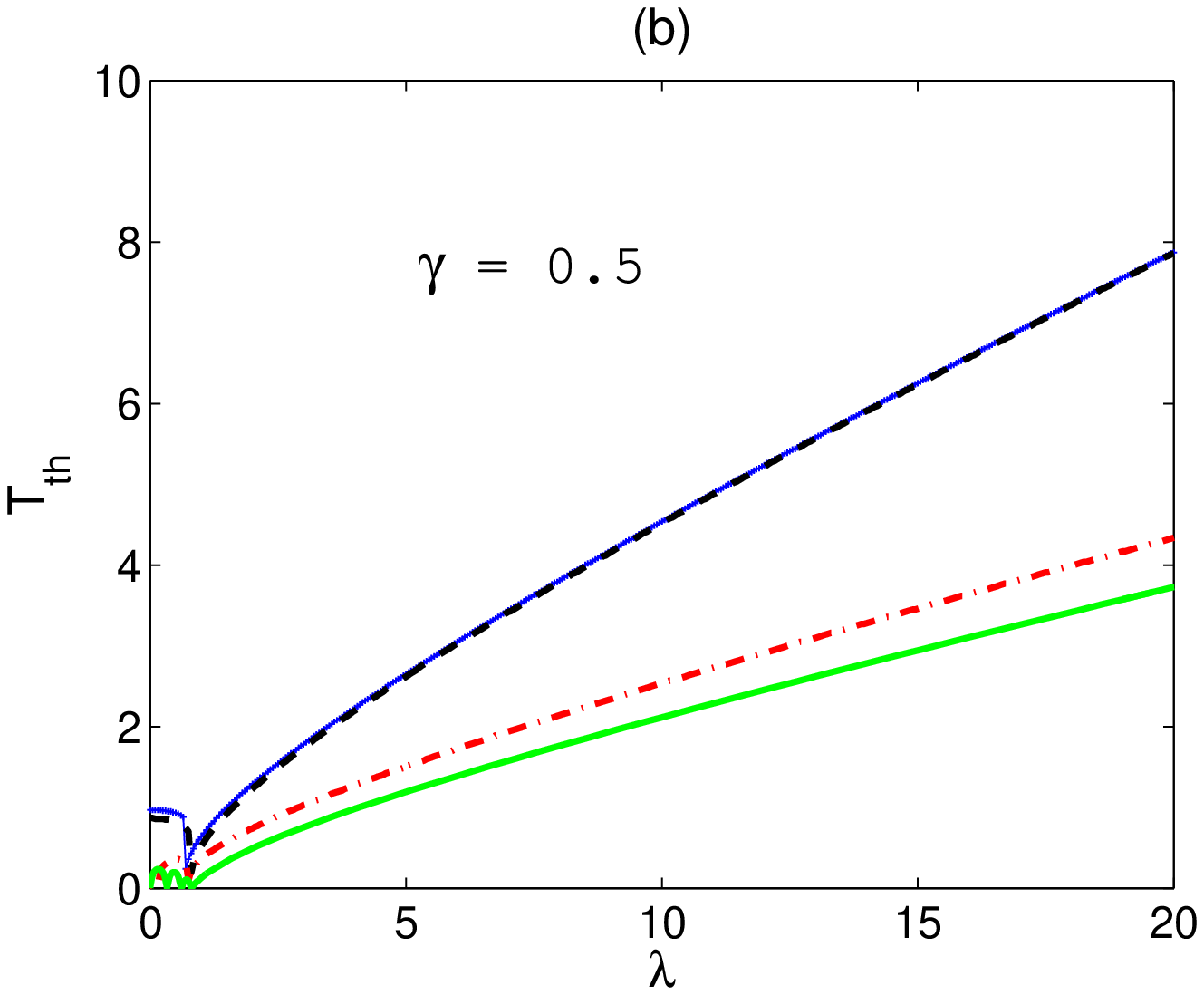}}\\
   \subfigure{\label{fig:1D_T_G_05_close}\includegraphics[width=7.5 cm]{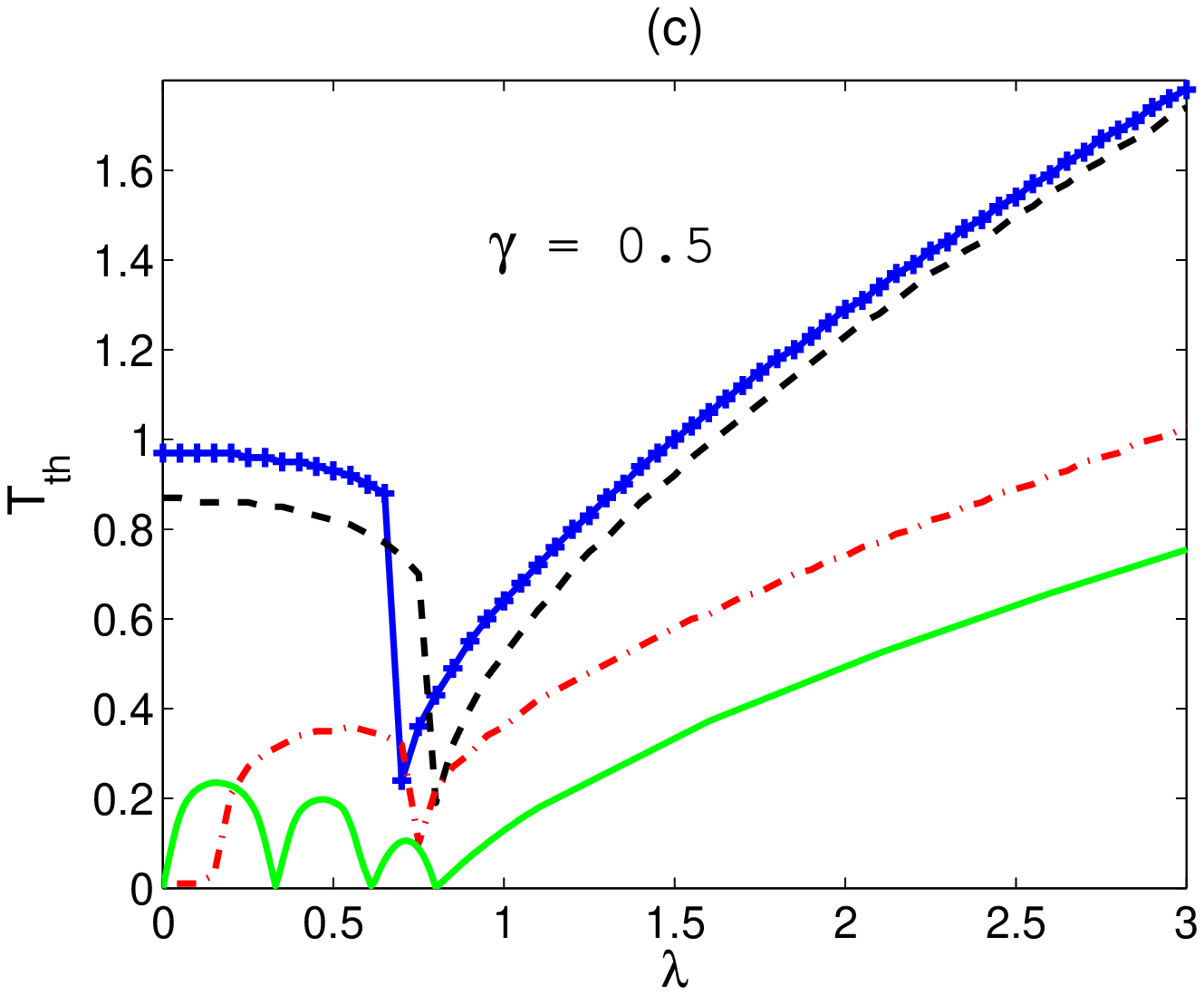}}\quad
   \subfigure{\label{fig:1D_T_G_0}\includegraphics[width=7.5 cm]{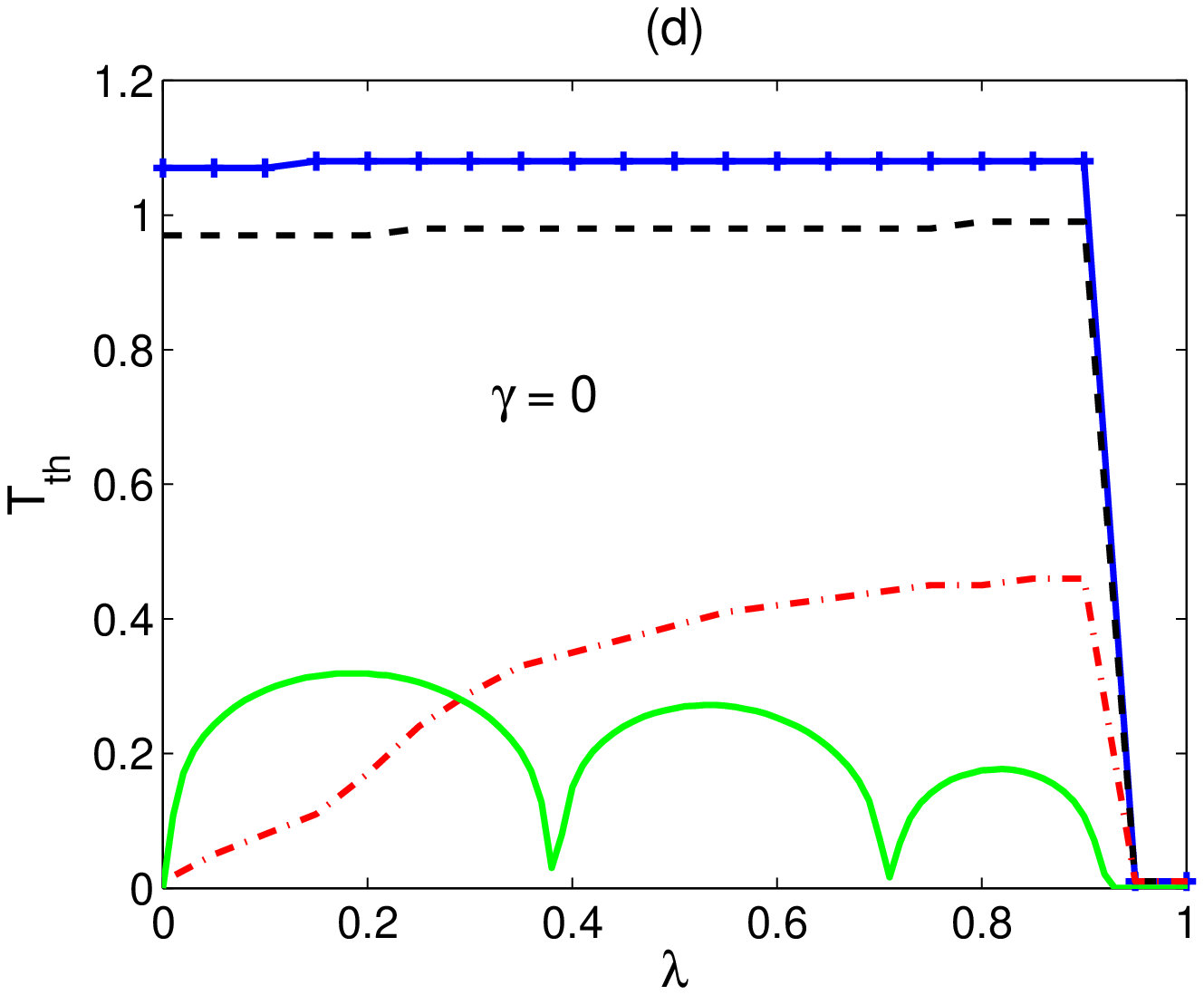}}\\
   \caption{{\protect\footnotesize (Color online) The threshold temperatures (in units of $J$) corresponding to the entanglements EF(1,2), EF(2,3), EF(1,3) and geometric entanglement of the 1D spin system versus $\lambda$ for $\gamma=0, \;0.5$ and 1. The legends are as shown in subfigure (a).}}
 \label{1D_T}
 \end{minipage}
\end{figure}
%fig_16
%%%%%%%%%%%%%%%%%%%%%%%%%%%%%%%%%%%%%%%%%%%%%%%%%%%%%%%%%%%%%%%%%%%%%%%%%%%%%%%%%%%%%
%%%%%%%%%%%%%%%%%%%%%%%%%%%%%%%%%%%%%%%%%%%%%%%%%%%%%%%%%%%%%%%%%%%%%%%%%%%%%%%%%%%%%
\begin{figure}[htbp]
\begin{minipage}[c]{\textwidth}
 \centering
   \subfigure{\label{fig:dE_g=0}\includegraphics[width=7.5 cm]{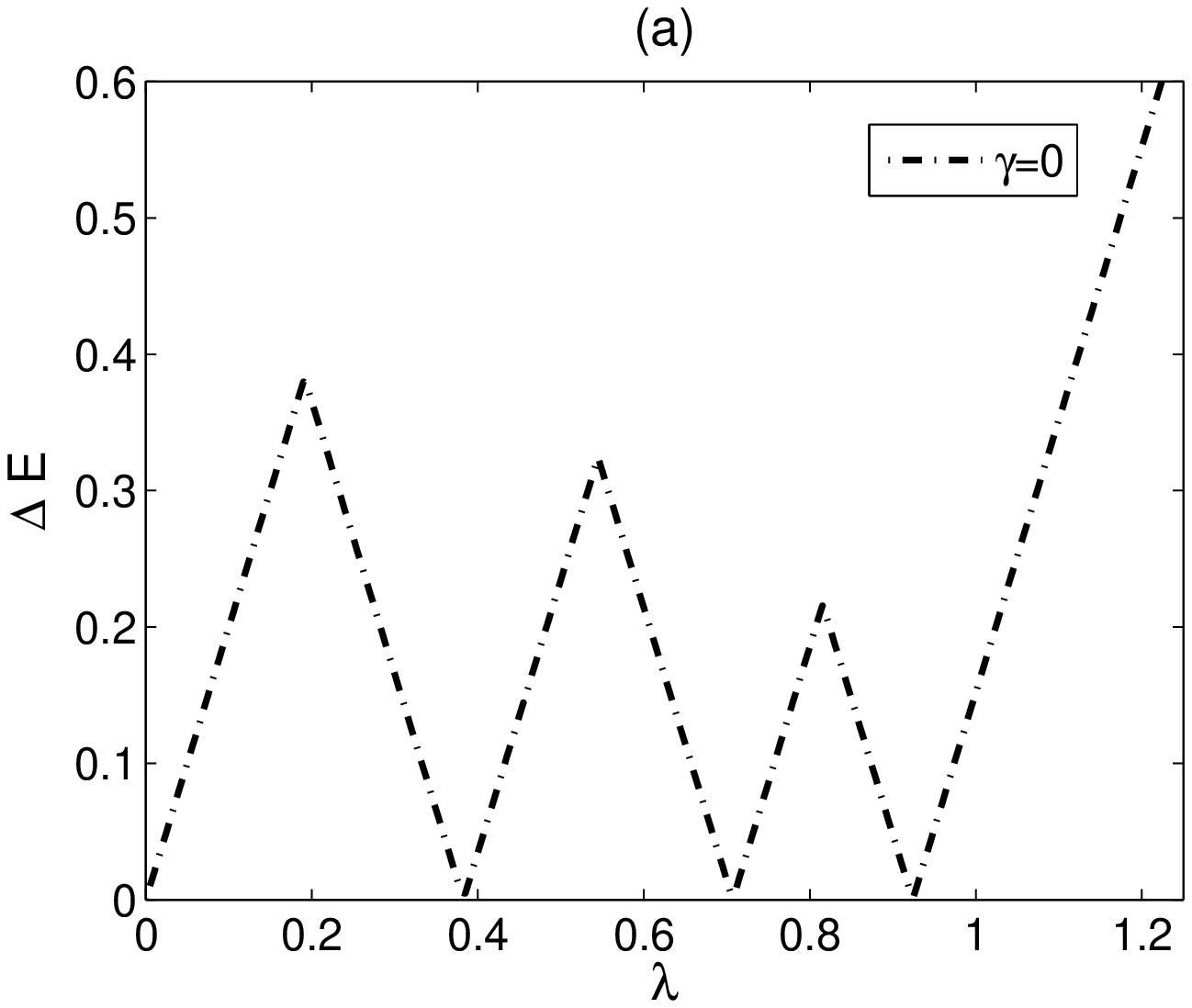}}\quad
   \subfigure{\label{fig:dE_g=05}\includegraphics[width=7.5 cm]{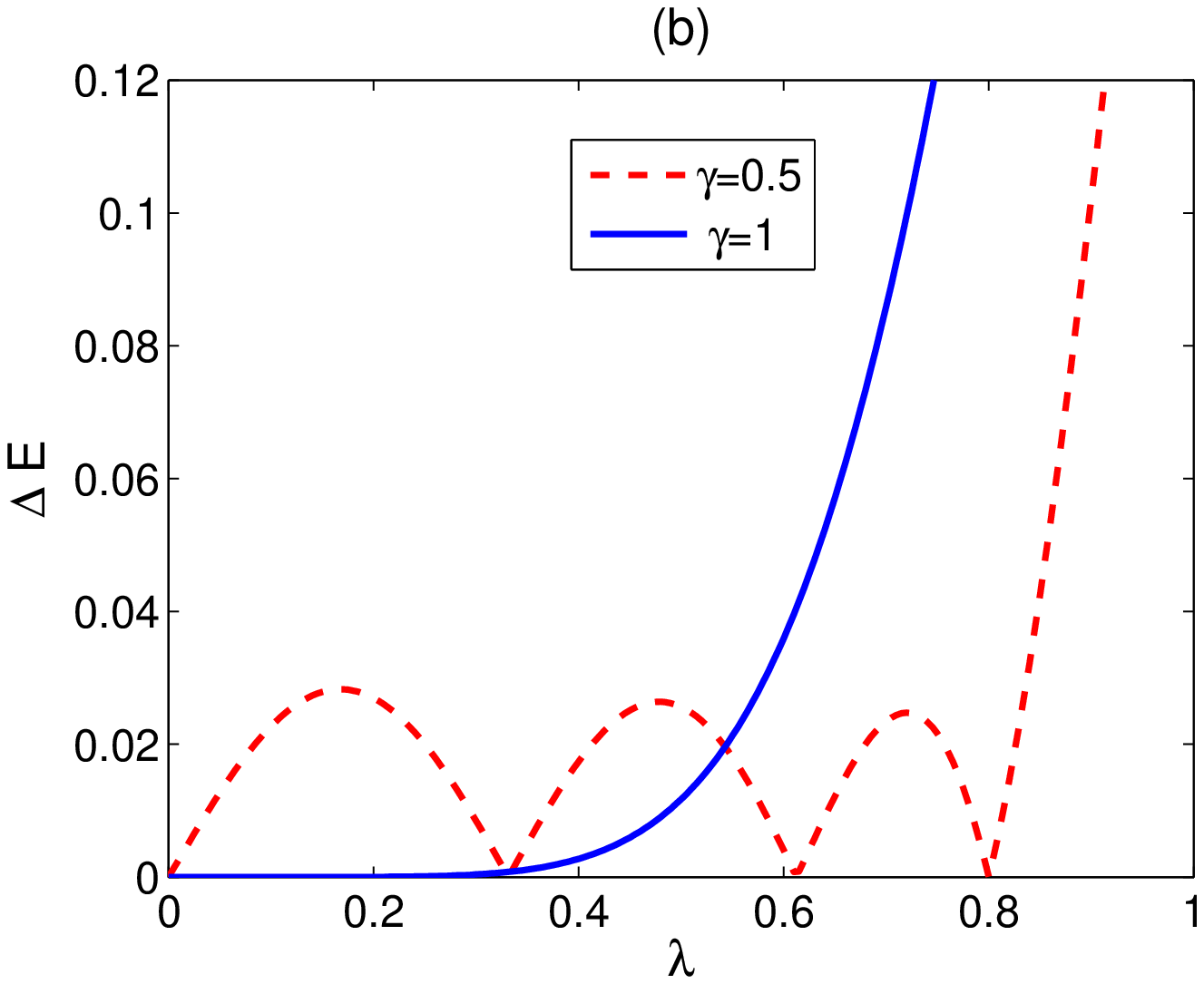}}\\
   \subfigure{\label{fig:dE_g=1}\includegraphics[width=7.5 cm]{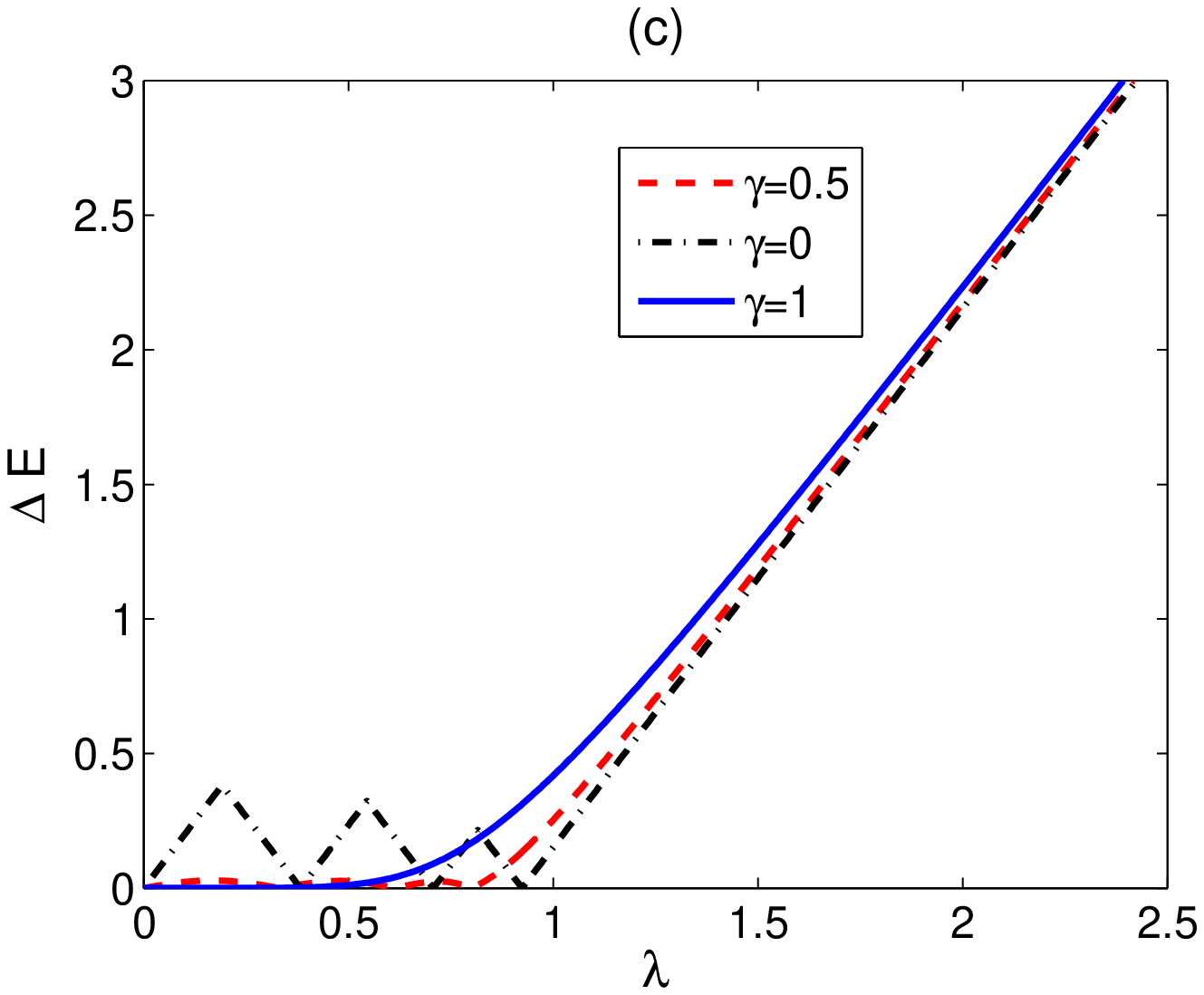}}\quad
   \caption{{\protect\footnotesize (Color online) The energy gap (in units of $J$) of the 1D spin system versus $\lambda$ for $\gamma=0, \; 0.5$ and 1 at zero temperature.}}
 \label{1D_dE}
 \end{minipage}
\end{figure}
%fig_17
%%%%%%%%%%%%%%%%%%%%%%%%%%%%%%%%%%%%%%%%%%%%%%%%%%%%%%%%%%%%%%%%%%%%%%%%%%%%%%%%%%%%%
Now let us consider a one dimensional $XY$ spin chain consisting of 7 spins, as sketched in fig.~\ref{Model}(b), which is described by the same Hamiltonian Eq. (\ref{Hamiltonian}) where in this case the exchange interaction $J_{i,j}$ exists only between each spin and its two nearest neighbor spins on the chain. The system shows a close behavior to what we have seen in the two-dimensional case.

In fig.~\ref{1D_E}(a) we compare the bipartite entanglements to the multipartite entanglement for the Ising system, where as can be seen the nearest neighbor entanglements $EF(1,2)$ and $EF(2,3)$ are very close as they start with zero magnitude at $\lambda=0$ and increase as $\lambda$ increases reaching a maximum value around $\lambda=1$ and then decay to zero at large $\lambda$. The next to nearest neighbor entanglement $EF(1,3)$ exhibits a similar behavior but with much smaller magnitude and in contrary to the two-dimensional case, it sustains at large values of the magnetic field.
The multipartite entanglement starts with a large value and abruptly decays approaching asymptotically the nearest neighbor bipartite entanglements at large magnetic field.
The behavior of the partially anisotropic system is very close to that of the Ising system as shown in fig.~\ref{1D_E}(b) except the quasi-oscillatory behavior of the bipartite entanglement at values of $\lambda < 1$ but again the nearest neighbor bipartite and the geometric entanglements become close asymptotically whereas the next to nearest neighbor entanglement sustains but with much smaller magnitude.
The entanglements of the isotropic system, similar to the two-dimensional case shows a step-like behavior and vanish at the same point, which is $\lambda \approx 0.92$ in the current case as depicted figs.~\ref{1D_E}(c). The threshold temperature of the different types of entanglements in the one dimensional chain is explored in fig.~\ref{1D_T}. Once more the behavior of the threshold temperatures of the nearest neighbor entanglements $EF(1,2)$ and $EF(2,3)$ are close at the different degrees of anisotropy of the system and the case is the same for the next to nearest neighbor bipartite and multipartite entanglements. Also as can be seen the isotropic system has zero threshold temperature at $\lambda \approx 0.92$. The behavior of both the entanglements and threshold temperatures in the one-dimensional spin chain can be explained in terms of the variation in the system energy gap at zero temperature which is depicted in fig.~\ref{1D_dE}. Similar to the two-dimensional case, one can see the strict correspondence between the variations in entanglements and threshold temperatures and the variations in the energy gap for all degrees of anisotropy of the system and at all $\lambda$ values.

Now it is clear that the general behavior of the multipartite and bipartite entanglements, in the one and two-dimensional systems, at the different degrees of anisotropy shows that the threshold temperature of the system, considered at the same magnetic field, increases with $\gamma$ within the range $0 < \gamma \leq 1$ and vanishes at a small value of the magnetic field for $\gamma=0$. 
In addition, the threshold temperatures increase monotonically with the magnetic field for $\lambda >> 1$. 

In fact, our results, particularly the one-dimensional case, are in a complete agreement with 
the findings of ref. \cite{Nakata2009} which is concerned with the threshold temperature corresponding to the global entanglement in one-dimensional $XY$ model at different degrees of anisotropy (see figs. 5 and 6 therein), though the results in those figures are for 80 spins. Also our results agree with those of ref. \cite{Patane2007} concerning the monotonic behavior and the relative magnitudes of nn to nnn of bipartite entanglement in the different pairs of three adjacent spins on a one-dimensional isotropic $XY$ spin chain utilizing the integrability of the model. However, the ME in their case, which is quantified using the negativity between one of the spins and the other two, shows higher threshold temperature compared to that of BE in that case (see fig. 3 therein). Furthermore, the minima that the entanglements go through at $\gamma=0.5$, as presented in figs. 8b, 9b, 9c, and 15b, show resemblance to the findings of refs. \cite{Kurmann1982, Amico2006} where minima of the entanglement were observed indicating the existence of factorizable (disentangled) ground states, which was also investigated in $XYX$ spin systems \cite{Roscilde2004,Roscilde2005}.

An estimate of the experimental values of the threshold temperatures for the typical spin systems of interest are due here. As the values of the threshold temperatures and energy gaps are all expressed in units of the exchange interaction constant $J$, which varies for the considered spin systems over a range of the order of $\mu eV$ to $m eV$ \cite{Ashcroft}, the corresponding range of the threshold temperature is $1.16\times 10^{-2}$ K to $11.6$ K. Also the typical value of the magnetic field $h$, which is also is expressed in units of J, can be evaluated here which goes over the range $1.7\times 10^{-2}$ T to $17$ T. Our results demonstrate that bigger energy gap would lead to higher threshold temperature but needs stronger magnetic field too. Using these results, one can come up with important estimates, where as can be concluded from fig.~\ref{2D_T}, the two-dimensional Ising system can reach a bipartite threshold temperature as high as $100$ K  which needs a magnetic field that is as high as $300$ T but the corresponding multipartite threshold temperature would be only about $50$ K. The isotropic system is entangled up to a magnetic field of about $30$ T where the maximum bipartite threshold temperature would be about $15$ K and the maximum reachable multipartite threshold temperature is $7.5$ K. For the same applied magnetic field, the threshold temperatures of the one dimensional spin chain would be slightly smaller than the corresponding ones in the two dimensional system as can be concluded from fig.~\ref{1D_T}.

\section{Quantum phase space of $XY$ spin systems}

Quantum phase transition in a many-body system happens either when an actual
crossing takes place between the excited state and the ground state or a limiting avoided level crossing between them exists, i.e., an energy gap between the two states that vanishes in the
infinite system size limit at the critical point \cite{Sachdev2001}. When a many-body system crosses a critical point, significant changes in both its wave function and ground-state energy take place, which can be realized in the behavior of the entanglement function and its derivatives \cite{Osterloh2002,Xu2010,Sadiek2012}. It is well known that an infinite many body system (i.e. at the thermodynamic limit) exhibits clear singularity at the critical point. However, finite size systems may still show strong tendency for being singular closer to the actual critical point of the system, which improves with the system size \cite{Xu2010, Sadiek2012}. 

The Hamiltonian Eq.(\ref{Hamiltonian}) describes a family of models with different distinct phases at the thermodynamic limit ($N \rightarrow \infty$). The quantum phase diagram for the one-dimensional system, in terms of $\gamma$ and $h$, is well determined and contains three different phases: Oscillatory, ferromagnetic and paramagnetic \cite{Henkel1999}. At the thermodynamic limit, the system reaches the isotropic and Ising limits at $\gamma=1$ and 0 respectively. The system belongs to a universality class, the istropic ($XX$), at $\gamma=0$ whereas it belongs to a different class, Ising (aniostropic $XY$), in the range $0 < \gamma \leq 1$. The system possesses a quantum critical point at $\lambda=\lambda_c=1$, where this point dictates the transition between different phases of the system depending on the value of $\gamma$. The system at all degrees of anisotropy exists in the paramagnetic phase for $\lambda > \lambda_c$. For $\gamma^2+\lambda^2 < \lambda_{c}^{2}$ and $\gamma \neq 0$ the system exits in the oscillatory phase whereas for $\gamma^2+\lambda^2 > \lambda_{c}^{2}$ and $\lambda < 1$ is paramagnetic. The phase diagram of the infinite two-dimensional system, though is very similar to the one-dimensional case, is not well established due to the lack of analytic solution, computational difficulty and the different structures that the system may have. Many efforts have been directed to the prediction of the critical value of $\lambda_c$ in the two dimensional spin system. For instance, the Renormalization group method has been applied to the two dimensional infinite triangular (square) lattice and estimated a critical point at $\lambda_c=4.75784 (2.62975)$ \cite{Penson1982}, whereas the finite size scaling method applied to the square lattice predicts $\lambda_c=3.044$ \cite{Hamer2000}. The point (with tendency to singularity) in the finite two-dimensional spin system with 7-sites considered in this paper (and also for 19-sites) was estimated previously and found to be $\lambda_c=1.64$ and $3.01$ for the 7-sites and 19-sites systems respectively by studying the pairwise concurrence and its derivative in the system \cite{Xu2010}. 

Though we emphasis here that the quantum phase transitions can take place only in the infinite system size (in the thermodynamic limit) we will try to draw a relation here between the behavior of the entanglements and the threshold temperatures in our considered systems and the different phases of the system. As one can notice for the Ising system ($\gamma=1$) the bipartite entanglements have one single peak, where in fact its derivative locates the point of strong tendency for being singular \cite{Xu2010, Sadiek2012}  before decaying montonically with $\lambda$ and the geometric entanglement decays montonically from large value. This behavior is reflected also in the threshold temperatures, which increase montonically with $\lambda$. This profile of the entanglements and temperatures can be related to transition from ferromagnetic to paramagnetic states for the Ising system as $\lambda$ increases crossing the expected critical point. In the partially anisotropic system $\gamma=0.5$, the entanglements (threshold temperatures) exhibit few maxima and minima with $\lambda$ before montonically decreasing (increasing) which can be explained in terms of the transition of the system from oscillatory to ferromagnetic and finally to the paramagnetic phase at the critical point. Finally the isotropic system shows sharp changes in the entanglements (threshold temperatures)as they decay before vanishing at $\lambda\approx 1.85$, which can be explained in terms of transition from the oscillatory phase to paramagnetic phase and as we mentioned before the vanishing of all entanglements and threshold temperatures is due to the fact that the isotropic system has a disentangled state for any magnetic field higher than this critical point.

%%%%%%%%%%%%%%%%%%%%%%%%%%%%%%%%%%%%%%%%%%%%%%%%%%%%%%%%%%%%%%%%%%%%%%%%%%%%%%%%%%%%%%%%%%%%%%%%%%%%%
\section{Conclusions}
%%%%%%%%%%%%%%%%%%%%%%%%%%%%%%%%%%%%%%%%%%%
We have investigated the robustness of bipartite and multipartite entanglement in one and two-dimensional $XY$ spin-1/2 lattices in an external magnetic field $h$ against thermal excitations. The spins are coupled to each other through nearest neighbor exchange interaction $J$. The number of spins in the lattice is 7, which are coupled to a heat bath at temperature $T$. We have compared the bipartite entanglement to the multipartite entanglement versus the external applied magnetic field and temperature. Also we compared the threshold temperature at which the entanglement vanishes in both cases. We used the entanglement as a measure of the bipartite entanglement and the geometric measure to evaluate the multipartite entanglement of the system.
 
In the one and two-dimensional cases for the anisotropic and partially anisotropic spin systems at zero temperature, the nearest neighbor bipartite and multipartite entanglement can be maintained at large magnetic fields, though would have very small values, which are still much higher than that of the next to nearest neighbor entanglements except in the two-dimensional Ising system where the latter vanishes at small value of the field. 
In the isotropic system case, all types of entanglement vanish at the same small value of the magnetic field. 
The nearest neighbor bipartite threshold temperature was found to be higher than that of the next to nearest neighbor bipartite and multipartite where the temperatures of the last two get closer asymptotically and the three of them increase monotonically as the magnetic field increases. The exception is the threshold temperature of the nearest neighbor entanglement in the two-dimensional Ising system which vanishes as small value of the magnetic field. Accordingly the bipartite entanglement of the far spins of the system and the multipartite entanglement are less resilient toward thermal excitations compared to the nearest neighbour entanglement.
All the threshold temperatures of the isotropic system vanish exactly at the same value of the magnetic field where all the entanglements vanish. Studying the different systems energy gaps as a function of the magnetic field showed that they have great correspondence to the behavior of the entanglements and the threshold temperatures, where large characteristic energy gap lead to stronger robustness of entanglement and higher threshold temperatures, while vanishing energy gap may cause zero threshold temperature. Nevertheless, the properties of the ground state of the system plays a major role in determining the behavior of the entanglement and the threshold temperature over the energy gap. This was particularly seen in the isotropic system case, which has a ground state that is entangled only below a threshold value of the magnetic field, which causes both the entanglement and the the threshold temperatures to vanish at this value and above regardless of the monotonic increase of the energy gap. The effect of a central impurity was found to be significant in enhancing the threshold temperatures and preserving all types of entanglements at high magnetic fields.
Furthermore, we have focused on examining the persistence of quantum effects at high temperatures where we found that the different types of entanglements (specially the bipartite), though would have very small values, can be maintained at high temperatures by applying sufficiently high magnetic fields. It is interesting in future to investigate the same systems coupled to a dissipative environment in presence of thermal excitations to test the interplay of the two environments. Also it is important to engineer similar systems with inserted impurities to examine their effect to tune the threshold temperatures. Furthermore we would like to investigate the same system with larger number of sites to test the system size effect on robustness of thermal entanglement and determine threshold temperatures using finite size scaling \cite{Kais2003,Kais2007}.
%%%%%%%%%%%%%%%%%%%%%%%%%%%%%%%%%%%%%%%%%%%%%%%%%%%%%%%%%%%%%%%%%%%%%%%%%%%%%%%%%%%%%%%%%%%%
\section*{Acknowledgments}
We are grateful to the Saudi NPST for support (project no. 11-MAT1492-02). We are also grateful to the National Science Foundation CCI center, "Quantum Information for Quantum Chemistry (QIQC)" (Award number CHE-1037992) for partial support of this work at Purdue.
%%%%%%%%%%%%%%%%%%%%%%%%%%%%%%%%%%%%%%%%%%%%%%%%%%%%%%%%%%%%%%%%%%%%%%%%%%%%%%%%%%%%%%%%%%%%%
%\bibliography{Reference_Database_20_12_12}

\begin{thebibliography}{10}

\bibitem{Peres1993} Peres 1993 {\it Quantum Theory: Concepts and Methods}, (The Netherlands: Kluwer, Dordrecht)

\bibitem{Benioff1981} Benioff 1981 {\it J. Math. Phys.} 22 495

\bibitem{Benioff1982} Benioff 1982 {\it Int. J. Theo. Phys.} 21 177

\bibitem{Bennett1985} Bennett C.~H. and Landauer R., 1985 {\it Scientifc American} 253 48

\bibitem{Deutsch1985} Deutsch D. 1985 {\it Proc. R. Roc. A} 400 97

\bibitem{Deutsch1989} Deutsch D. 1989 {\it Proc. R. Soc. A} 425 73

\bibitem{Feynman1982} Feynman R.~P. 1982 {\it Int. J. Theoret. Phyiscs} 21 467

\bibitem{Landauer1961} Landauer R. 1961 {\it IBM J. Res. Develop.} 3 183

\bibitem{Nielsen2000-book} Nielsen M.~A. and I.~L. Chuang 2000 {\it Quantum Computation and Quantum Information} (Cambridge: Cambridge University press)

\bibitem{HorodeckiM2001} Horodecki M. 2001 {\it Quan. Inf. and Comp.} 1 (1) 3

\bibitem{HorodeckiP2001} Horodecki P. and Horodecki R. 2001 {\it Quant. Inf. and Comp.} 1 (1) 45

\bibitem{Wootters2001-2} Wootters W.~K. 2001 {\it Quantum Inf. Comput.} 1 (1) 27

\bibitem{Vedral1998} Vedral V. and Plenio M.~B. 1998 {\it Phys. Rev.} A 57 1619

\bibitem{Wei2003} Wei T.-C. and Goldbart P.~M. 2003 {\it Phys. Rev.} A 68 042307

\bibitem{Nielsen2000} Nielsen M.~A. (2000) {\it phys. Rev.} A 61 064301

\bibitem{Sachdev2001} Sachdev S. (2001) {\it Quantum Phase Transitions} (Cambridge: Cambridge Univ. Press)

\bibitem{Osborne2002} Osborne T.~J. and Nielsen M.~A. (2002) {\it Phys. Rev.} A 66 032110

\bibitem{Wei2005} Wei T.-C., Das D., Mukhopadyay S., Vishveshwara S. and Goldbart P.~M. (2005) {\it Phys. Rev.} A 71 060305

\bibitem{Sadiek2010} Sadiek G., Alkurtass B. and Aldossary O. (2010) {\it Phys. Rev. A} 82 052337

\bibitem{Amico2008} Amico L., Fazio R., Vedral V. (2008) {\it Rev. Mod. Phys.} 80 517 

\bibitem{Latorre2009} Latorre J. I. and Riera A. (2009) {\it J. Phys. A} 42 504002 

\bibitem{Kurmann1982} Kurmann J., Thomas H. and Müller G. (1982) {\it Physica A} 112 235

\bibitem{Roscilde2004} Roscilde T., Verrucchi P., Fubini A., Haas S. and Tognetti V. (2004) {\it Phys. Rev. Lett.} 93 167203

\bibitem{Amico2006} Amico L., Baroni F., Fubini A., Patanè D., Tognetti V. and Verrucchi P. (2006) {\it Phys. Rev. A} 74 022322

\bibitem{Patane2007} Patane D., Fazio R. and Amico L. (2007) {\it N. J. Phys.} 9 322

\bibitem{Alkurtass2011} Alkurtass B., Sadiek G. and Kais S. (2011) {\it Phys. Rev. A} 84 022314

\bibitem{Coffman2000} Coffman V., Kundu J. and Wootters W. K. (2000) {\it Phys. Rev. A} 61 052306

\bibitem{Amico2004} Amico L., Osterloh A., Plastina F., Fazio R. and Palma G. M. {\it Phys. Rev. A} 69 022304

\bibitem{Wootters1998} Wootters W.~K. (1998) {\it Phys. Rev. Lett.} 80 2245

\bibitem{Vedral1997} Vedral V., Plenio M. B., Rippin M. A. and Knight P. L. (1997) {\it Phys. Rev. Lett.} 78 2275

\bibitem{Vidal1999} Vidal G. and Tarrach R. (1999) {\it Phys. Rev. A} 59 141

\bibitem{Meyer2002} Meyer D. A. and Wallach N. R. (2002) {\it J. Math. Phys.} 43 4273

\bibitem{Markham2008} Markham D., Anders J., Vedral V., Murao M. and Miyake A. (2008) {\it Europhy. Lett.} 81 40006

\bibitem{Tamaryan2009} Tamaryan S., Wei T.-C. and Park D. (2009) {\it Phys. Rev. A} 80 52315

\bibitem{Nakata2009} Nakata Y., Markham D. and Murao M. (2009) {\it Phys. Rev. A} 79 042313

\bibitem{Hubener2009} Hübener R., Kleinmann M., Wei T.-C., González-Guillén C. and Gühne O. (2009) {\it Phys. Rev. A} 80 032324

\bibitem{Blasone2009} Blasone M., Dell'Anno F., De Siena S., Giampaolo S. M. and Illuminati F. (2009) {\it J. Phys.} (conference series) 174 012062
 
\bibitem{Wei2011} Wei T., Vishveshwara S. and Goldbart P. M. (2011) {\it Quant. Info. and Comp.} 11 0326

\bibitem{Zurek1991} Zurek W. 1991 {\it Phys. Today} 44 36

\bibitem{Bacon2000} Bacon D. ,Kempe  J., A. L.~D. and Whaley K.~B. (2000) {\it Phys. Rev. Lett.} 85 1758

\bibitem{Dmitriev2002} Dmitriev D. V., Krivnov V. Ya., Ovchinnikov A. A., Langari A. (2002) {\it J. Exp. Theor. Phys.} 95 538

\bibitem{Vidal2002} Vidal G. and Werner R.F. (2002) {\it Phys. Rev. A} 65 032314

\bibitem{Hide2012} Hide J., Y. Nakata Y. and Murao M. (2012) {\it Phys. Rev. A} 85 042303

\bibitem{Syljuasen2003} Syljuasen O. F. (2004) {\it Phys. Rev. A} 68 60301

\bibitem{Roscilde2005} Roscilde T., Verrucchi P., Fubini A., Haas S. and Tognett V. (2005) {\it Phys. Rev. Lett.} 94 147208

\bibitem{Zhou2008} Zhou H-Q., Orus R. and Vidal G. (2008) {\it Phys. Rev. Lett.} 100 080601

\bibitem{Orus2009} Orus R., Doherty A. C. and Vidal G. (2009) {\it Phys. Rev. Lett.} 102 077203

\bibitem{Li2009} Li B., Li S-H. and Zhou H-Q. (2009) {\it Phys. Rev. E} 79 060101(R)

\bibitem{Murg2007} Murg V., Verstraete F. and Cirac J. I. (2007) {\it Phys. Rev. A} 75 033605

\bibitem{Affleck2009} Affleck I., Lafforencie N. and Sorensen E. S. (2009) {\it J. Phys. A} 42 504009

\bibitem{Andrei1980}  Andrei N. (1980) {\it Phys. Rev. Lett.} 45 379

\bibitem{Vigman1980} Vigman P. B. (1980) {\it JETP Lett.} 31 (7) 364

\bibitem{Sadiek2012}  Sadiek G., Xu Q. and Kais S. (2012) {\it Phys. Rev. A} 85 042313

\bibitem{Huang2006} Huang Z., Sadiek G. and Kais S. (2006) {\it J. Chem. Phys.} 124 144513

\bibitem{Apollaro2008} Apollaro T. J., Cuccoli A., Fubini A., Plastina F. and Verrucchi P. (2008) {\it Phys. Rev. A} 77 062314


\bibitem{Burkard1999} Burkard G. and Loss D. (1999) {\it Phys. Rev. B} 59 2070

\bibitem{Ashcroft} Ashcroft N. W. and Mermin N. D.  (1976) {\it Solid State Physics} (New York: Saunders College Publishing)

\bibitem{Osterloh2002} Osterloh A., Amico L., Falci G. and Fazio R. (2002) {\it Nature} 416 608

\bibitem{Xu2010} Xu Q., Kais S., Naumov M. and Sameh A. (2010) {\it Phys. Rev. A} 81 022324

\bibitem{Henkel1999} Henkel M. (1999) {\it Conformal Invariance and critical Phenomena} (Berlin: Springer)

\bibitem{Penson1982} Penson K., Jullien R. and Pfeuty P. (1982) {\it Phys. Rev. B} 25, 1837

\bibitem{Hamer2000} Hamer C.  (2000) {\it J. Phys. A} 33 6683


\bibitem{Kais2003} Kais S. and Serra P. (2003) {\it Advances in Chemical Physics} 125, pp. 1--99 (New York: John Wiley \& Sons Inc)

\bibitem{Kais2007} S. Kais (2007) {\it Advances in Chemical Physics} 134, pp. 493--535 (New York: John Wiley \& Sons Inc)
\end{thebibliography}

%%%%%%%%%%%%%%%%%%%%%%%%%%%%%%%%%%%%%%%%%%%%%%%%%%%%%%%%%%%%%%%%%%%%%%%%%%%%%%%%%%%%%%%%%%%%%
\end{document}